\documentclass[11pt]{article}
\newcounter{ourcount}
\usepackage{amsmath,amsfonts,amssymb,graphicx,epsfig,pdflscape,multirow,theorem,gensymb}
\usepackage[matrix,arrow,curve]{xy} 
\numberwithin{equation}{section}
\graphicspath{{./eps/}}
\usepackage[nosort]{cite}
\usepackage[usenames]{color}
\usepackage{pstricks}
\usepackage[backref=false]{hyperref}
\hypersetup{
colorlinks=true,
citecolor=red,
linkcolor=darkblue
}
\definecolor{darkblue}{rgb}{0,0,.8}
\definecolor{red}{rgb}{1,0,0}

\theorembodyfont{\itshape} 
\theoremheaderfont{\scshape}
\theoremstyle{plain}

\numberwithin{equation}{section}

\newcommand{\nc}{\newcommand}
\def\ch{\mbox{ch}}
\nc{\fh}{\hat{f}}
\nc{\muh}{\hat{\mu}}
\nc{\nuh}{\hat{\nu}}
\nc{\disp}{\displaystyle}
\nc{\cosec}{\mathop{\mbox{cosec}}}
\nc{\ir}{\mathrm{i}}
\def\Re{\mathop{\mbox{Re}}}

\nc{\bib}{\bibitem}
\nc{\al}{\alpha}
\nc{\g}{\gamma}
\nc{\G}{\Gamma}
\nc{\D}{\Delta}
\nc{\eps}{\epsilon}
\nc{\la}{\lambda}
\nc{\La}{\Lambda}
\nc{\var}{\varphi}
\nc{\pa}{\partial}
\nc{\nn}{\nonumber \\ }
\nc{\hf}{\frac{1}{2}}
\nc{\dz}{\frac{dz}{2\pi i}}
\nc{\bin}[2]{\left(\!\!\!\begin{array}{c} {#1}\\ {#2} \end{array}\!\!\!\right)}
\nc{\be}{\begin{equation}}
\nc{\ee}{\end{equation}}
\nc{\bea}{\begin{eqnarray}}
\nc{\eea}{\end{eqnarray}}
\nc{\bra}[1]{\langle {#1}|}
\nc{\ket}[1]{|{#1}\rangle}
\nc{\chit}{\raisebox{0.25ex}{$\chi$}}
\nc{\Dbh}{\mbox{\boldmath $\hat D$}}
\nc{\Dbb}{\mbox{\boldmath $\bar D$}}
\nc{\Dbm}{\mbox{\boldmath $\mathcal D$}}
\nc{\db}{\mbox{\boldmath $d$}}
\nc{\Ab}{\mbox{\boldmath $A$}}
\nc{\Bb}{\mbox{\boldmath $B$}}
\nc{\Cb}{\mbox{\boldmath $C$}}
\nc{\Db}{\mbox{\boldmath $D$}}
\nc{\eb}{\mbox{\boldmath $e$}}
\nc{\Fb}{\mbox{\boldmath $F$}}
\nc{\Fbt}{\mbox{\boldmath $\tilde{F}$}}
\nc{\fb}{\mbox{\boldmath $f$}}
\nc{\fbt}{\mbox{\boldmath $\tilde{f}$}}
\nc{\Gb}{\mbox{\boldmath $G$}}
\nc{\Hb}{\mbox{\boldmath $H$}}
\nc{\Jb}{\mbox{\boldmath $J$}}
\nc{\Kb}{\mbox{\boldmath $K$}}
\nc{\Mb}{\mbox{\boldmath $M$}}
\nc{\Pb}{\mbox{\boldmath $P$}}
\nc{\Qb}{\mbox{\boldmath $Q$}}
\nc{\Tb}{\mbox{\boldmath $T$}}
\nc{\Tbb}{\mbox{\boldmath $\bar T$}}
\nc{\Tbm}{\mbox{\boldmath $\mathcal T$}}
\nc{\tb}{\mbox{\boldmath $t$}}
\nc{\Ub}{\mbox{\boldmath $U$}}
\nc{\Vb}{\mbox{\boldmath $V$}}
\nc{\Wb}{\mbox{\boldmath $W$}}
\nc{\Xb}{\mbox{\boldmath $X$}}
\nc{\yb}{\mbox{\boldmath $y$}}
\nc{\Zb}{\mbox{\boldmath $Z$}}
\nc{\Hc}{{\cal H}}
\nc{\Rc}{{\cal R}}
\nc{\Lc}{{\cal L}}
\nc{\Vc}{{\cal V}}
\nc{\Ib}{\mbox{\boldmath $I$}}
\nc{\qb}{\bar{q}}
\nc{\oN}{\mathbb{N}}
\nc{\oZ}{\mathbb{Z}}
\nc{\oR}{\mathbb{R}}
\newrgbcolor{darkgreen}{0., 0.733333, 0.0621042}
\def\vvdots{\mathinner{\mkern1mu\raise1pt\vbox{\kern7pt\hbox{.}}\mkern2mu
  \raise4pt\hbox{.}\mkern2mu\raise7pt\hbox{.}\mkern1mu}}
\nc{\gauss}[2]{\left[\!\!\begin{array}{c} {#1}\\ {#2} \end{array}\!\!\right]}
\nc{\sbin}[2]{\left\{\!\!\!\begin{array}{c} {#1}\\ {#2} 
\end{array}\!\!\!\right\}}
\nc{\sbinlr}[2]{\Big\langle\!\!\begin{array}{c} {#1}\\ {#2} 
\end{array}\!\!\Big\rangle}
\nc{\bino}[2]{\left(\!\!\begin{array}{c} {#1}\\ {#2} \end{array}\!\!\right)}
\def\half {\mbox{$\textstyle \frac{1}{2}$}}
\def\vec#1{\mbox {\boldmath $#1$}}

\definecolor{lightblue}{rgb}{.7,.7,1}
\definecolor{lightestblue}{rgb}{.95,.95,1}
\definecolor{lightlightblue}{rgb}{.85,.85,1}
\definecolor{midblue}{rgb}{.7,.7,1}

\def\loopa{
\psframe[linewidth=.25pt](0,0)(1,1)
\psarc[linewidth=1.5pt,linecolor=blue](1,0){.5}{90}{180}
\psarc[linewidth=1.5pt,linecolor=blue](0,1){.5}{-90}{0}
}
\def\loopb{
\psframe[linewidth=.25pt](0,0)(1,1)
\psarc[linewidth=1.5pt,linecolor=blue](0,0){.5}{0}{90}
\psarc[linewidth=1.5pt,linecolor=blue](1,1){.5}{180}{270}
}

\def\facegrid#1#2{
\psframe[fillstyle=solid,fillcolor=lightlightblue,linewidth=0pt]#1#2
\psgrid[gridlabels=0pt,subgriddiv=1]#1#2}

\def\bfloor#1{{\big\lfloor #1\big\rfloor}}

\newcommand{\statei}{{\ket{-\frac{3}{32}}}\ar@{};[0,0];}
\newcommand{\stateii}{{\ket{\frac{5}{32}}}\ar@{};[0,0];}
\newcommand{\stateiii}{{\ket{\frac{21}{32}}}\ar@{};[0,0];}
\newcommand{\stateiv}{{\ket{\frac{45}{32}}}\ar@{};[0,0];}
\newcommand{\statev}{{\ket{\frac{77}{32}}}\ar@{};[0,0];}
\newcommand{\statevi}{{\ket{\frac{117}{32}}}\ar@{};[0,0];}
\newcommand{\stated}{{\dots}\ar@{};[0,0];}
\newcommand{\statee}{{}\ar@{};[0,0];}

 \def\Re{{\rm Re ~}}

 \def\a{\alpha}
 \def\b{\beta}
 \def\g{\gamma}

 \def\eps{\varepsilon}
 \def\th{\theta}
 
 \def\k{\kappa}
 \def\l{\lambda}

 \def\th{\theta}

 \def\G{\Gamma}
 \def\D{\Delta}

 \def\calE{\mathcal{E}}

\newcommand{\superp}[2]{\genfrac{}{}{0pt}{}{#1}{#2}}

\def \Monoid{
\rput(0,0){\MonoidCap}
\rput(0,1){\MonoidCup}
}

\def \MonoidCap{
\psbezier[linewidth=1.5pt,linecolor=blue](0,0)(0,.55)(1,.55)(1,0)
}

\def \MonoidCup{
\psbezier[linewidth=1.5pt,linecolor=blue](0,0)(0,-.55)(1,-.55)(1,0)
}

\nc\drtm{{\vec D}}      		
\nc\face{\mathbb{X}} 		
\nc\faceK{\mathbb{K}} 		
\nc\genface[1]{\mathbb{X}^{(#1)}} 
\nc \ham{{\mathcal H}}			

\nc \nface{\mathbb{\hat X}}	

\nc\TLw{\omega} 			
\nc\TLb{\beta} 				

\nc\BMWw{\omega_2}		
\nc\BMWb{\beta_2}			

\nc\fTLw{\omega_2} 			
\nc\fTLb{\beta_2} 			

\nc{\genw}[1]{\omega_{#1}}  	
\nc{\Genw}[2]{\omega_{#1}^{(#2)}}
\nc{\genb}[1]{\beta_{#1}}       	

\def \trinomial[#1][#2][#3][#4]{\left[{#1\atop #2,#3,#4}\right]}

\def \superTrinomial[#1][#2]{
\left({#1 \atop #2} \right)_{\!2}
}

\def \qTrinomial[#1][#2][#3][#4]{\left[{#1\atop #2,#3,#4}\right]_{\!q}}

\def\Re{\mathop{\mbox{Re}}}

\nc{\XiShift}{\overline{\xi}_\rho}

\nc{\fws}{\small}

\def\dddots{\mathinner{\mkern1mu\raise9pt\vbox{\kern7pt\hbox{.}}\mkern2mu
  \raise5pt\hbox{.}\mkern2mu\raise1pt\hbox{.}\mkern1mu}}

\nc{\smbin}[2]{\Big(\!\!\!\begin{array}{c} {#1}\\[-3pt] {#2} \end{array}\!\!\!\Big)}
  
\begin{document}

\topmargin -5mm
\oddsidemargin 5mm

\vspace*{-2cm}

\setcounter{page}{1}

\vspace{2mm}
\begin{center}
{\huge {\bf  Logarithmic Minimal Models\\[6pt] with Robin Boundary Conditions}}

\vspace{10mm}
{\Large Jean-Emile Bourgine$^\dagger$, Paul A. Pearce$^\ast$, Elena Tartaglia$^\ast$}
\\[.4cm]
{\em {}$^\dagger$INFN Bologna, Universit\`a di Bologna}\\
{\em Via Irnerio 46, 40126 Bologna, Italy}\\
{\it on leave from}\\
{\em Asia Pacific Center for Theoretical Physics (APCTP)}\\
{\em Pohang, Gyeongbuk 790-784, Republic of Korea}
\\[.4cm]
{\em {}$^\ast$School of Mathematics and Statistics, University of Melbourne}\\
{\em Parkville, Victoria 3010, Australia}
\\[.4cm]
{\tt bourgine\,@\,bo.infn.it,\quad \tt p.pearce\,@\,ms.unimelb.edu.au,\quad \tt elena.tartaglia\,@\,unimelb.edu.au}
\end{center}


\vspace{8mm}
\centerline{{\bf{Abstract}}}
\vskip.4cm
\noindent 
We consider general logarithmic minimal models ${\cal LM}(p,p')$, with $p,p'$ coprime, on a strip of $N$ columns with the $(r,s)$ Robin boundary conditions introduced by Pearce, Rasmussen and Tipunin. 
On the lattice, these models are Yang-Baxter integrable loop models that are described algebraically by the one-boundary Temperley-Lieb algebra. The $(r,s)$ Robin boundary conditions are a class of integrable boundary conditions satisfying the boundary Yang-Baxter equations which allow loop segments to either reflect or terminate on the boundary. The associated conformal boundary conditions are organized into infinitely extended Kac tables labelled by the Kac labels $r\in{\Bbb Z}$ and $s\in{\Bbb N}$. The Robin vacuum boundary condition, labelled by $(r,s\!-\!\frac{1}{2})=(0,\half)$, is given as a linear combination of Neumann and Dirichlet boundary conditions. 
The general $(r,s)$ Robin boundary conditions are constructed, using fusion, by acting on the Robin vacuum boundary with an $(r,s)$-type seam consisting of an $r$-type seam of width $w$ columns and an $s$-type seam of width $d=s-1$ columns. 
The $r$-type seam admits an arbitrary boundary field which we fix to the special value $\xi=-\tfrac{\lambda}{2}$ where $\lambda=\frac{(p'-p)\pi}{p'}$ is the crossing parameter. The $s$-type boundary introduces $d$ defects into the bulk. We consider the commuting  double-row transfer matrices and their associated quantum Hamiltonians and calculate analytically the boundary free energies of the $(r,s)$ Robin boundary conditions. Using finite-size corrections and sequence extrapolation out to system sizes $N+w+d\le 26$, the conformal spectrum of boundary operators is accessible by numerical diagonalization of the Hamiltonians. Fixing the parity of $N$ for $r\ne 0$ and restricting to the ground state sequences $w=\bfloor{\frac{|r|p'}{p}}$, $r\in{\Bbb Z}$ with the inverse $r=(-1)^{N+w+d}\big\lceil \frac{p w}{p'}\big\rceil$, we find that the conformal weights take the values $\Delta^{p,p'}_{r,s-\frac12}$ where $\Delta^{p,p'}_{r,s}$ is given by the usual Kac formula. The $(r,s)$ Robin boundary conditions are thus conjugate to scaling operators with half-integer values for the Kac label $s-\half$. Level degeneracies support the conjecture that the characters of the associated (reducible or irreducible) representations are given by Virasoro Verma characters.

\vspace{.5cm}
\noindent\textbf{Keywords:} Exactly solvable models, logarithmic conformal field theory, loop models\\

\newpage
\tableofcontents

\newpage
\hyphenpenalty=30000

\setcounter{footnote}{0}

\section{Introduction}

It is well established that,  in the continuum scaling limit, Conformal Field Theories (CFTs) such as the family of minimal models ${\cal M}(m,m')$~\cite{BPZ} describe the universal scaling properties of two-dimensional lattice models. In the case of unitary~\cite{FQS} minimal models, $m=m'-1$, the associated lattice models are the multicritical Ising models~\cite{Huse}. The simplest CFTs fall into the class of rational CFTs~\cite{MooreSeiberg,FMS} which are characterized by a finite number of primary scaling operators. The representations associated with these operators are irreducible and close among themselves under fusion. For rational boundary CFTs~\cite{BPPZ}, the conformal boundary conditions are conjugate (in one-to-one correspondence) to the primary scaling operators. It follows that the conformal data, including the central charges and conformal weights, can be obtained from the lattice by studying the finite-size corrections~\cite{BCN,Aff} to the eigenvalues of the transfer matrices, or equivalently, the associated one-dimensional quantum Hamiltonians. 

In fact, many CFTs are realized as the continuum scaling limit of Yang-Baxter integrable lattice models~\cite{BaxterBook}. For example, the minimal models describe the continuum scaling limit of the Restricted Solid-On-Solid (RSOS) lattice models~\cite{RSOS}. A boundary condition on the lattice is integrable if it satisfies the boundary Yang-Baxter equation~\cite{Cherednik,Sklyanin,BPO}. Invariably, if a rational CFT is associated with a lattice model that is Yang-Baxter integrable in the bulk, it seems possible to construct a representative integrable lattice boundary condition to realize each of the conformal boundary conditions in the continuum scaling limit. This program has been carried to completion~\cite{BehrendP} for the critical $A$-$D$-$E$\/ RSOS models~\cite{Pasquier} associated with the $A$-$D$-$E$ minimal models~\cite{CIZ}. 
The RSOS models have degrees of freedom in the form of local heights. The continuum scaling limit of such theories with local degrees of freedom are described by rational CFTs. 

In this paper, we study the logarithmic minimal models 
${\cal LM}(p,p')$~\cite{PRZ2006} with Virasoro conformal symmetry (as opposed to ${\cal W}$-extended conformal symmetry~\cite{FGST}). The first members of this family are critical dense polymers ${\cal LM}(1,2)$~\cite{PRpolymers} and the loop version ${\cal LM}(2,3)$~\cite{TL} of the square lattice bond percolation model. 
Like the minimal models~\cite{GKO}, the logarithmic minimal models are coset CFTs~\cite{PR2011,PRcoset}. 
But unlike the minimal models, these theories describe systems with nonlocal degrees of freedom in the form of loop segments. The continuum scaling limit of such theories with nonlocal degrees of freedom are described by logarithmic CFTs~\cite{specialissue} that are nonunitary and non-rational.  The family of logarithmic minimal models plays the same role for logarithmic CFTs that the minimal models play for rational CFTs in the sense that they provide prototypical examples of logarithmic CFTs. The properties of logarithmic CFTs, however, are profoundly different to those of rational CFTs. Most importantly, these theories are characterized~\cite{Gurarie} by the existence of reducible yet indecomposable representations but, as yet, there is no exhaustive classification of all possible representations.  Instead, the focus has been on identifying different kinds of Virasoro representations such as irreducible, fully reducible, projective and their ${\cal W}$-algebra counterparts. In the logarithmic context, it is no longer true in general that each representation is conjugate to a conformal boundary condition.  For the logarithmic minimal models, however, it is known~\cite{PRZ2006,RasKac,PRV,PTC,logKac} that there are a countably infinite number of Virasoro Kac representations with conjugate $(r,s)$ boundary conditions organized into infinitely extended Kac tables. The central charges and conformal weights, given by the Kac formula, are
\be
c=c^{p,p'}=1-\frac{6(p'-p)^2}{pp'},\qquad \Delta_{r,s}^{p,p'}=\frac{(rp'-sp)^2-(p'-p)^2}{4pp'},\qquad 1\le p<p',\qquad r,s\in \oN
\label{KacFormula}
\ee
All of these models are nonunitary with effective central charge $c_{\text{eff}}=1$.

The conformal properties of polymers and percolation have been studied~\cite{Saleur87a,Duplantier86,Saleur87b,SaleurHalfInt87} since the late eighties. 
Remarkably, there are indications~\cite{Duplantier86,SaleurHalfInt87} that certain representations occur with conformal weights given by the Kac formula (\ref{KacFormula}) with half-integer Kac labels. 
Indeed, it has been suggested~\cite{Rid0808} that a field with conformal weight $\Delta^{2,3}_{2,\frac{5}{2}}=0$ plays the role of Watts'~\cite{Watts} primary field in the description of critical percolation. 
Moreover, the existence of a family of spin fields with conformal weights $\Delta^{p,p+1}_{r-\frac{1}{2},0}$ for $r\in\oN$ has recently been posited~\cite{Delfino}. We restrict $r$ and $s$ to be integers and use $r,s$ as integer Kac labels and $r-\half,s-\half$ as half-integer Kac labels throughout this paper.

Robin boundary conditions~\cite{Robin} are linear combinations of Neumann and Dirichlet boundary conditions. In the context of loop models, they allow loop segments to either reflect or terminate at the boundary. 
In this paper, we apply the $(r,s)$ Robin boundary conditions of Pearce, Rasmussen and Tipunin~\cite{PRT14} to the general minimal models and confirm numerically that the associated conformal weights are $\Delta^{p,p'}_{r,s-\frac{1}{2}}$ where $r\in{\Bbb Z}$ and $s\in {\Bbb N}$. The $(r,s)$ Robin boundary conditions are thus conjugate to scaling operators with half-integer values for the Kac label $s-\half$. In particular, the Robin vacuum $(r,s\!-\!\half)=(0,\half)$ has conformal weight
\bea
\Delta^{p,p'}_{0,\frac{1}{2}}=-\frac{(2p'-p)(2p'-3p)}{16p p'},\qquad 1\le p<p',\qquad p,p'\in \oN
\eea
These results were established analytically in \cite{PRT14} for critical dense polymers ${\cal LM}(1,2)$ with $\beta=0$. 
The $(r,s)$ Robin boundary conditions are so-named because of the $(r,s)$-type seam by which they are constructed.
The $(r,s)$ Robin boundary conditions have similarities to the so-called JS boundary conditions~\cite{JSbdy}. The main differences are that the $(r,s)$ Robin boundary conditions are (i) manifestly Yang-Baxter integrable and (ii) they are  constructed using  standard $r$- and $s$-type seams, which behave as topological defects. These seams propagate freely along the row due to the generalized Yang-Baxter equation and ensure the expected $su(2)$ fusion rules for Cardy fusion~\cite{Cardy1989} of boundary conditions on the lattice.

The layout of this paper is as follows. In Section~2, we describe the logarithmic minimal lattice models on the strip in terms of the one-boundary Temperley-Lieb algebra. Following \cite{PRT14}, we recall the definition of the Robin link states and the construction of the integrable $(r,s)$ Robin boundary conditions. The commuting double-row transfer matrices and their associated quantum Hamiltonians are revisited in Section~3. The analytic derivation of the exact bulk and boundary free energies and their Hamiltonian limits are also given in this section. The details of the derivation and solution of the inversion relation are relegated to Appendices~A and B. Our numerical results for the finite-size spectra are presented in Section~4 along with a discussion of the finitized conformal partition functions and the logarithmic limit. Properties of the Robin boundaries as representations of the one-boundary Temperley-Lieb algebra are discussed in Appendix~C. Some final remarks are given in the conclusion.

\section{Logarithmic Minimal Models and Robin Boundary Conditions}
\subsection{Temperley-Lieb algebra and local relations}
We study the logarithmic minimal models ${\cal LM}(p,p')$~\cite{PRZ2006} on a square lattice with the geometry of the strip and apply Robin boundary conditions. Here $p, p'$ are coprime integers with $1\leq p<p'$. The models are built using the one-boundary Temperley-Lieb (TL) or blob algebra~\cite{TL,MaSa93,MaWood2000,NRG2005,Nichols2006a,Nichols2006b} and describe the statistical interaction of densely packed self- and mutually-avoiding loops. The strip has $N$ bulk columns and $2M$ rows and is built by $M$ applications of the double-row transfer matrix. Closed loops in the bulk have a fugacity 
$\beta=2\cos\lambda$ where the crossing parameter is
\begin{equation}\label{def_l}
\lambda=\dfrac{(p'-p)\pi}{p'},\qquad 1\leq p<p',\quad p,p'\in \oN
\end{equation}

Loop configurations on the square lattice are built on two face tiles which are the generators of the planar TL algebra~\cite{Jones}
\be
\psset{unit=.77cm}
\begin{pspicture}[shift=-.45](1,1)
\facegrid{(0,0)}{(1,1)}
\rput[bl](0,0){\loopa}
\end{pspicture}
\quad\text{and}\quad
\begin{pspicture}[shift=-.42](1,1)
\facegrid{(0,0)}{(1,1)}
\rput[bl](0,0){\loopb}
\end{pspicture}
\label{tiles}
\ee
These tiles, corresponding to the identity $I$ and monoids $e_j$ respectively, can be viewed as operators acting from the bottom-left edges to the top-right edges of an elementary face. Together they generate the linear TL algebra 
\be
{TL_N(\beta):=\big\langle I,\, e_j;\ j=1,\ldots, N-1\big\rangle}
\ee
satisfying the relations
\begin{equation}
e_je_{j\pm1}e_j=e_j,\qquad e_j^2=\b e_j,\qquad e_je_k=e_ke_j,\quad |j-k|\geq2
\end{equation}
The diagrammatic action of the tiles on a set of $N$ parallel strings, where $e_j$ acts between string $j$ and $j+1$, gives a faithful representation of the algebra
\be
I = \;
\psset{unit=.7cm}
\begin{pspicture}[shift=-.75](0,-.45)(3.5,1)
{\scriptsize
\rput[B](0,-.5){1}\rput[B](1,-.5){2}\rput[B](2.5,-.5){$N\!-\!1$}\rput[B](3.5,-.5){$N$}}
\rput(1.75,.5){\ldots}
\multirput(0,0)(2.5,0){2}{\multirput(0,0)(1,0){2}{\psline[linewidth=1.5pt,linecolor=blue](0,0)(0,1)}}
\end{pspicture}\qquad\qquad
e_j=\;
\begin{pspicture}[shift=-.75](0,-.45)(8,1)
{\scriptsize
\rput[B](0,-.5){1}\rput[B](1,-.5){2}\rput[B](3.5,-.5){$j$}\rput[B](4.5,-.5){$j\!+\!1$}\rput[B](7,-.5){$N\!-\!1$}\rput[B](8,-.5){$N$}}
\multirput(1.75,.5)(4.5,0){2}{\ldots}
\multirput(0,0)(7,0){2}{\multirput(0,0)(1,0){2}{\psline[linewidth=1.5pt,linecolor=blue](0,0)(0,1)}}
\multirput(2.5,0)(3,0){2}{\psline[linewidth=1.5pt,linecolor=blue](0,0)(0,1)}
\rput(3.5,0){\Monoid}
\end{pspicture}
\ee

To build commuting transfer matrices~\cite{PRZ2006}, we introduce a spectral parameter $u$ related to spatial anisotropy~\cite{KimP}. Face operators are defined as the linear combinations 
\be
\psset{unit=.77cm}
\begin{pspicture}[shift=-.42](1,1)
\facegrid{(0,0)}{(1,1)}
\psarc[linewidth=0.025,linecolor=red]{-}(0,0){0.16}{0}{90}
\rput(.5,.5){$_u$}
\end{pspicture}
\, :=\ 
s_1(-u)\ \begin{pspicture}[shift=-.42](1,1)
\facegrid{(0,0)}{(1,1)}
\rput[bl](0,0){\loopa}
\end{pspicture}
\;+\,s_0(u)\
\begin{pspicture}[shift=-.45](1,1)
\facegrid{(0,0)}{(1,1)}
\rput[bl](0,0){\loopb}
\end{pspicture}\;,\qquad\quad X_j(u)=s_1(-u)I+s_0(u) e_j
\label{1x1}
\ee
where
\be 
s(u)=s_0(u),\qquad s_k(u):=\frac{\sin (u+k\lambda)}{\sin\lambda},\qquad k\in\oZ
\label{sk}
\ee
and the marked corner fixes the orientation of the face.
The face operators satisfy the Yang-Baxter equation
\begin{equation}
X_j(v)X_{j+1}(u+v)X_j(u)=X_{j+1}(u)X_j(u+v)X_{j+1}(v)
\end{equation}
or diagrammatically
\be
\begin{array}{rcl}
\psset{unit=.36cm}
\begin{pspicture}[shift=-1.8](0,0)(7,3.8)
\pspolygon[linewidth=1pt,linecolor=black,fillstyle=solid,fillcolor=lightlightblue](0,2)(2,0)(5,0)(7,2)(5,4)(2,4)(0,2)
\pspolygon[linewidth=1pt,linecolor=black,fillstyle=solid,fillcolor=lightlightblue](0,2)(3,2)(5,0)(7,2)(5,4)(3,2)
\psarc[linewidth=.75pt,linecolor=red](2,0){.3}{0}{135}
\psarc[linewidth=.75pt,linecolor=red](0,2){.5}{0}{45}
\psarc[linewidth=.75pt,linecolor=red](3,2){.4}{-45}{45}
\rput(2.5,1){\small $u$}
\rput(2.5,3){\small$v$}
\rput(5,2){\small $v\!-\!u$}
\end{pspicture}\ &=&
\psset{unit=.375cm}
\begin{pspicture}[shift=-1.8](0,0)(7,3.8)
\pspolygon[linewidth=1pt,linecolor=black,fillstyle=solid,fillcolor=lightlightblue](0,2)(2,0)(5,0)(7,2)(5,4)(2,4)(0,2)
\pspolygon[linewidth=1pt,linecolor=black,fillstyle=solid,fillcolor=lightlightblue](7,2)(4,2)(2,0)(0,2)(2,4)(4,2)
\rput(4.5,1){\small $v$}
\rput(4.5,3){\small $u$}
\rput(2,2){\small $v\!-\!u$}
\psarc[linewidth=.75pt,linecolor=red](0,2){.4}{-45}{45}
\psarc[linewidth=.75pt,linecolor=red](2,0){.5}{0}{45}
\psarc[linewidth=.75pt,linecolor=red](4,2){.3}{0}{135}
\end{pspicture}
\end{array}
\label{YBE}
\ee
The face operators $X_j$ also satisfy the local inversion and crossing relations
\be
\psset{unit=0.65cm}
\begin{pspicture}[shift=-1.13](-.5,0.75)(4,3.25)
\pspolygon[fillstyle=solid,fillcolor=lightlightblue](0,2)(1,1)(2,2)(1,3)(0,2)
\pspolygon[fillstyle=solid,fillcolor=lightlightblue](2,2)(3,1)(4,2)(3,3)(2,2)
\psarc[linewidth=.75pt,linecolor=red](0,2){.2}{-45}{45}
\psarc[linewidth=.75pt,linecolor=red](2,2){.2}{-45}{45}
\psarc[linecolor=blue,linewidth=1.5pt](2,2){.7}{45}{135}
\psarc[linecolor=blue,linewidth=1.5pt](2,2){.7}{-135}{-45}
\rput(1,2){\small $u$}
\rput(3,2){\small $\!-u$}
\end{pspicture}
=s(\lambda-u)s(\lambda+u)\
\begin{pspicture}[shift=-1.13](1,0.75)(3.2,3.25)
\pspolygon[fillstyle=solid,fillcolor=lightlightblue](1,2)(2,1)(3,2)(2,3)(1,2)
\psarc[linecolor=blue,linewidth=1.5pt](2,1){.7}{45}{135}
\psarc[linecolor=blue,linewidth=1.5pt](2,3){.7}{-135}{-45}
\end{pspicture},\qquad \qquad 
\psset{unit=.95cm}
\begin{pspicture}[shift=-.42](1,1)
\facegrid{(0,0)}{(1,1)}
\rput(.5,.5){\small $u$}
\psarc[linewidth=.75pt,linecolor=red](0,0){.15}{0}{90}
\end{pspicture}\ =\ 
\begin{pspicture}[shift=-.42](1,1)
\facegrid{(0,0)}{(1,1)}
\rput(.5,.5){\small $\lambda\!-\!u$}
\psarc[linewidth=.75pt,linecolor=red](1,0){.15}{90}{180}
\end{pspicture}
\label{Invrel}
\ee

\subsection{Robin link states}
\label{Sec:LinkStates}

In this section we describe the link states associated with $(r,s)$ Robin boundary conditions. 
We denote the vector space of Robin link states by ${\cal V}_d^{(N,w)}$. 
A link state on $N$ bulk and $w$ boundary nodes is a planar diagram of non-crossing arc segments. The nodes are evenly spaced on a horizontal line and the arc segments lie above this line. 
A link state contains $d=s-1\ge 0$ {\em defects} (vertical line segments) attaching individual (bulk or boundary) nodes to a point above at infinity and 
$b\ge 0$ {\em boundary links} linking individual nodes to the right edge of the rectangle containing the links state. The remaining $\half(N+w-d-b)$ nodes 
are connected pairwise by half-arcs with
\be
w=\mbox{width of $r$-type seam},\qquad\quad  N+w-d-b=0\mod 2
\label{Ndb02}
\ee
A half-arc is an arc segment connecting two nodes.

 A {\em Robin link state} satisfies the three properties:\\[-20pt]
\begin{enumerate}
\item[(i)] no half-arc joins a pair of boundary nodes\\[-20pt]
\item[(ii)] no boundary link emanates from a boundary node\\[-20pt]
\item[(iii)] every boundary node is either a defect or is linked to a bulk node\\[-16pt]
\end{enumerate}
In the Robin vacuum sector with $w=d=0$, the vector space of link states with $N=4$ is
\bea
{\cal V}_0^{(4,0)}&\!\!=\!\!&\mbox{span}\,\Big\{
\begin{pspicture}[shift=-0.15](-0.0,-0.1)(1.7,0.8)
\psline[linewidth=0.5pt]{-}(0,0)(1.6,0)
\psarc[linewidth=1.5pt,linecolor=darkgreen](0.4,0){0.2}{0}{180}
\psarc[linewidth=1.5pt,linecolor=darkgreen](1.2,0){0.2}{0}{180}
\end{pspicture},
\begin{pspicture}[shift=-0.15](-0.0,-0.1)(1.7,0.8)
\psline[linewidth=0.5pt]{-}(0,0)(1.6,0)
\psarc[linewidth=1.5pt,linecolor=darkgreen](0.8,0){0.2}{0}{180}
\psarc[linewidth=1.5pt,linecolor=darkgreen](0.8,0){0.6}{0}{180}
\end{pspicture},
\begin{pspicture}[shift=-0.15](-0.0,-0.1)(1.7,0.8)
\psline[linewidth=0.5pt]{-}(0,0)(1.6,0)
\psarc[linewidth=1.5pt,linecolor=darkgreen](0.4,0){0.2}{0}{180}
\psarc[linewidth=1.5pt,linecolor=darkgreen](1.6,0){0.2}{90}{180}
\psbezier[linecolor=darkgreen,linewidth=1.5pt](1,0)(1.1,0.4)(1.6,0.4)(1.6,0.4)
\end{pspicture},
\begin{pspicture}[shift=-0.15](-0.0,-0.1)(1.7,0.8)
\psline[linewidth=0.5pt]{-}(0,0)(1.6,0)
\psarc[linewidth=1.5pt,linecolor=darkgreen](1.2,0){0.2}{0}{180}
\psbezier[linecolor=darkgreen,linewidth=1.5pt](0.2,0)(0.3,0.6)(0.8,0.6)(1.6,0.6)
\psbezier[linecolor=darkgreen,linewidth=1.5pt](0.6,0)(0.7,0.4)(1,0.4)(1.6,0.4)
\end{pspicture},
\begin{pspicture}[shift=-0.15](-0.0,-0.1)(1.7,0.8)
\psline[linewidth=0.5pt]{-}(0,0)(1.6,0)
\psarc[linewidth=1.5pt,linecolor=darkgreen](0.8,0){0.2}{0}{180}
\psbezier[linecolor=darkgreen,linewidth=1.5pt](0.2,0)(0.3,0.6)(0.8,0.6)(1.6,0.6)
\psarc[linewidth=1.5pt,linecolor=darkgreen](1.6,0){0.2}{90}{180}
\end{pspicture},
\begin{pspicture}[shift=-0.15](-0.0,-0.1)(1.7,0.8)
\psline[linewidth=0.5pt]{-}(0,0)(1.6,0)
\psarc[linewidth=1.5pt,linecolor=darkgreen](1.6,0){0.2}{90}{180}
\psbezier[linecolor=darkgreen,linewidth=1.5pt](0.2,0)(0.3,0.8)(0.8,0.8)(1.6,0.8)
\psbezier[linecolor=darkgreen,linewidth=1.5pt](0.6,0)(0.7,0.6)(1.2,0.6)(1.6,0.6)
\psbezier[linecolor=darkgreen,linewidth=1.5pt](1,0)(1.1,0.4)(1.6,0.4)(1.6,0.4)
\end{pspicture} \Big\}\ \ \ \ 
\eea
The number of these link states is $ \dim {\cal V}_0^{(N,0)}=\big(\superp{N}{\lfloor N/2\rfloor}\big)$. Examples of vector spaces of Robin link states with defects are
\bea
{\cal V}_1^{(3,1)}&\!\!=\!\!&\mbox{span}\,\Big\{\ \
\begin{pspicture}[shift=-0.2](0,0)(1.5,.7)
\psline[linewidth=0.5pt](-0.2,0)(1.4,0)
\psline[linewidth=0.5pt,linestyle=dashed, dash=1.5pt 1.5pt](1,-0.1)(1,0.8)
\psline[linewidth=1.5pt,linecolor=darkgreen](0,0)(0,0.6)
\psbezier[linecolor=darkgreen,linewidth=1.5pt](0.4,0)(0.45,0.5)(0.9,0.45)(1.4,0.45)
\psarc[linewidth=1.5pt,linecolor=darkgreen](1,0){0.2}{0}{180}
\end{pspicture}
 \Big\}
 =\mbox{span}\,\Big\{\ \
\begin{pspicture}[shift=-0.2](-.4,0)(1.5,.7)
\psline[linewidth=0.5pt](-0.6,0)(1.4,0)
\psline[linewidth=0.5pt,linestyle=dashed, dash=1.5pt 1.5pt](-.2,-0.1)(-.2,0.8)
\psline[linewidth=0.5pt,linestyle=dashed, dash=1.5pt 1.5pt](1,-0.1)(1,0.8)
\psarc[linewidth=1.5pt,linecolor=darkgreen](-.2,0){0.2}{0}{180}
\psbezier[linecolor=darkgreen,linewidth=1.5pt](0.4,0)(0.45,0.5)(0.9,0.45)(1.4,0.45)
\psarc[linewidth=1.5pt,linecolor=darkgreen](1,0){0.2}{0}{180}
\end{pspicture}
 \Big\}
 =\mbox{span}\,\Big\{\ \
\begin{pspicture}[shift=-0.2](0,0)(1.9,.7)
\psline[linewidth=0.5pt](-0.2,0)(1.8,0)
\psline[linewidth=0.5pt,linestyle=dashed, dash=1.5pt 1.5pt](1,-0.1)(1,0.8)
\psline[linewidth=0.5pt,linestyle=dashed, dash=1.5pt 1.5pt](1.4,-0.1)(1.4,0.8)
\psarc[linewidth=1.5pt,linecolor=darkgreen](0.8,0){.8}{0}{180}
\psbezier[linecolor=white,linewidth=4.5pt](0.4,0)(0.45,0.5)(0.9,0.45)(1.8,0.45)
\psbezier[linecolor=darkgreen,linewidth=1.5pt](0.4,0)(0.45,0.5)(0.9,0.45)(1.8,0.45)
\psarc[linewidth=1.5pt,linecolor=darkgreen](1,0){0.2}{0}{180}
\end{pspicture}
 \Big\}
 \nn
 {\cal V}_1^{(3,2)}&\!\!=\!\!&\mbox{span}\,\Big\{\ \
\begin{pspicture}[shift=-0.2](0,0)(1.9,1)
\psline[linewidth=0.5pt](-0.2,0)(1.8,0)
\psline[linewidth=0.5pt,linestyle=dashed, dash=1.5pt 1.5pt](1,-0.1)(1,0.8)
\psline[linewidth=1.5pt,linecolor=darkgreen](0,0)(0,0.6)
\psbezier[linecolor=darkgreen,linewidth=1.5pt](0.4,0)(0.45,0.5)(1.55,0.5)(1.6,0)
\psarc[linewidth=1.5pt,linecolor=darkgreen](1,0){0.2}{0}{180}
\rput(2,0){,}
\end{pspicture}
\qquad
\begin{pspicture}[shift=-0.2](0,0)(1.9,1)
\psline[linewidth=0.5pt](-0.2,0)(1.8,0)
\psline[linewidth=0.5pt,linestyle=dashed, dash=1.5pt 1.5pt](1,-0.1)(1,0.8)
\psarc[linewidth=1.5pt,linecolor=darkgreen](0.2,0){0.2}{0}{180}
\psarc[linewidth=1.5pt,linecolor=darkgreen](1,0){0.2}{0}{180}
\psline[linewidth=1.5pt,linecolor=darkgreen](1.6,0)(1.6,0.6)
\rput(2,0){,}
\end{pspicture}
\qquad
\begin{pspicture}[shift=-0.2](0,0)(1.9,1)
\psline[linewidth=0.5pt](-0.2,0)(1.8,0)
\psline[linewidth=0.5pt,linestyle=dashed, dash=1.5pt 1.5pt](1,-0.1)(1,0.8)
\psbezier[linecolor=darkgreen,linewidth=1.5pt](0,0)(0.05,0.5)(1.15,0.5)(1.2,0)
\psarc[linewidth=1.5pt,linecolor=darkgreen](0.6,0){0.2}{0}{180}
\psline[linewidth=1.5pt,linecolor=darkgreen](1.6,0)(1.6,0.6)
\end{pspicture}
\Big\} 
\label{LinkStates}
\\
 {\cal V}_0^{(4,2)}&\!\!=\!\!&\mbox{span}\,\Big\{\ \
\begin{pspicture}[shift=-0.2](0,0)(2.3,1)
\psline[linewidth=0.5pt](-0.2,0)(2.2,0)
\psline[linewidth=0.5pt,linestyle=dashed, dash=1.5pt 1.5pt](1.4,-0.1)(1.4,0.8)
\psarc[linewidth=1.5pt,linecolor=darkgreen](0.2,0){0.2}{0}{180}
\psbezier[linecolor=darkgreen,linewidth=1.5pt](0.8,0)(0.85,0.5)(1.95,0.5)(2,0)
\psarc[linewidth=1.5pt,linecolor=darkgreen](1.4,0){0.2}{0}{180}
\rput(2.4,0){,}
\end{pspicture}
\qquad
\begin{pspicture}[shift=-0.2](0,0)(2.3,1)
\psline[linewidth=0.5pt](-0.2,0)(2.2,0)
\psline[linewidth=0.5pt,linestyle=dashed, dash=1.5pt 1.5pt](1.4,-0.1)(1.4,0.8)
\psbezier[linecolor=darkgreen,linewidth=1.5pt](0,0)(0.05,0.7)(1,0.75)(2.2,0.75)
\psbezier[linecolor=darkgreen,linewidth=1.5pt](0.4,0)(0.45,0.5)(0.8,0.55)(2.2,0.55)
\psbezier[linecolor=darkgreen,linewidth=1.5pt](0.8,0)(0.85,0.5)(1.95,0.5)(2,0)
\psarc[linewidth=1.5pt,linecolor=darkgreen](1.4,0){0.2}{0}{180}
\rput(2.4,0){,}
\end{pspicture}
\qquad
\begin{pspicture}[shift=-0.2](0,0)(2.3,1)
\psline[linewidth=0.5pt](-0.2,0)(2.2,0)
\psline[linewidth=0.5pt,linestyle=dashed, dash=1.5pt 1.5pt](1.4,-0.1)(1.4,0.8)
\psbezier[linecolor=darkgreen,linewidth=1.5pt](0,0)(0.1,0.75)(1.9,0.75)(2,0)
\psarc[linewidth=1.5pt,linecolor=darkgreen](0.6,0){0.2}{0}{180}
\psarc[linewidth=1.5pt,linecolor=darkgreen](1.4,0){0.2}{0}{180}
\rput(2.4,0){,}
\end{pspicture}
\qquad
\begin{pspicture}[shift=-0.2](0,0)(2.3,1)
\psline[linewidth=0.5pt](-0.2,0)(2.2,0)
\psline[linewidth=0.5pt,linestyle=dashed, dash=1.5pt 1.5pt](1.4,-0.1)(1.4,0.8)
\psbezier[linecolor=darkgreen,linewidth=1.5pt](0,0)(0.1,0.75)(1.9,0.75)(2,0)
\psbezier[linecolor=darkgreen,linewidth=1.5pt](0.4,0)(0.45,0.5)(1.55,0.5)(1.6,0)
\psarc[linewidth=1.5pt,linecolor=darkgreen](1,0){0.2}{0}{180}
\end{pspicture}
\Big\}
\nonumber
\eea
where the dashed lines separate the bulk and the $r$- and $s$-type seams on the left or right boundary. The number of these link states is
\be
 \dim {\cal V}_d^{(N,w)}=\begin{pmatrix}N \\ \left\lfloor\frac{N-d}{2}\right\rfloor+(-1)^{N-d-w}\left\lceil\frac{w}{2}\right\rceil\end{pmatrix}\label{countstates}
\ee
By adding an additional $d$ nodes on the left, the defects can be closed to an $s$-type seam on the left. Equivalently, by adding $d$ nodes on the right, the defects can be closed to an $s$-type seam on the right by passing under any boundary links at the cost of introducing braided crossings and losing the planarity of the link states. The link states for ${\cal V}_1^{(3,1)}$ with the defect closed on the left and the right respectively are shown in (\ref{LinkStates}). The latter is used in the construction of Robin boundaries using $(r,s)$-type seams but the former implementation is convenient in our numerical calculations. 
The dimension of the vector space ${\cal V}^{(N,w)}$ of Robin link states with an arbitrary number of defects is {\em independent} of $w$
\be
 \dim {\cal V}^{(N,w)}=\sum_{d=0}^{N+w}\dim{\cal V}_d^{(N,w)}=2^N
\label{dimVNw}
\ee

\subsection{One-boundary Temperley-Lieb algebra}

The one-boundary TL or blob algebra~\cite{MaSa93,MaWood2000,NRG2005,Nichols2006a,Nichols2006b} is a planar diagrammatic algebra generated by two bulk face tiles and one boundary triangle
\bea
\psset{unit=.9cm}
\begin{pspicture}[shift=-.42](1,1)
\facegrid{(0,0)}{(1,1)}
\rput[bl](0,0){\loopa}
\end{pspicture}
\qquad \qquad
\begin{pspicture}[shift=-.42](1,1)
\facegrid{(0,0)}{(1,1)}
\rput[bl](0,0){\loopb}
\end{pspicture}\qquad\qquad
\psset{unit=.7cm}
\begin{pspicture}[shift=-0.89](0,0)(1,2)
\pspolygon[fillstyle=solid,fillcolor=lightlightblue](0,1)(1,2)(1,0)(0,1)
\psline[linecolor=blue,linewidth=1.5pt]{-}(0.4,0.6)(1,0.6)
\psline[linecolor=blue,linewidth=1.5pt]{-}(0.4,1.4)(1,1.4)
\end{pspicture}
\eea
Fixing the direction of transfer gives the linear one-boundary TL algebra
\be
{TL_N(\beta;\beta_1,\beta_2):=\big\langle I,\, e_j,\, f_N;\ j=1,\ldots, N-1\big\rangle}
\label{1bdyTL}
\ee
where $\beta_1, \beta_2$ are fugacities of loops that terminate on the boundary. The generators 
\psset{unit=1cm}
\be
 I:=\!
\begin{pspicture}[shift=-0.55](-0.1,-0.65)(2.0,0.45)
\rput(1.4,0.0){\small$...$}
\psline[linecolor=blue,linewidth=1.5pt](0.2,0.35)(0.2,-0.35)\rput(0.2,-0.55){$_1$}
\psline[linecolor=blue,linewidth=1.5pt](0.6,0.35)(0.6,-0.35)
\psline[linecolor=blue,linewidth=1.5pt](1.0,0.35)(1.0,-0.35)
\psline[linecolor=blue,linewidth=1.5pt](1.8,0.35)(1.8,-0.35)\rput(1.8,-0.55){$_N$}
\end{pspicture} 
 ,\qquad
 e_j:=\!
 \begin{pspicture}[shift=-0.55](-0.1,-0.65)(3.2,0.45)
\rput(0.6,0.0){\small$...$}
\rput(2.6,0.0){\small$...$}
\psline[linecolor=blue,linewidth=1.5pt](0.2,0.35)(0.2,-0.35)\rput(0.2,-0.55){$_1$}
\psline[linecolor=blue,linewidth=1.5pt](1.0,0.35)(1.0,-0.35)
\psline[linecolor=blue,linewidth=1.5pt](2.2,0.35)(2.2,-0.35)
\psline[linecolor=blue,linewidth=1.5pt](3.0,0.35)(3.0,-0.35)\rput(3.0,-0.55){$_{N}$}
\psarc[linecolor=blue,linewidth=1.5pt](1.6,0.35){0.2}{180}{0}\rput(1.4,-0.55){$_j$}
\psarc[linecolor=blue,linewidth=1.5pt](1.6,-0.35){0.2}{0}{180}
\end{pspicture} 
 ,\qquad
 f_N:=\!
\begin{pspicture}[shift=-0.55](-0.1,-0.65)(2.0,0.45)
\rput(1.4,0.0){\small$...$}
\psline[linecolor=blue,linewidth=1.5pt](0.2,0.35)(0.2,-0.35)\rput(0.2,-0.55){$_1$}
\psline[linecolor=blue,linewidth=1.5pt](0.6,0.35)(0.6,-0.35)
\psline[linecolor=blue,linewidth=1.5pt](1.0,0.35)(1.0,-0.35)
\psline[linecolor=blue,linewidth=1.5pt](1.8,0.35)(1.8,-0.35)
\psarc[linecolor=blue,linewidth=1.5pt](2.4,-0.35){0.2}{90}{180}
\psarc[linecolor=blue,linewidth=1.5pt](2.4,0.35){0.2}{180}{-90}\rput(2.2,-0.55){$_{N}$}
\end{pspicture} 
\label{Ief}
\ee
satisfy\vspace{-.1in}
\bea
 {\begin{array}{rcll}
 [e_i,e_j]&\!\!\!=\!\!\!&0,\qquad\quad &|i-j|>1\\[.15cm]
 e_ie_je_i&\!\!\!=\!\!\!&e_i,\qquad\quad &|i-j|=1\\[.15cm]
 e_j^2&\!\!\!=\!\!\!&\beta e_j,\qquad\quad &j=1,\ldots, N-1\\[.15cm]
 [e_j,f_N]&\!\!\!=\!\!\!&0,\qquad\quad &j=1,\ldots, N-2\\[.15cm]
 e_{N-1}f_Ne_{N-1}&\!\!\!=\!\!\!&\beta_1 e_{N-1},\qquad\quad 
 &f_N^2=\beta_2 f_N
 \end{array}}
\label{1TLN}
\eea
Multiplication is by vertical concatenation of diagrams. For example
\be
 f_N^2=
\begin{pspicture}[shift=-0.55](-0.1,-0.65)(2.5,.8)
\rput(1.4,-0.35){\small$...$}
\rput(1.4,0.35){\small$...$}
\psline[linecolor=blue,linewidth=1.5pt](0.2,0.7)(0.2,-0.7)\rput(0.2,-0.9){$_1$}
\psline[linecolor=blue,linewidth=1.5pt](0.6,0.7)(0.6,-0.7)
\psline[linecolor=blue,linewidth=1.5pt](1.0,0.7)(1.0,-0.7)
\psline[linecolor=blue,linewidth=1.5pt](1.8,0.7)(1.8,-0.7)
\psarc[linecolor=blue,linewidth=1.5pt](2.4,-0.7){0.2}{90}{180}
\psarc[linecolor=blue,linewidth=1.5pt](2.4,0){0.2}{90}{270}
\psarc[linecolor=blue,linewidth=1.5pt](2.4,0.7){0.2}{180}{-90}
\psline[linewidth=0.5pt,linestyle=dashed, dash=1pt 1pt](0,0)(2.4,0)
\rput(2.2,-0.9){$_{N}$}
\end{pspicture} 
\;=
\beta_2\times\!\!
\begin{pspicture}[shift=-0.55](-0.1,-0.65)(2.5,0.45)
\rput(1.4,0.0){\small$...$}
\psline[linecolor=blue,linewidth=1.5pt](0.2,0.35)(0.2,-0.35)\rput(0.2,-0.55){$_1$}
\psline[linecolor=blue,linewidth=1.5pt](0.6,0.35)(0.6,-0.35)
\psline[linecolor=blue,linewidth=1.5pt](1.0,0.35)(1.0,-0.35)
\psline[linecolor=blue,linewidth=1.5pt](1.8,0.35)(1.8,-0.35)
\psarc[linecolor=blue,linewidth=1.5pt](2.4,-0.35){0.2}{90}{180}
\psarc[linecolor=blue,linewidth=1.5pt](2.4,0.35){0.2}{180}{-90}\rput(2.2,-0.55){$_{N}$}
\end{pspicture} 
\;=\beta_2f_N\\[4pt]
\ee
In this paper, for the numerics, we fix $\beta_1=\beta_2=1$.

For $\beta_2\ne 0$ the one-boundary TL algebra is equivalent (\cite{PRT14} Appendix~A), up to a rescaling, to the blobbed TL algebra that consists of the elements $I,e_j$ and the blob projector $b=f_N/\b_2$ that commutes with $e_j$ for $j<N-1$ and satisfies
\begin{equation}
 e_{N-1}be_{N-1}=\dfrac{\b_1}{\b_2}e_{N-1},\qquad b^2=b
\end{equation}
The blobbed TL algebra was used by Jacobsen and Saleur in \cite{JSbdy} to define boundary conditions with similarities to the Robin boundary conditions. 
In both cases, the $(r,s)$ boundary conditions are described by the one-boundary Temperley-Lieb algebra but their constructions are different and they act on different vector spaces of link states.

\subsection{Robin vacuum boundary condition}

The Robin vacuum boundary condition, which we often call simply {\em Robin vacuum}, was introduced in the vertex model context in \cite{Doikou2002}.  It was further developed in the loop context in \cite{PRT14} and is a solution to the boundary Yang-Baxter equation~\cite{Cherednik,Sklyanin,BPO} (\ref{BYBE}). It is a linear combination of Neumann and Dirichlet boundary conditions
\be
\psset{unit=.6cm}
\begin{pspicture}[shift=-0.89](0,0)(1,2)
\pspolygon[fillstyle=solid,fillcolor=lightlightblue](0,1)(1,2)(1,0)(0,1)
\rput(0.65,1){$_u$}
\end{pspicture}
\ =\Gamma(u)\
\begin{pspicture}[shift=-0.89](0,0)(1,2)
\pspolygon[fillstyle=solid,fillcolor=lightlightblue](0,1)(1,2)(1,0)(0,1)
\psarc[linewidth=1.5pt,linecolor=blue](0,1){.7}{-45}{45}
\end{pspicture}
\ +s_0(2u)\
\begin{pspicture}[shift=-0.89](0,0)(1,2)
\pspolygon[fillstyle=solid,fillcolor=lightlightblue](0,1)(1,2)(1,0)(0,1)
\psline[linecolor=blue,linewidth=1.5pt]{-}(0.4,0.6)(1,0.6)
\psline[linecolor=blue,linewidth=1.5pt]{-}(0.4,1.4)(1,1.4)
\end{pspicture}
\label{twist}
\ee
With Neumann boundary conditions, the loop segments are reflected at the boundary. With Dirichlet boundary conditions, loops are allowed to terminate on the boundary. In agreement with (\ref{1TLN}), boundary half-loops terminating on the boundary receive a different fugacity $\beta_1$ or $\beta_2$ depending on the parity of the (single) row where the lower rows in each double row are designated as odd
\be
\psset{unit=.6cm}
\beta_1:\qquad
\begin{pspicture}[shift=-2.39](0,0)(1,5)
\pspolygon[fillstyle=solid,fillcolor=lightlightblue](0,1)(1,2)(1,0)(0,1)
\pspolygon[fillstyle=solid,fillcolor=lightlightblue](0,4)(1,5)(1,3)(0,4)
\rput(1,2.7){$\vdots$}
\psline[linecolor=blue,linewidth=1.5pt](0.4,0.6)(1,0.6)
\psline[linecolor=blue,linewidth=1.5pt](0.4,4.4)(1,4.4)
\psbezier[linewidth=1.5pt,linecolor=blue](0.4,0.6)(-1.4,0.45)(-1.4,4.55)(0.4,4.4)
\rput(1.8,0.65){\small odd}
\rput(1.8,4.4){\small even}
\end{pspicture}
\hspace{2cm}
\beta_2:\quad\
\begin{pspicture}[shift=-2.39](0,0)(1,5)
\pspolygon[fillstyle=solid,fillcolor=lightlightblue](0,1)(1,2)(1,0)(0,1)
\pspolygon[fillstyle=solid,fillcolor=lightlightblue](0,4)(1,5)(1,3)(0,4)
\rput(1,2.7){$\vdots$}
\psline[linecolor=blue,linewidth=1.5pt](0.4,3.6)(1,3.6)
\psline[linecolor=blue,linewidth=1.5pt](0.4,1.4)(1,1.4)
\psbezier[linewidth=1.5pt,linecolor=blue](0.4,1.4)(-0.8,1.5)(-0.8,3.5)(0.4,3.6)
\rput(1.8,1.4){\small even}
\rput(1.8,3.65){\small odd}
\end{pspicture}
\qquad\qquad 
\ee

In \cite{PRT14}, the function $\G(u)$ is written as
\begin{equation}
\G(u)=s_1(\xi-u)[\b_1s_1(\xi+u)-\b_2s_0(\xi+u)],\qquad \G(\xi+\lambda)=0
\label{tuned}
\end{equation}
where the boundary field $\xi$ is an arbitrary parameter defined modulo $\pi$. 
Most importantly $\G(\xi+\lambda)=0$ is required~\cite{PRT14}, in the presence of an $r$-type seam, to impose a drop-down property (equation (3.23) of \cite{PRT14}) ensuring the rule which disallows boundary links emanating from the lower edge of the $r$-type seam to propagate up to the upper edge. This relation, reflecting a coupling between the Robin vacuum and the $r$-type seam, fixes the coincidence of the boundary field $\xi$ in the $r$-type seam and the Robin vacuum. It is convenient to parameterize the boundary loop fugacities by the variables $R\in\mathbb{R}^+$ and $\a\in[0,2\pi)$ as 
\begin{equation}\label{param_b12}
\b_1=\dfrac{R\sin\a}{\sin\l},\qquad \b_2=\dfrac{R\sin(\a-\l)}{\sin\l}
\end{equation}
so that $\G(u)$ takes a factorized form
\begin{equation}\label{Param_R}
\G(u)=\G(u|\xi,\alpha)=Rs_1(\xi-u)s_0(\xi+u+\a)
\end{equation}
This form of $\G(u)$ and (\ref{twist}) agree with (3.46) of the second reference of \cite{BehrendP} with $\xi_\text{BP}=\xi+\lambda$ and $\alpha=r\lambda$ so that our later specialization to $\xi=-\frac{\lambda}{2}$, for the numerics, coincides with the specialization $\xi_\text{BP}=\frac{\lambda}{2}$. 
For the numerics, we also specialize to the case $\b_1=\b_2=1$, for which $R=2\sin\tfrac{\lambda}{2}$ and
\begin{equation}
\a=\frac{\lambda+\pi}{2},\qquad \G(u)=\dfrac{\sin(\xi+\l-u)\cos(\xi+\tfrac{\lambda}{2}+u)}{\sin\lambda\cos\tfrac{\lambda}{2}}
\end{equation}
For later reference, in the case $\beta_1=\beta_2=1$, we note that
\bea
\frac{\Gamma(0)}{s_1(\xi)}=s_1(\xi)-s_0(\xi)=\dfrac{\cos(\xi+\tfrac{\lambda}{2})}{\cos\tfrac{\lambda}{2}}
\eea
In the notation of Doikou and Martin~\cite{Doikou2002}, $\mu_\text{DM}=\l$, $\th_\text{DM}=-iu/\l$, 
$m_\text{DM}\l=\l-\a$ and $\zeta_\text{DM}\l=\xi+(\l+\a)/2$ with $R=1$.

At first sight, the Robin boundary conditions depend on the three parameters $R$, $\xi$ and $\a$ (in addition to the bulk parameters $u,\lambda$). Upon renormalization of the boundary operator $f_N$, it can be shown that the parameter $R\ne 0$ corresponds to an overall scale factor.
Let us introduce the quantities $\tilde{\b}_1$, $\tilde{\b}_2$ independent of $R$ as $\b_1=R\tilde{\b}_1$ and $\b_2=R\tilde{\b}_2$ and define the operator $\tilde{f}_N$ as $f_N=R\tilde{f}_N$ such that $e_{N-1}\tilde{f}_Ne_{N-1}=\tilde{\b}_1e_{N-1}$ and $\tilde{f}_N^2=\tilde{\b}_2\tilde{f}_N$. Noticing that $\tilde{\G}(u)=R^{-1}\G(u)$ is independent of $R$, we deduce that the $R$-dependence of the Robin vacuum factorizes
\begin{equation}
\G(u)I+s_0(2u)f_N=R\big[\tilde{\G}(u)I+s_0(2u)\tilde{f}_N\big]
\end{equation}
We deduce that the thermodynamic and conformal properties are independent of $R$ and 
only depend on the angle $\a$ or, equivalently, on the ratio $\b_1/\b_2$ for $\beta_2\ne 0$.

\subsection{Integrable $(r,s)$ Robin boundary conditions}

Following~\cite{BehrendP,PRZ2006,PRT14}, we dress the Robin vacuum boundary condition by fusing with $r$- and $s$-type seams on the right boundary 
\setlength{\unitlength}{1.1cm}
\psset{unit=1.1cm}
\thicklines
\nc{\spos}[2]{\makebox(0,0)[#1]{$\sm{#2}$}}
\nc{\sm}[1]{{\scriptstyle #1}}
\definecolor{lightpurple}{rgb}{1,.65,1}
\smallskip
\bea
\mbox{}\hspace{-.75in}
\raisebox{-1.4\unitlength}[1.6\unitlength][1.2\unitlength]
{\begin{picture}(5.,2.0)
\put(1,.5){\color{lightlightblue}\rule{5\unitlength}{2\unitlength}}
\put(0.45,1.5){\makebox(0,0)[]{$=$}}
\put(.49,2.83){\spos{bc}{\color{blue}=}}
\put(-.16,2.75){\spos{bc}{\color{blue}(r,s-\frac{1}{2})}}
\put(3.9,2.8){\spos{bc}{\color{blue}(r,1)}}
\put(6,2.8){\spos{bc}{\color{blue}\otimes}}
\put(-0.1,0.5){\line(0,1){2}}
\multiput(1,0.5)(1,0){4}{\line(0,1){2}}
\multiput(5,0.5)(1,0){2}{\line(0,1){2}}
\multiput(1,0.5)(0,1){3}{\line(1,0){5}}
\put(-0.6,1.5){\line(1,2){0.5}}
\put(-0.6,1.5){\line(1,-2){0.5}}
\pspolygon[linewidth=1pt,fillstyle=solid, fillcolor=lightlightblue](-0.6,1.5)(-.1,2.5)(-.1,.5)
\put(1.5,1){\spos{}{u\!-\!\xi_{w}}}
\put(2.5,1){\spos{}{u\!-\!\xi_{w\!-\!1}}}
\put(5.5,1){\spos{}{u\!-\!\xi_1}}
\put(1.5,2){\spos{}{-\!u\!-\!\xi_{w\!-\!1}}}
\put(2.5,2){\spos{}{-\!u\!-\!\xi_{w\!-\!2}}}
\put(5.5,2){\spos{}{-\!u\!-\!\xi_0}}
\put(-0.34,1.5){\spos{}{u,\xi}}
\multiput(0,0)(1,0){5}{\psarc[linewidth=1.5pt,linecolor=darkgreen](.5,2.5){1}{0}{40}}
\multiput(0,0)(1,0){5}{\psline[linewidth=1.5pt](1.5,.3)(1.5,.5)}
\psline[linewidth=1.5pt](.8,1)(1,1)
\psline[linewidth=1.5pt](.8,2)(1,2)
\psline[linewidth=1.5pt](-.55,1)(-.35,1)
\psline[linewidth=1.5pt](-.55,2)(-.35,2)
{\color{red}
\put(5,.5){\oval(.2,.2)[tr]}
\put(5,1.5){\oval(.2,.2)[tr]}
\multiput(1,.5)(1,0){2}{\oval(.2,.2)[tr]}
\multiput(1,1.5)(1,0){2}{\oval(.2,.2)[tr]}}
\end{picture}}
\raisebox{-1.4\unitlength}[1.6\unitlength][1.2\unitlength]
{\begin{pspicture}(4.5,2.0)
\put(1,.5){\color{lightpurple}\rule{5\unitlength}{2\unitlength}}
\psline[linewidth=1.5pt](1,.5)(1,2.5)
\put(2.9,2.8){\spos{bc}{\color{blue}(1,s)}}
\put(6.5,2.7){\spos{bc}{\color{blue}(0,\half)}}
\put(6,2.8){\spos{bc}{\color{blue}\otimes}}
\put(6.5,0.5){\line(0,1){2}}
\multiput(1,0.5)(1,0){4}{\line(0,1){2}}
\multiput(5,0.5)(1,0){2}{\line(0,1){2}}
\multiput(1,0.5)(0,1){3}{\line(1,0){5}}
\put(6,1.5){\line(1,2){0.5}}
\put(6,1.5){\line(1,-2){0.5}}
\pspolygon[linewidth=1pt,fillstyle=solid, fillcolor=lightlightblue](6,1.5)(6.5,2.5)(6.5,.5)
\multiput(0,0)(1,0){5}{\psarc[linewidth=1.5pt,linecolor=darkgreen](.5,2.5){1}{0}{40}}
\multiput(0,0)(1,0){5}{\psline[linewidth=1.5pt](1.5,.3)(1.5,.5)}
\multiput(6,0.5)(0,2){2}{\makebox(0.5,0){\dotfill}}
\psline[linewidth=1.5pt,linecolor=blue](1,1)(6.25,1)
\psline[linewidth=1.5pt,linecolor=blue](1,2)(6.25,2)
\multirput(0,0)(1,0){5}{\psline[linewidth=1.5pt,linecolor=blue](1.5,.5)(1.5,.9)
\psline[linewidth=1.5pt,linecolor=blue](1.5,1.1)(1.5,1.9)
\psline[linewidth=1.5pt,linecolor=blue](1.5,2.1)(1.5,2.5)}
\psline[linewidth=3pt,linecolor=Blue](-3.75,0.5)(.75,0.5)
\psline[linewidth=3pt,linecolor=Blue](1.25,0.5)(5.75,0.5)
\rput(-1.5,0){$\underbrace{\hspace{5\unitlength}}_{\mbox{\small $w$ columns}}$}
\rput(3.5,0){$\underbrace{\hspace{5\unitlength}}_{\mbox{\small $d$ columns}}$}
\rput(6.27,1.5){\scriptsize $u$}
{\color{red}\put(5,.5){\oval(.2,.2)[tr]}
\put(5,1.5){\oval(.2,.2)[tr]}
\multiput(1,.5)(1,0){2}{\oval(.2,.2)[tr]}
\multiput(1,1.5)(1,0){2}{\oval(.2,.2)[tr]}}
\end{pspicture}}\label{rsRobinBdy}
\\[6pt]
\nonumber
\eea
The fusion of the $w$ or $d$ columns to form a seam is implemented diagrammatically~\cite{PRV,PTC,PRTsuper} (as opposed to using  Wenzl-Jones projectors) by forbidding closed half-arcs along the lower edge. The drop-down property (equation (3.23) of \cite{PRT14}) ensures that no closed half-arcs can occur along the top edge. With the parameters appropriately tuned (\ref{tuned}) to give the drop-down property of boundary arcs, this gives the general integrable $(r,s)$ Robin boundary conditions. 
The column inhomogeneities are $\xi_k=\xi+k\lambda$ and the dependence on $\xi$ in the Robin vacuum is suppressed but is coupled to coincide with the value of $\xi$ in the $r$-type seam. 
The action of the double-row transfer matrix must map ${\cal V}_d^{(N,w)}$ back to itself. This is implemented by the diagrammatic fusion projection (described above) in the $r$- and $s$-type seams and the drop-down property of boundary arcs. In the $s$-type seam, this prevents the closing of two defects to form a closed loop ensuring that the number of defects is conserved and $s$ or $d$ is a good quantum number. 
In the $r$-type seam, the action also kills~\cite{PRV} certain words among the boundary TL generators. 
We also note that, due to the generalized Yang-Baxter equation, the $r$- and $s$-type boundary seams commute and freely propagate along the row under a similarity transformation that preserves the spectrum. In particular, they obey the fusion rules
\bea
(r,s-\half)=(r,1)\otimes (1,s)\otimes(0,\half)=(r,s)\otimes (0,\half)
\eea

Using the fact that the Robin vacuum satisfies the boundary Yang-Baxter equation, and following \cite{BPO}, it follows that the $(r,s)$ Robin boundary conditions satisfy the boundary Yang-Baxter equation
\be
\psset{unit=.7cm}
\begin{pspicture}[shift=-1.89](0,0)(3,4)
\pspolygon[fillstyle=solid,fillcolor=lightlightblue](0,1)(1,2)(2,1)(1,0)(0,1)
\pspolygon[fillstyle=solid,fillcolor=lightlightblue](1,2)(2,3)(3,2)(2,1)(1,2)
\pspolygon[fillstyle=solid,fillcolor=lightlightblue](2,1)(3,2)(3,0)(2,1)
\pspolygon[fillstyle=solid,fillcolor=lightlightblue](2,3)(3,4)(3,2)(2,3)
\psarc[linewidth=1.5pt,linecolor=blue](2,1){.7}{-135}{-45}
\psarc[linewidth=0.025,linecolor=red]{-}(0,1){0.16}{-45}{45}
\psarc[linewidth=0.025,linecolor=red]{-}(1,2){0.16}{-45}{45}
\rput(1,1){$_{u-v}$}
\rput(2.05,2){$_{\lambda-u-v}$}
\rput(2.65,1){$_{u,\xi}$}
\rput(2.65,3){$_{v,\xi}$}
\end{pspicture}
\quad =\ \
\begin{pspicture}[shift=-1.89](0,0)(3,4)
\pspolygon[fillstyle=solid,fillcolor=lightlightblue](0,3)(1,4)(2,3)(1,2)(0,3)
\pspolygon[fillstyle=solid,fillcolor=lightlightblue](1,2)(2,3)(3,2)(2,1)(1,2)
\pspolygon[fillstyle=solid,fillcolor=lightlightblue](2,1)(3,2)(3,0)(2,1)
\pspolygon[fillstyle=solid,fillcolor=lightlightblue](2,3)(3,4)(3,2)(2,3)
\psarc[linewidth=1.5pt,linecolor=blue](2,3){.7}{45}{135}
\psarc[linewidth=0.025,linecolor=red]{-}(0,3){0.16}{-45}{45}
\psarc[linewidth=0.025,linecolor=red]{-}(1,2){0.16}{-45}{45}
\rput(1,3){$_{u-v}$}
\rput(2.05,2){$_{\lambda-u-v}$}
\rput(2.65,1){$_{v,\xi}$}
\rput(2.65,3){$_{u,\xi}$}
\end{pspicture}
\label{BYBE}
\ee
where the boundary conditions are represented diagrammatically by the $u$- or $v$-dependent right boundary triangles. Although we do not do so in this paper, similar integrable boundary conditions can be constructed on the left boundary. 

As an operator acting on ${\cal V}^{(N,w)}$, the boundary operator implementing the $(r,s)$ Robin boundary condition is
\bea
 K_N^{(w)}(u,\xi)&\!\!=\!\!&X_N(u-\xi_w)X_{N+1}(u-\xi_{w-1})\ldots X_{N+w-1}(u-\xi_1)\big[\Gamma(u)I+s_0(2u)f_{N+w}\big]\nn
  &&X_{N+w-1}(u+\xi_1)X_{N+w-2}(u+\xi_2)\ldots X_N(u+\xi_w)
\eea
When restricted to acting from ${\cal V}_d^{(N,w)}$ to itself, the boundary operator reduces~\cite{PRT14} to
\bea
 K_N^{(w)}(u,\xi)&\!\!\simeq\!\!&\prod_{j=1}^w s_{-1}(u+\xi_j)s_{-1}(u-\xi_j)\Big[\alpha_0^{(w)}I+\alpha_1^{(w)}e_N+\alpha_2^{(w)}e_Ne_{N+1}+\ldots\nn
   &\!\!&\quad\mbox{}+\alpha_w^{(w)}e_Ne_{N+1}\ldots e_{N+w-1}+\alpha_{w+1}^{(w)}e_Ne_{N+1}\ldots e_{N+w-1}f_{N+w}\Big]
\eea
where
\bea
 \alpha_0^{(w)}&\!\!\!=\!\!\!&\Gamma(u)\nn
 \alpha_k^{(w)}&\!\!\!=\!\!\!&\frac{(-1)^{k}s_0(2u)}{s_0(u+\xi)s_{w+1}(\xi-u)}
   \Big(s_{w-k}(\lambda)\,\Gamma(u)-\beta_1s_{w-k+1}(u+\xi)s_{1}(\xi-u)\Big),\qquad k=1,2,\ldots,w \nn
 \alpha_{w+1}^{(w)}&\!\!\!=\!\!\!&\frac{(-1)^ws_0(2u)s_{1}(\xi-u)}{s_{w+1}(\xi-u)}
\label{al}
\eea
All other TL words are killed. The similarity sign $\,\simeq\,$ (instead of an equality sign) indicates that the actions agree only when restricted to the vector space ${\cal V}_d^{(N,w)}$.

\section{Transfer Matrices, Hamiltonians and Non-Universal Quantities}

\subsection{Commuting double-row transfer matrices}\label{DTMs}

In this paper, we study the double row transfer matrix $\Db(u)$ with a Neumann or $(1,1)$ Kac boundary condition on the left edge of the strip and an $(r,s)$ Robin boundary condition (\ref{rsRobinBdy}) on the right edge of the strip
\be
 \Db(u):=\;\frac{1}{{\cal N}^{(w)}(u,\xi)}\qquad
\psset{unit=.95cm}
\setlength{\unitlength}{.95cm}
\begin{pspicture}[shift=-0.89](0.4,0)(6.5,2)
\facegrid{(1,0)}{(5,2)}
\pspolygon[fillstyle=solid,fillcolor=lightlightblue](0,0)(1,1)(0,2)(0,0)
\pspolygon[fillstyle=solid,fillcolor=lightlightblue](5,1)(6,2)(6,0)(5,1)
\psarc[linewidth=0.025,linecolor=red](1,0){0.16}{0}{90}
\psarc[linewidth=0.025,linecolor=red](1,1){0.16}{0}{90}
\psarc[linewidth=0.025,linecolor=red](2,0){0.16}{0}{90}
\psarc[linewidth=0.025,linecolor=red](2,1){0.16}{0}{90}
\psarc[linewidth=0.025,linecolor=red](4,0){0.16}{0}{90}
\psarc[linewidth=0.025,linecolor=red](4,1){0.16}{0}{90}
\rput(1.5,0.5){$_{u}$}
\rput(2.5,0.5){$_{u}$}
\rput(4.5,0.5){$_{u}$}
\rput(1.5,1.5){$_{\lambda-u}$}
\rput(2.5,1.5){$_{\lambda-u}$}
\rput(4.5,1.5){$_{\lambda-u}$}
\rput(3.5,0.5){$\ldots$}
\rput(3.5,1.5){$\ldots$}
\rput(5.6,1){$_{u,\xi}$}
\psarc[linewidth=1.5pt,linecolor=blue](1,1){.5}{90}{270}
\psline[linecolor=blue,linewidth=1.5pt]{-}(5,0.5)(5.5,0.5)
\psline[linecolor=blue,linewidth=1.5pt]{-}(5,1.5)(5.5,1.5)
\rput(3,-0.3){\scriptsize $\underbrace{\hspace{4\unitlength}}_{N}$}
\end{pspicture}
\vspace{0.5cm}
\label{Duxi}
\ee
This transfer matrix acts on the link states described in Section~\ref{Sec:LinkStates} with the fixed $d=s-1$ defects closed to the right, under the boundary arcs, and onto the $s$-type seam. 
Following the diagrammatic arguments of \cite{BPO}, the Yang-Baxter and boundary Yang-Baxter equations (\ref{YBE}) and (\ref{BYBE}) ensure the commutation of the transfer matrices $\Db(u)$ and $\Db(v)$. As in \cite{PRT14}, it also follows that the transfer matrices are crossing symmetric
\bea
\vec D(\lambda-u)=\vec D(u)
\eea
The double row transfer matrix is normalized by a crossing symmetric factor (c.f. (3.4) of \cite{PTC})
\be 
{\cal N}^{(w)}(u,\xi)=\begin{cases}
\disp (-1)^w \beta\,\Gamma(0) s(\xi)s(\xi_{w+1})\prod_{j=1}^{w-1}s(\xi_j+u)s(\xi_{j+1}-u),& w>0\\[6pt]
\disp\beta\Gamma(0)\frac{s(\xi)s(\xi_1)}{s(\xi+u)s(\xi_1-u)},&w=0
\end{cases}
\label{normalization}
\ee
so that, in addition, it satisfies $\vec D(0)=\vec D(\lambda)=I$. For $w\ge 2$, the product removes the common factors  resulting from fusion in the seam, otherwise, this product takes the value 1. Note, by convention~\cite{BPO}, the (normalized) weight associated with an $r$-type seam of width $w=0$ is not unity and is effectively fixed to 
\bea
\tilde\kappa_0(u,\xi)=\tilde\kappa_0(\lambda-u,\xi)=\frac{s(\xi+u)s(\xi_1-u)}{s(\xi)s(\xi_1)}
\label{zeroWidth}
\eea

\subsection{Quantum Hamiltonians}

In this paper we work with the normalized transfer matrix $\vec D(u)$ given in (\ref{Duxi}). But, to take the Hamiltonian limit, it is convenient  to use a transfer matrix normalized as in \cite{PRT14}
\begin{equation}
\vec d(u)=\frac{{\cal N}^{(w)}(u,\xi)}{\eta(u)}\,\vec D(u),\qquad \eta(u)=\dfrac{\b\,\G(0)s(u+\xi_w)s(\xi)\eta^{(w)}(u,\xi)}{s(u+\xi)s(\xi_w)}
\label{d(u)}
\end{equation}
where
\begin{equation}
\eta^{(w)}(u,\xi)=\prod_{j=1}^w s_{-1}(u+\xi_j)s_{-1}(u-\xi_j),\qquad \frac{\eta(u)}{{\cal N}^{(w)}(u,\xi)}=\frac{s(u+\xi_w)s(\xi_{w+1}-u)}{s(\xi_w)s(\xi_{w+1})}
\label{expr_eta}
\end{equation}
With this normalization the transfer matrix $\vec d(u)$ is crossing symmetric $\vec d(\lambda-u)=\vec d(u)$ and satisfies the intial condition $\vec d(0)=I$.

The commuting family $\vec D(u)$ of double-row transfer matrices produces an infinite set of commuting conserved quantities by expansion in the spectral parameter $u$. The Hamiltonian ${\cal H}$ arises as the first non-trivial operator in the limit $u\to0$
\be
\vec d(u)=\frac{{\cal N}^{(w)}(u,\xi)}{\eta(u)}\,\vec D(u)=I-\frac{2u}{\sin\lambda}({\cal H}+hI)+{\cal O}(u^2)
\label{dH}
\ee
where $h$ measures a convenient shift in the groundstate energy. Explicitly, as an operator acting on the vector space ${\cal V}_d^{(N,w)}$ with $\beta_1=\beta_2=1$ and $\lambda\ne\frac{\pi}{2}$, the Hamiltonian is~\cite{PRT14}
\begin{subequations}
\begin{align}
 {\cal H}={\cal H}^{w,d}&\simeq
  -\sum_{j=1}^{N-1}e_j-\sum_{k=1}^{w}\frac{(-1)^k}{s_0(\xi)s_{w+1}(\xi)}
  \Big(s_{w-k}(\lambda)-\frac{s_1(\xi)s_{w-k+1}(\xi)}{\Gamma(0)}\Big) e_Ne_{N+1}\ldots e_{N+k-1}\nn
 &\qquad\qquad\qquad -(-1)^w\frac{s_1(\xi)}{s_{w+1}(\xi)\Gamma(0)} e_Ne_{N+1}\ldots e_{N+w-1}f_{N+w}\\[6pt]
&= -\sum_{j=1}^{N-1}e_j-\frac{\cos\frac{\lambda}{2}}{s_{w+1}(\xi)\cos(\xi+\frac{\lambda}{2})}\,\sum_{k=0}^{w} (-1)^k
 \frac{\cos((w-k-\frac{1}{2})\lambda)}{\cos \frac{\lambda}{2}}\, e_Ne_{N+1}\ldots e_{N+k}\\[6pt]
&= -\sum_{j=1}^{N-1}e_j-\frac{\cos\frac{\lambda}{2}}{s_{w+1}(\xi)\cos(\xi+\frac{\lambda}{2})}\,F_N
\label{Hbeta}
\end{align}
\end{subequations}
where, by convention, we set $e_{N+w}=f_{N+w}$ and $F_N$ is a generalized projector (that is a projector up to rescaling) described in Appendix~\ref{AppC1}. We observe that the boundary term is singular at $\xi+\alpha=0$ mod $\pi$ and at $\xi+(w+1)\lambda=0$ mod~$\pi$. So it is anticipated that the boundary energies will be discontinuous at these points. 
Fixing $\lambda,w$ and varying $\xi$, it follows that the singularities occur at one or two points in the interval $[0,\pi)$. The two points coincide whenever $(w+\half)\lambda=\tfrac{\pi}{2}$ mod $\pi$. 
Allowing for periodicity, the boundary energy $\calE_R(\xi)$ will accordingly consist of one or two analytic branches separated by the points of discontinuity.

Note that ${\cal H}^{w,d}$ is independent of $d$ which only enters via the vector space of link states on which the Hamiltonian acts. 
For $\lambda=\frac{\pi}{2}$, that is critical dense polymers ${\cal LM}(1,2)$, the Hamiltonian is different. In this case, since the ${\cal O}(u)$ term vanishes, the Hamiltonian is  given by the ${\cal O}(u^2)$ term in the expansion~\cite{PRT14}. 
In the absence of an $r$-type seam ($w=0$), the Hamiltonian (\ref{Hbeta}) reduces to 
\begin{equation}\label{def_H}
\mathcal{H}=-\sum_{j=1}^{N-1}e_j-\dfrac1{\G(0)}f_N
\end{equation}
This is the Hamiltonian studied by Jacobsen and Saleur~\cite{JSbdy}. In this case, they argue that the conformal properties depend on the sign of $\b_2/\G(0)$ or, equivalently, the value of the parameter $\xi\in [0,\pi)$. The JS critical phase corresponds to
\begin{equation}
0<\dfrac{\b_2}{\G(0)}=\dfrac{\sin \l\sin(\a-\l)}{\sin(\xi+\l)\sin(\xi+\a)}
\end{equation}

\subsection{Bulk and boundary free energies}
\label{sec_bulk_bdy_energy}

The logarithmic minimal models ${\cal LM}(p,p')$~\cite{PRZ2006} are exactly solvable on the lattice. The key to this exact integrability are functional equations in the form of $T$- and $Y$-systems~\cite{Zamolodchikov,KlumperP,KNS,BPO}. 
Indeed, it is known~\cite{MDPR} that the logarithmic minimal models satisfy $T$- and $Y$-systems on the strip and cylinder. Since the structure of these equations is universal~\cite{universal} in the sense that the same equations hold independent of the topology and boundary conditions, it is expected that the same equations will also hold for logarithmic minimal models with $(r,s)$ Robin boundary conditions. In principle, these equations could be solved for the conformal weights using dilogarithm identities~\cite{KlumperPHSq,KlumperP}. In practice, the required methodologies have not yet been developed so, in this paper, we calculate the conformal weights numerically. 
In this section, we use an inversion identity~\cite{InvId}, which is the first functional equation in the $T$-system, to calculate the bulk and boundary free energies exactly. The bulk free energy is obtained using the inversion relation method~\cite{Stroganov,BaxterInv}. The boundary free energies are calculated using the boundary inversion relation methods of \cite{OPB95,PTC}. Knowing the bulk and boundary free energies exactly greatly improves the accuracy of the numerically extrapolated estimates for the conformal weights in Section~\ref{sec:numerics}.

The transfer matrix $\Db(u)$ (\ref{Duxi}) satisfies an inversion identity~\cite{InvId} of the form
\begin{equation}
\Db(u)\Db(u+\l)=\phi(\lambda+u)\phi(\lambda-u)I +\phi(u) \vec D^2(u)
\label{InvId}
\end{equation}
where $\vec D^2(u)$ denotes the transfer matrix at fusion level 2. 
In calculating the bulk and boundary free energies we observe that, in the physical region, the second term can be neglected since it is exponentially small compared to the first term. 
This yields an inversion relation, derived in Appendix~\ref{AppA}, for each eigenvalue
\bea
& D(u)D(u+\lambda)=D(u)D(-u)=\phi(\lambda+u)\phi(\lambda-u)&\label{InvRel}\\[4pt]
&=\disp[s_1(u)s_1(-u)]^{2N}\,\frac{s_2(2u)s_2(-2u)}{s_1(2u)s_1(-2u)}\,\frac{\G(u)\G(-u)}{\b^2\,\G(0)^2}\,
\dfrac{s(\xi+u)s(\xi-u)s(\xi_{w+1}+u)s(\xi_{w+1}-u)}{s(\xi)^2 s(\xi_{w+1})^2}&\nonumber
\eea 
The inversion relation is sufficient to determine analytically the bulk and boundary free energies in the thermodynamic limit $N\to\infty$. 
Subsequently, taking the Hamiltonian limit $u\to0$, yields the exact bulk and boundary energies. 

The partition function per site $\k(u,\xi)$, given by the largest eigenvalue $D(u)$ of the transfer matrix $\vec D(u)$ in a given $(r,s)$ sector, factorizes into contributions of order ${\cal O}(N)$, ${\cal O}(1)$ and ${\cal O}(1/N)$ 
\begin{equation}
\frac{D(u)}{\tilde\k_0(u,\xi)}=\k(u,\xi)=\k_\text{bulk}(u)^{2N}\k_0(u)\,\kappa_R(u,w,\xi)\,\ell(u),\qquad \kappa_R(u,w,\xi)=\k_R(u,\xi){\k_w(u,\xi)}
\label{kappaDecomp}
\end{equation}
The non-universal contributions, common to all eigenvalues in the given sector, are denoted by $\k_\text{bulk}(u)$, $\k_0(u)$, $\k_R(u,\xi)$, $\k_w(u,\xi)$ and $\tilde\k_0(u,\xi)$ for  the bulk, the left Kac vacuum, the right Robin vacuum, the $r$-type seam of width $w>0$ and the zero width seam respectively. 
They will be determined exactly by solving the inversion relation (\ref{InvRel}). 
In the case $w=0$, the trivial extra normalization of the transfer matrix removes the contribution (\ref{zeroWidth}) from the seam of zero width.
For compatibility with $w=0$, we also require
\bea
\kappa_R(u,0,\xi)=\k_R(u,\xi)\quad\Rightarrow\quad\kappa_0(u,\xi)=1
\eea
The boundary contribution is thus
\bea
\kappa_\text{bdy}(u,w,\xi)=\k_0(u)\,\kappa_R(u,w,\xi)=\k_0(u)\,\k_R(u,\xi)\,{\k_w(u,\xi)}
\eea
The remaining ${\cal O}(1/N)$ contribution $\ell(u)$ is different for each eigenvalue. It encodes the universal conformal properties of the model and can only be obtained by solving the full $Y$-system. 

As a result of the factorization of the inversion identity (\ref{InvRel}), the non-universal quantities satisfy
\begin{subequations}
\begin{align}
&\k_\text{bulk}(u)\k_\text{bulk}(u+\l)=\dfrac{\sin(\l+u)\sin(\l-u)}{\sin^2\l}\label{eqs_kappa1}\\
&\k_0(u)\k_0(u+\l)=\dfrac{\sin^2\l\sin(2\l+2u)\sin(2\l-2u)}{\sin^22\l\sin(\l+2u)\sin(\l-2u)}\label{eqs_kappa2}\\
&\k_R(u,\xi)\k_R(u+\l,\xi)=\G(u)\G(-u)\label{eqs_kappa3}/\G(0)^2\\
& \kappa_w(u,\xi)\kappa_w(u+\lambda,\xi)=\frac{\sin^2 \xi_1\sin(\xi_{w+1}+u)\sin(\xi_{w+1}-u)}{\sin^2\xi_{w+1}\sin(\xi_1+u)\sin(\xi_1-u)},\qquad w>0\label{eqs_kappa4}
\end{align}
\label{eqs_kappa}
\end{subequations}
$\!\!$together with the crossing symmetry $\k(\l-u,\xi)=\k(u,\xi)$. The last inversion relation is related to $\kappa^\text{PTC}_\rho(u,\xi)$ and (3.21) of \cite{PTC} by the trivial rescaling
\bea
\kappa^\text{PTC}_\rho(u,\xi)=\tilde{\kappa}_0(u,\xi)\kappa_w(u,\xi)=\frac{s(\xi+u)s(\xi_1-u)}{s(\xi)s(\xi_1)}\,\kappa_w(u,\xi),\qquad \rho=w+1
\eea

The ${\cal O}(N)$ inversion relation for the bulk free energy $f_{\text{bulk}}(u)$ has been solved by Baxter~\cite{BaxterInv} using Fourier/Laplace transforms to give
\bea
-f_{\text{bulk}}(u)=\log\kappa_{\text{bulk}}(u)=\int_{-\infty}^{\infty}\frac{\cosh(\pi-2\lambda)t\sinh ut\sinh(\lambda-u)t}{t\sinh\pi t\cosh\lambda t}\,dt\label{bulkfreeenergy}
\eea
The ${\cal O}(1)$ inversion relation for the boundary free energy $f_0(u)$ of the left Kac vacuum has been similarly solved in \cite{PTC}
\begin{align}
-f_0(u)=\log\kappa_0(u)&=\log\frac{\cos u\cos(\lambda-u)}{2\cos\lambda\cos(u-\tfrac{\lambda}{2})}-\int_{-\infty}^{\infty}\frac{\sinh ut\sinh(\lambda-u)t}{t\sinh\pi t\cosh\lambda t}dt
\nonumber\\
&\quad-\int_{-\infty}^{\infty}\frac{\sinh{\lambda t\over2}\sinh({3\lambda\over2}-\pi)t\cosh(\lambda-2u)t}{t\sinh\pi t\cosh\lambda t}dt,\qquad 0\le\lambda\le\pi
\label{finalKappa0}
\end{align}
Further modifications to the method lead to the derivation, given in Appendix~\ref{AppB}, of the new contribution $f_R(u,\xi)$ arising from the Robin vacuum
\begin{equation}
-f_R(u,\xi)=\log\k_R(u,\xi)=\int_{-\infty}^\infty\dfrac{\sinh ut\sinh(\l-u)t}{t\sinh\pi t\cosh\l t}(\cosh \th_{\xi+\l}t+\cosh \th_{\xi+\a}t)dt
\label{sol_fR}
\end{equation}
For $x\in\mathbb{R}$, the angle $\th_x\in[-\pi,\pi]$ is defined by 
\bea
\th_x+\pi=2x \mbox{\ mod $2\pi$},\qquad x\in\mathbb{R}\label{thx}
\eea
with jump discontinuities at $x=k\pi$, $k\in{\Bbb Z}$. The function $\th_x$ is $\pi$-periodic in $x$ on ${\Bbb R}$. 

In accord with (\ref{dH}), the bulk and boundary energies of the Hamiltonian are obtained from these expressions by taking a derivative with respect to $u$ and setting $u=0$
\begin{subequations}
\begin{align}
\calE_\text{bulk}&=-\sin\lambda\int_{-\infty}^\infty\dfrac{\tanh\l t \cosh(2\l-\pi)t}{\sinh\pi t}\,dt-\cos\l,\qquad 0<\lambda<\pi\\
\calE_0&=\sin\l\int_{-\infty}^\infty\dfrac{\tanh\l t \sinh{(\frac{\pi}{2}-\frac{3\l}{2})t}\sinh\frac{\l t}{2}}{\sinh\frac{\pi t}{2}}\,dt+\dfrac1{2\cos\l},\qquad 0<\lambda<\pi/2\\
\calE_R(\xi,\alpha)&=-\half\sin\l\Big[\int_{-\infty}^{\infty}\dfrac{\tanh\l t}{\sinh\pi t}(\cosh\th_{\xi+\l}t+\cosh\th_{\xi+\a}t)dt+\cot(\xi\!+\!\l)-\!\cot(\xi\!+\!\a)\Big]\label{RvacEnergy}
\end{align}
\end{subequations}
The Kac boundary energy $\calE_0$ exhibits a pole at $\l=\tfrac{\pi}{2}$ and the integral formula requires an analytic continuation in $\lambda$ performed in \cite{PTC}
\begin{equation}
\calE_0=\half+\half\sin\l\int_{-\infty}^\infty\big[1-2\sinh\tfrac{\l t}2\sinh\big(\tfrac{3\l}2-\pi\big)t\big]\dfrac{\tanh\l t}{\sinh\pi t}\,dt,\qquad 0<\lambda<\pi
\end{equation}
The cotangents in the Robin boundary energy $\calE_R(\xi,\alpha)$ can be absorbed into the integral formula (\ref{RvacEnergy}) using
\begin{equation}
\cot x=-\int_{-\infty}^\infty\dfrac{\sinh\th_x t}{\sinh \pi t}\,dt,\qquad x\in{\Bbb R}
\end{equation}
which yields
\begin{equation}\label{calER}
\calE_R(\xi,\alpha)=\half\sin\l\int_{-\infty}^\infty\frac{\sinh(\th_{\xi+\l}-\l)t-\sinh(\th_{\xi+\a}+\l)t}{\sinh\pi t\cosh\l t}\,dt,\qquad 0<\lambda<\pi
\end{equation}

\begin{figure}[p] 
 \centering
   \includegraphics[width=3.25in]{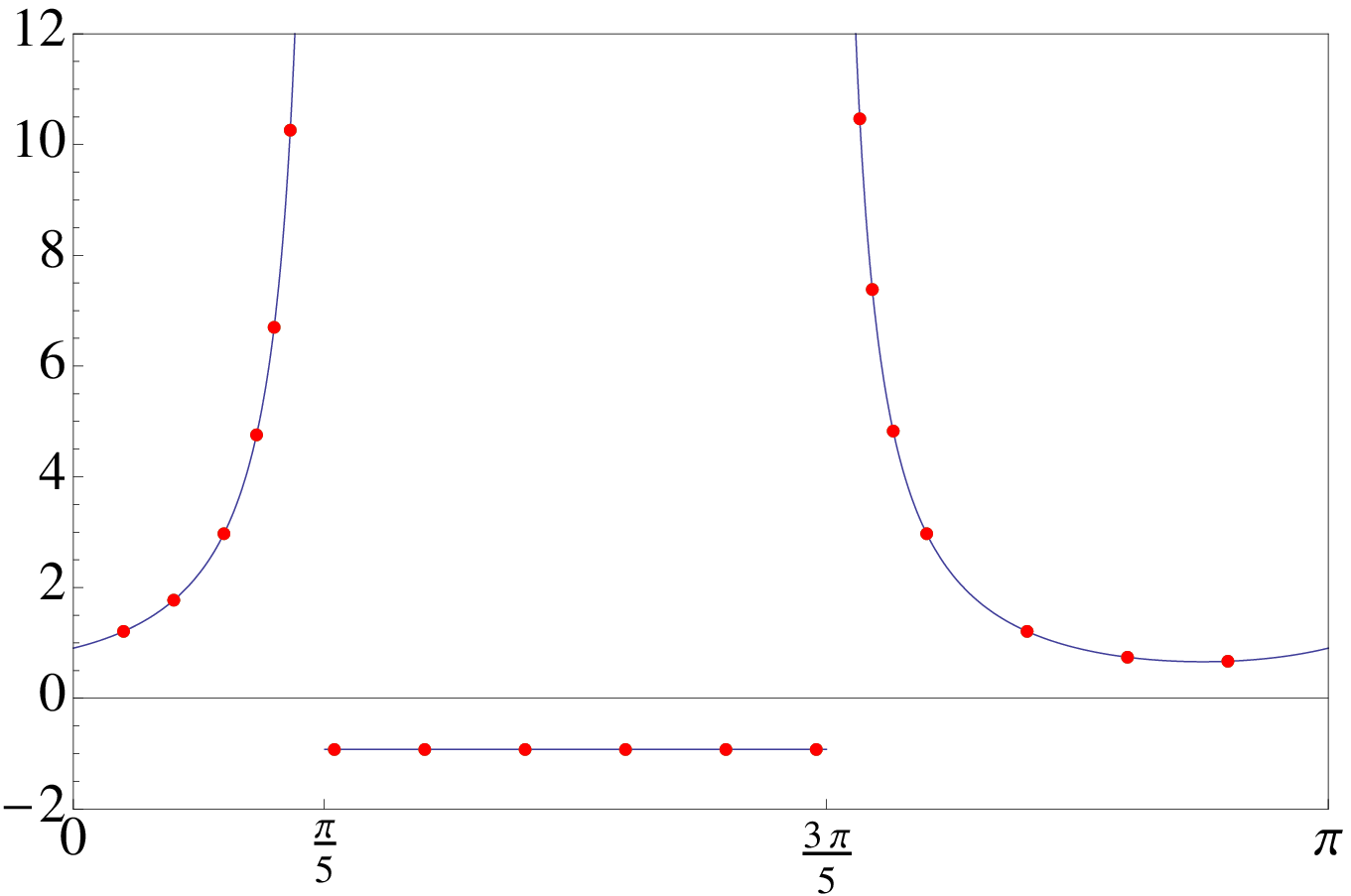}\quad \includegraphics[width=3.25in]{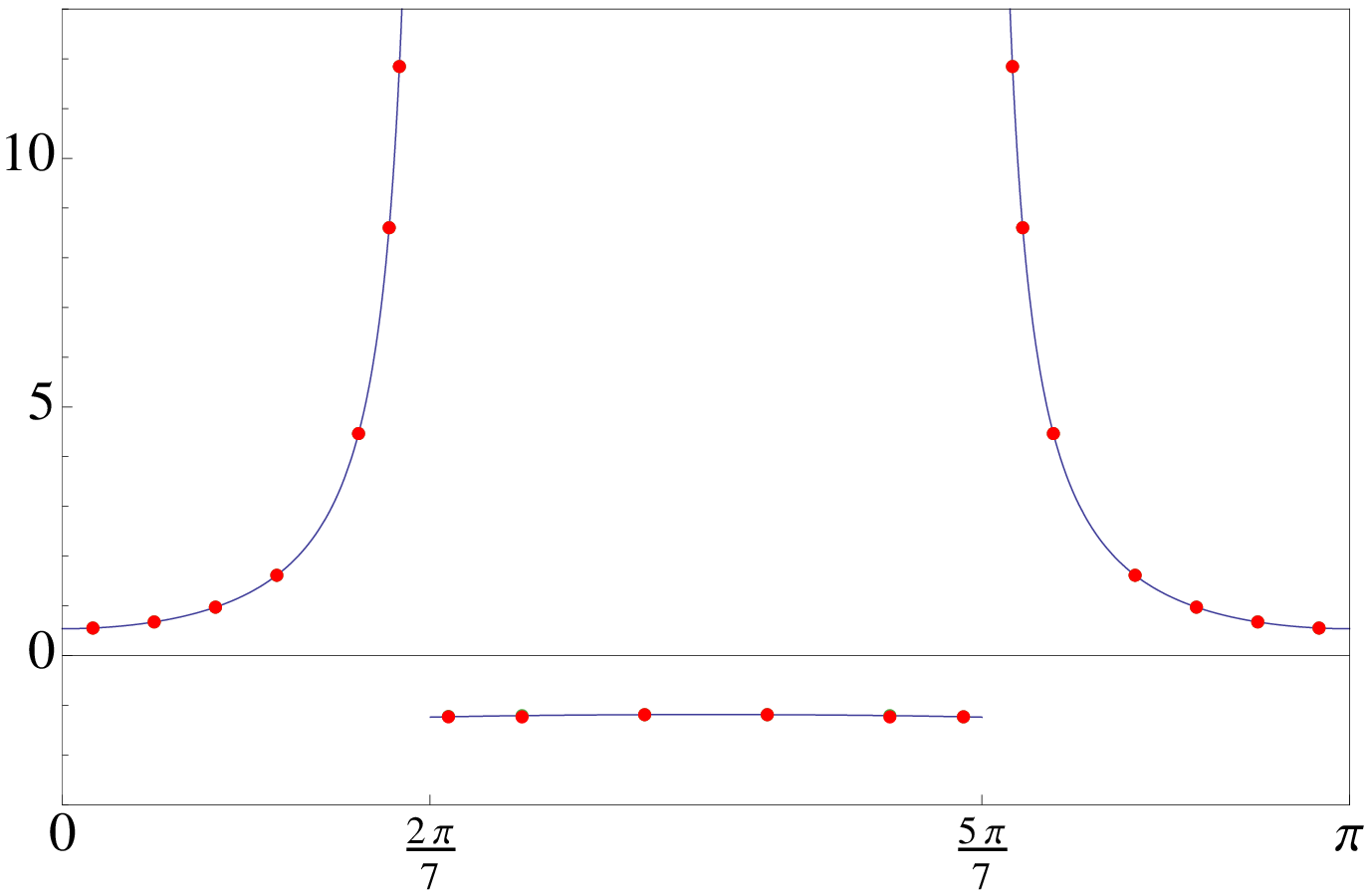}\
   \caption{Plots of the boundary energy ${\cal E}_{\text{bdy}}(w,\xi)$ against $\xi$ for (i) ${\cal LM}(2,5)$ with $\lambda=\frac{3\pi}{5}$, $w=3$ and (ii) ${\cal LM}(4,7)$ with $\lambda=\frac{3\pi}{7}$, $w=2$.}
   \label{BdyEnergy1}
\end{figure}

\begin{figure}[p] 
 \centering
   \ \,\includegraphics[width=3.15in]{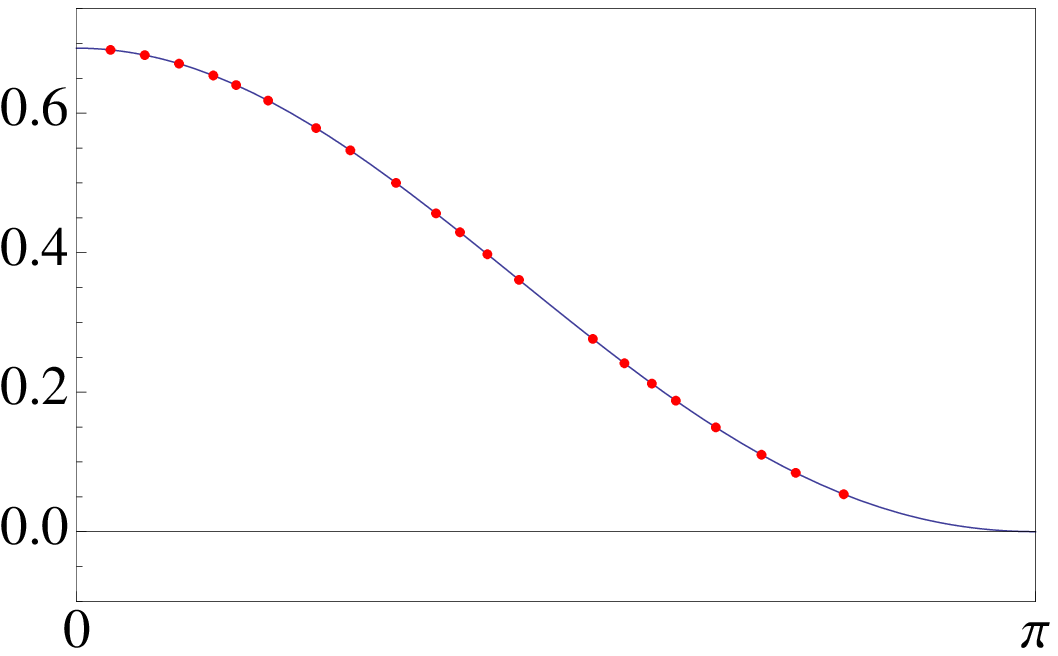} \ \ \includegraphics[width=3.2in]{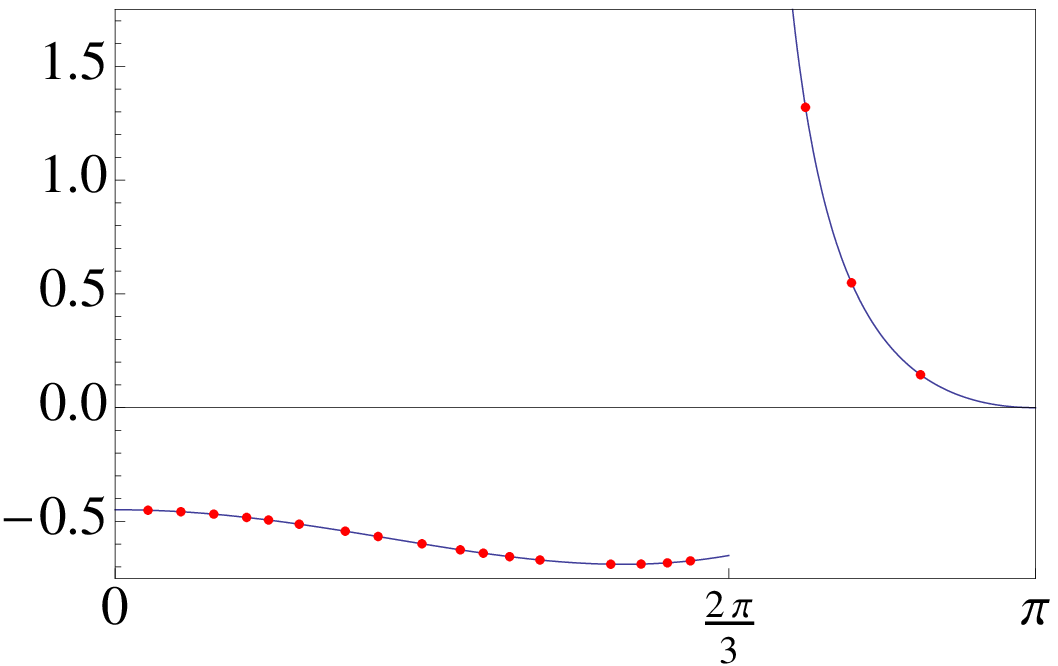}\ \ \ \ \\
   \includegraphics[width=3.3in]{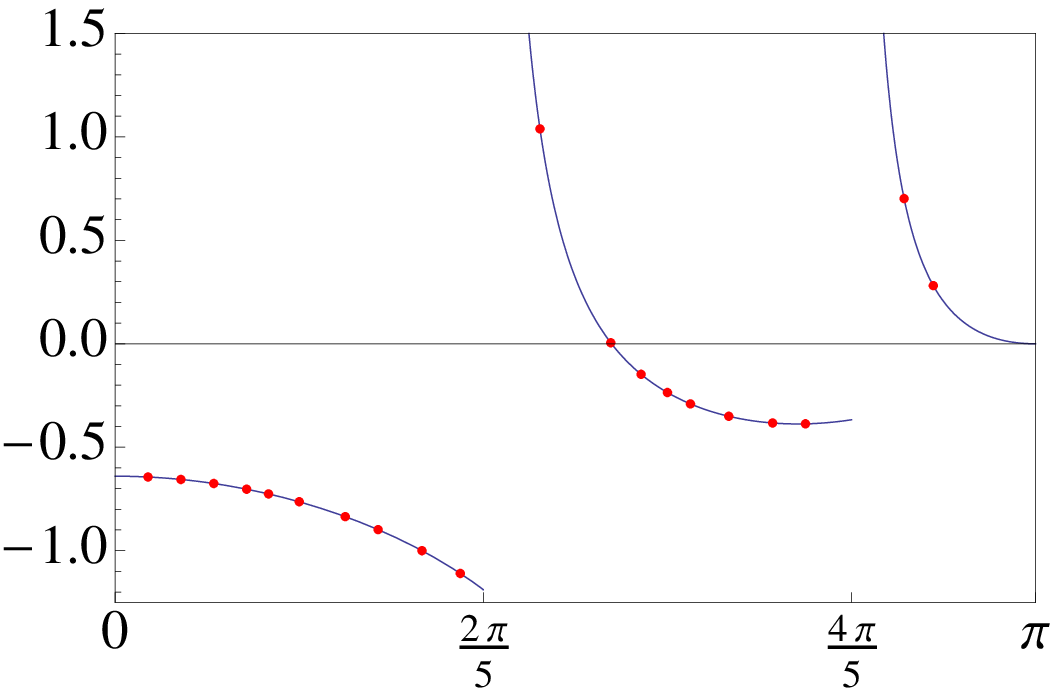}\ \ \includegraphics[width=3.2in]{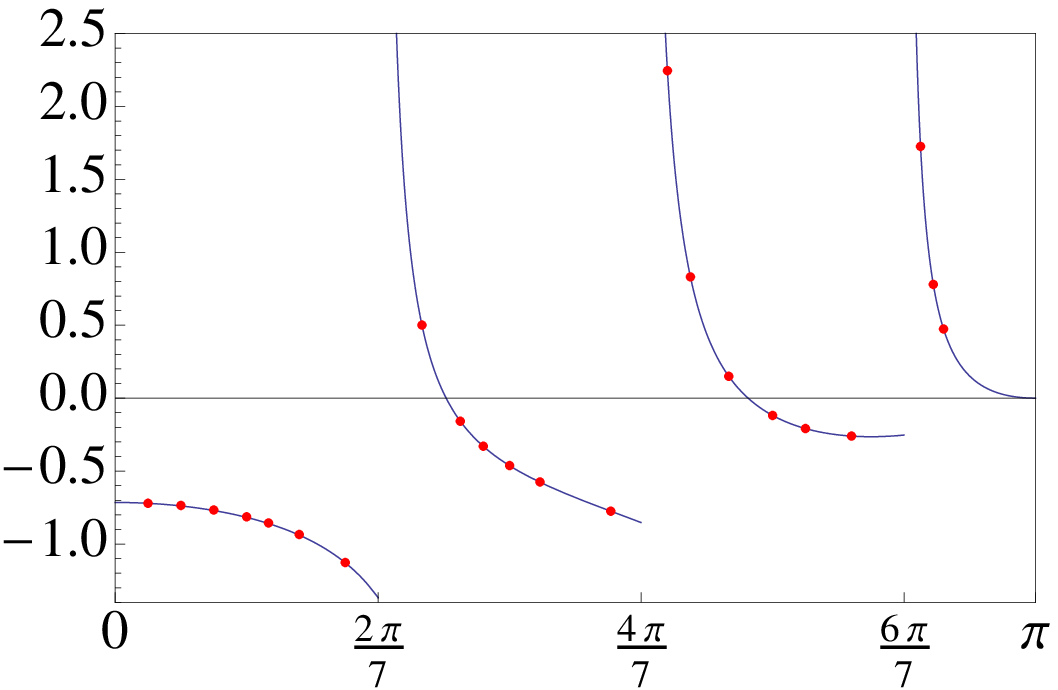}\ \ \
   \caption{Plots of the boundary energy ${\cal E}_{\text{bdy}}(w,\xi)$ against $\lambda$ for $\xi=-\half\lambda$ with $w=0,1,2,3$.}
   \label{BdyEnergyPlots}
\end{figure}

So far, we have only treated the case of the Robin vacuum. Note that there are no contributions to the boundary energies arising from the $s$-type seam. So, finally, we need to find the contribution from the $r$-type seam. In fact, this has been calculated in \cite{PTC} but it is convenient to alternatively include this contribution with the contribution from the Robin vacuum. The expression for the Robin boundary energy (dressed by an $r$-type seam) is suggested as a consequence of an observation made in Appendix~\ref{AppC}. Namely, as an operator, the $r$-type Robin operator is algebraically equivalent (is an image of the same element in the one-boundary TL algebra but under a different representation) to the vacuum Robin operator with the shifted parameters $\xi_w=\xi+w\l$ and $\alpha_{-w}=\a-w\l$. The corresponding Hamiltonians are algebraically related in a similar manner. So it actually suffices to apply these shifts to the values of $\xi$ and $\a$ in (\ref{calER}) to obtain the $(r,s)$ Robin boundary energies
\bea
{\cal E}_{\text{bdy}}(w,\xi)=\calE_0+\calE_R(\xi)+\calE_w(\xi)=\calE_0+\calE_R(\xi_w,\alpha_{-w}),\qquad \xi_w=\xi+w\l,\quad \alpha_{-w}=\a-w\l
\eea
A proper derivation of this formula, based on the inversion relations (\ref{eqs_kappa3}) and (\ref{eqs_kappa4}), is given in Appendix~\ref{AppB}. We have checked numerically that these shifted formulas agree with the expressions coming from \cite{PTC}. 
Example plots of the analytic Robin boundary energies ${\cal E}_{\text{bdy}}(w,\xi)$ against numerical values are shown in Figures~\ref{BdyEnergy1} and \ref{BdyEnergyPlots}. The analytic formulas for the boundary energies, including the jump discontinuities, are numerically confirmed with high accuracy.

\section{Finite-Size Spectra}
\label{sec:numerics}

\subsection{Finitized characters}

\begin{figure}[p]
\begin{center}
\begin{pspicture}(-6,0)(7,8)
\psframe[linewidth=0pt,fillstyle=solid,fillcolor=lightestblue](-6,0)(7,8)
\multiput(-6,0)(2,0){7}{\psframe[linewidth=0pt,fillstyle=solid,fillcolor=lightlightblue](0,0)(1,8)}
\multiput(0,0)(0,3){3}{\psframe[linewidth=0pt,fillstyle=solid,fillcolor=lightlightblue](-6,0)(7,1)}
\multiput(0,0)(0,3){3}{\multiput(-6,0)(2,0){7}{\psframe[linewidth=0pt,fillstyle=solid,fillcolor=lightblue](0,0)(1,1)}}
\psgrid[gridlabels=0pt,subgriddiv=1]
\multiput(-5.75,.4)(0,1){7}{$\cdots$}
\multiput(-4.55,7.35)(1,0){11}{$\vdots$}
\multiput(6.25,.4)(0,1){7}{$\cdots$}
\rput(-5.65,7.65){.}\rput(-5.5,7.5){.}\rput(-5.35,7.35){.}
\rput(6.35,7.35){.}\rput(6.5,7.5){.}\rput(6.65,7.65){.}
\rput(-4.5,6.5){$\frac {261}{8}$}\rput(-3.5,6.5){$26$}\rput(-2.5,6.5){$\frac {161}{8}$}\rput(-1.5,6.5){$15$}\rput(-0.5,6.5){$\frac {85}{8}$}
\rput(.5,6.5){$7$}\rput(1.5,6.5){$\frac {33}{8}$}\rput(2.5,6.5){$2$}\rput(3.5,6.5){$\frac{5}{8}$}\rput(4.5,6.5){$0$}\rput(5.5,6.5){$\frac{1}{8}$}
\rput(-4.5,5.5){$\frac {225}{8}$}\rput(-3.5,5.5){$22$}\rput(-2.5,5.5){$\frac {133}{8}$}\rput(-1.5,5.5){$12$}\rput(-0.5,5.5){$\frac {65}{8}$}
\rput(.5,5.5){$5$}\rput(1.5,5.5){$\frac {21}{8}$}\rput(2.5,5.5){$1$}\rput(3.5,5.5){$\frac{1}{8}$}\rput(4.5,5.5){$0$}\rput(5.5,5.5){$\frac{5}{8}$}
\rput(-4.5,4.5){$\frac {575}{24}$}\rput(-3.5,4.5){$\frac {55}{3}$}\rput(-2.5,4.5){$\frac {323}{24}$}\rput(-1.5,4.5){$\frac {28}{3}$}\rput(-0.5,4.5){$\frac {143}{24}$}
\rput(.5,4.5){$\frac {10}{3}$}\rput(1.5,4.5){$\frac {35}{24}$}\rput(2.5,4.5){$\frac{1}{3}$}\rput(3.5,4.5){$-\frac{1}{24}$}\rput(4.5,4.5){$\frac{1}{3}$}\rput(5.5,4.5){$\frac{35}{24}$}
\rput(-4.5,3.5){$\frac {161}{8}$}\rput(-3.5,3.5){$15$}\rput(-2.5,3.5){$\frac {85}{8}$}\rput(-1.5,3.5){$7$}\rput(-0.5,3.5){$\frac {33}{8}$}
\rput(.5,3.5){$2$}\rput(1.5,3.5){$\frac 5{8}$}\rput(2.5,3.5){$0$}\rput(3.5,3.5){$\frac{1}{8}$}\rput(4.5,3.5){$1$}\rput(5.5,3.5){$\frac{21}{8}$}
\rput(-4.5,2.5){$\frac {133}{8}$}\rput(-3.5,2.5){$12$}\rput(-2.5,2.5){$\frac {65}{8}$}\rput(-1.5,2.5){$5$}\rput(-0.5,2.5){$\frac {21}{8}$}
\rput(.5,2.5){$1$}\rput(1.5,2.5){$\frac 1{8}$}\rput(2.5,2.5){$0$}\rput(3.5,2.5){$\frac{5}{8}$}\rput(4.5,2.5){$2$}\rput(5.5,2.5){$\frac{33}{8}$}
\rput(-4.5,1.5){$\frac {323}{24}$}\rput(-3.5,1.5){$\frac {28}{3}$}\rput(-2.5,1.5){$\frac {143}{24}$}\rput(-1.5,1.5){$\frac {10}{3}$}\rput(-0.5,1.5){$\frac {35}{24}$}
\rput(.5,1.5){$\frac 1{3}$}\rput(1.5,1.5){$-\frac {1}{24}$}\rput(2.5,1.5){$\frac{1}{3}$}\rput(3.5,1.5){$\frac{35}{24}$}\rput(4.5,1.5){$\frac{10}{3}$}\rput(5.5,1.5){$\frac{143}{24}$}
\rput(-4.5,0.5){$\frac {85}{8}$}\rput(-3.5,0.5){$7$}\rput(-2.5,0.5){$\frac {33}{8}$}\rput(-1.5,0.5){$2$}\rput(-0.5,0.5){$\frac {5}{8}$}
\rput(.5,.5){$0$}\rput(1.5,.5){$\frac 1{8}$}\rput(2.5,.5){$1$}\rput(3.5,.5){$\frac{21}{8}$}\rput(4.5,.5){$5$}\rput(5.5,.5){$\frac{65}{8}$}
{\color{blue}
\rput(-4.5,-.5){$-5$}
\rput(-3.5,-.5){$-4$}
\rput(-2.5,-.5){$-3$}
\rput(-1.5,-.5){$-2$}
\rput(-0.5,-.5){$-1$}
\rput(.5,-.5){$0$}
\rput(1.5,-.5){$1$}
\rput(2.5,-.5){$2$}
\rput(3.5,-.5){$3$}
\rput(4.5,-.5){$4$}
\rput(5.5,-.5){$5$}
\rput(6.5,-.5){$r$}
\rput(-6.5,.5){$1$}
\rput(-6.5,1.5){$2$}
\rput(-6.5,2.5){$3$}
\rput(-6.5,3.5){$4$}
\rput(-6.5,4.5){$5$}
\rput(-6.5,5.5){$6$}
\rput(-6.5,6.5){$7$}
\rput(-6.5,7.5){$s$}}
\end{pspicture}
\vspace{1cm}
\caption{Robin Kac table of conformal weights $\Delta_{r,s-\frac{1}{2}}$ for critical percolation ${\cal LM}(2,3)$.\label{23Kac}}
\end{center}
\end{figure}
\begin{figure}[p]
\begin{center}
\begin{pspicture}(-6,0)(7,8)
\psframe[linewidth=0pt,fillstyle=solid,fillcolor=lightestblue](-6,0)(7,8)
\multiput(-6,0)(2,0){7}{\psframe[linewidth=0pt,fillstyle=solid,fillcolor=lightlightblue](0,0)(1,8)}
\multiput(0,0)(0,5){2}{\psframe[linewidth=0pt,fillstyle=solid,fillcolor=lightlightblue](-6,0)(7,1)}
\multiput(0,0)(0,5){2}{\multiput(-6,0)(2,0){7}{\psframe[linewidth=0pt,fillstyle=solid,fillcolor=lightblue](0,0)(1,1)}}
\psgrid[gridlabels=0pt,subgriddiv=1]
\multiput(-5.75,.4)(0,1){7}{$\cdots$}
\multiput(-4.55,7.35)(1,0){11}{$\vdots$}
\multiput(6.25,.4)(0,1){7}{$\cdots$}
\rput(-5.65,7.65){.}\rput(-5.5,7.5){.}\rput(-5.35,7.35){.}
\rput(6.35,7.35){.}\rput(6.5,7.5){.}\rput(6.65,7.65){.}
\rput(-4.5,6.5){$\frac {287}{8}$}\rput(-3.5,6.5){$27$}\rput(-2.5,6.5){$\frac {155}{8}$}\rput(-1.5,6.5){$13$}\rput(-0.5,6.5){$\frac {63}{8}$}
\rput(.5,6.5){$4$}\rput(1.5,6.5){$\frac {11}{8}$}\rput(2.5,6.5){$0$}\rput(3.5,6.5){$-\frac{1}{8}$}\rput(4.5,6.5){$1$}\rput(5.5,6.5){$\frac{27}{8}$}
\rput(-4.5,5.5){$\frac {1287}{40}$}\rput(-3.5,5.5){$\frac {119}{5}$}\rput(-2.5,5.5){$\frac {667}{40}$}\rput(-1.5,5.5){$\frac {54}{5}$}\rput(-0.5,5.5){$\frac {247}{40}$}
\rput(.5,5.5){$\frac {14}{5}$}\rput(1.5,5.5){$\frac {27}{40}$}\rput(2.5,5.5){$-\frac{1}{5}$}\rput(3.5,5.5){$\frac{7}{40}$}\rput(4.5,5.5){$\frac{9}{5}$}\rput(5.5,5.5){$\frac{187}{40}$}
\rput(-4.5,4.5){$\frac {1147}{40}$}\rput(-3.5,4.5){$\frac {104}{5}$}\rput(-2.5,4.5){$\frac {567}{40}$}\rput(-1.5,4.5){$\frac {44}{5}$}\rput(-0.5,4.5){$\frac {187}{40}$}
\rput(.5,4.5){$\frac {9}{5}$}\rput(1.5,4.5){$\frac {7}{40}$}\rput(2.5,4.5){$-\frac{1}{5}$}\rput(3.5,4.5){$\frac{27}{40}$}\rput(4.5,4.5){$\frac{14}{5}$}\rput(5.5,4.5){$\frac{247}{40}$}
\rput(-4.5,3.5){$\frac {203}{8}$}\rput(-3.5,3.5){$18$}\rput(-2.5,3.5){$\frac {95}{8}$}\rput(-1.5,3.5){$7$}\rput(-0.5,3.5){$\frac {27}{8}$}
\rput(.5,3.5){$1$}\rput(1.5,3.5){$-\frac 1{8}$}\rput(2.5,3.5){$0$}\rput(3.5,3.5){$\frac{11}{8}$}\rput(4.5,3.5){$4$}\rput(5.5,3.5){$\frac{63}{8}$}
\rput(-4.5,2.5){$\frac {891}{40}$}\rput(-3.5,2.5){$\frac {77}{5}$}\rput(-2.5,2.5){$\frac {391}{40}$}\rput(-1.5,2.5){$\frac {27}{5}$}\rput(-0.5,2.5){$\frac {91}{40}$}
\rput(.5,2.5){$\frac {2}{5}$}\rput(1.5,2.5){$-\frac 9{40}$}\rput(2.5,2.5){$\frac{2}{5}$}\rput(3.5,2.5){$\frac{91}{40}$}\rput(4.5,2.5){$\frac{27}{5}$}\rput(5.5,2.5){$\frac{391}{40}$}
\rput(-4.5,1.5){$\frac {155}{8}$}\rput(-3.5,1.5){$13$}\rput(-2.5,1.5){$\frac {63}{8}$}\rput(-1.5,1.5){$4$}\rput(-0.5,1.5){$\frac {11}{8}$}
\rput(.5,1.5){$0$}\rput(1.5,1.5){$-\frac 1{8}$}\rput(2.5,1.5){$1$}\rput(3.5,1.5){$\frac{27}{8}$}\rput(4.5,1.5){$7$}\rput(5.5,1.5){$\frac{95}{8}$}
\rput(-4.5,0.5){$\frac {667}{40}$}\rput(-3.5,0.5){$\frac {54}{5}$}\rput(-2.5,0.5){$\frac {247}{40}$}\rput(-1.5,0.5){$\frac {14}{5}$}\rput(-0.5,0.5){$\frac {27}{40}$}
\rput(.5,.5){$-\frac 1{5}$}\rput(1.5,.5){$\frac 7{40}$}\rput(2.5,.5){$\frac{9}{5}$}\rput(3.5,.5){$\frac{187}{40}$}\rput(4.5,.5){$\frac{44}{5}$}\rput(5.5,.5){$\frac{567}{40}$}
{\color{blue}
\rput(-4.5,-.5){$-5$}
\rput(-3.5,-.5){$-4$}
\rput(-2.5,-.5){$-3$}
\rput(-1.5,-.5){$-2$}
\rput(-0.5,-.5){$-1$}
\rput(.5,-.5){$0$}
\rput(1.5,-.5){$1$}
\rput(2.5,-.5){$2$}
\rput(3.5,-.5){$3$}
\rput(4.5,-.5){$4$}
\rput(5.5,-.5){$5$}
\rput(6.5,-.5){$r$}
\rput(-6.5,.5){$1$}
\rput(-6.5,1.5){$2$}
\rput(-6.5,2.5){$3$}
\rput(-6.5,3.5){$4$}
\rput(-6.5,4.5){$5$}
\rput(-6.5,5.5){$6$}
\rput(-6.5,6.5){$7$}
\rput(-6.5,7.5){$s$}}
\end{pspicture}
\vspace{1cm}
\caption{Robin Kac table of conformal weights $\Delta_{r,s-\frac{1}{2}}$ for the logarithmic Yang-Lee model ${\cal LM}(2,5)$.\label{25Kac}}
\label{Kac}
\end{center}
\end{figure}

We conjecture that, if $\Delta_{r,s-\frac{1}{2}}\ne \Delta_{r'\!,s'}$ for any $r'\!,s'\in{\Bbb N}$, the $(r,s-\tfrac{1}{2})$ Robin representations correspond to irreducible (highest weight) Virasoro Verma modules. 
If $\Delta_{r,s-\frac{1}{2}}= \Delta_{r'\!,s'}$ for some $r'\!,s'\in{\Bbb N}$, the $(r,s-\tfrac{1}{2})$ Robin representation may be reducible. These modules may exhibit Feigin-Fuchs structures~\cite{logKac}. 
In all cases, the associated conformal characters 
\bea
\ch_\Delta^{p,p'}(q)=q^{-\frac{c}{24}+\Delta}\,\widehat{\ch}_\Delta^{p,p'}(q)=q^{-\frac{c}{24}+\Delta} \sum_E q^E
\eea
are the spectrum generating functions of the integer conformal energies $E\ge 0$ with $c=c^{p,p'}$ as in (\ref{KacFormula}). The finitized characters, given by the finitized conformal partition functions in the $(r,s)$ sector, are
\bea
\ch_{r,s-\frac{1}{2}}^{p,p';(N)}(q)=Z_{(1,1)|(r,s)}^{p,p';(N)}(q)=q^{-\frac{c}{24}+\Delta^{p,p'}_{r,s-1/2}} \bigg[\begin{matrix}N \\ \left\lfloor\frac{N-d}{2}\right\rfloor+(-1)^{N-d-w}\left\lceil\frac{w}{2}\right\rceil\end{matrix}\bigg]_q,\qquad r\in {\Bbb Z},\ s\in {\Bbb N}
\eea
These encode finitely truncated sets of the integer conformal energies of the infinite system. 
Setting $q=1$ in the $q$-binomial gives the correct counting of states (\ref{countstates}). 
Taking the thermodynamic limit $N\to\infty$ gives the Virasoro Verma characters
\bea
\ch_{r,s-\frac{1}{2}}^{p,p'}(q)=\frac{q^{-\frac{c}{24}+\Delta^{p,p'}_{r,s-1/2}}}{(q)_\infty},\qquad r\in {\Bbb Z},\ s\in {\Bbb N},\qquad (q)_\infty=\prod_{n=1}^\infty (1-q^n)\label{VirChars}
\eea

Example Robin Kac tables of conformal weights $\Delta^{p,p'}_{r,s-\frac{1}{2}}$ are shown for ${\cal LM}(2,3)$ and ${\cal LM}(2,5)$ in Figures~\ref{23Kac} and \ref{25Kac}. These Kac tables can be extended to $s\le 0$ using the periodicity
\bea
\Delta^{p,p'}_{r,s}=\Delta^{p,p'}_{r+p,s+p'},\qquad (r,s)\equiv (r+p,s+p')
\eea
In the case that these conformal weights correspond to irreducible (highest weight) Virasoro Verma modules, any two modules with the same conformal weight are isomorphic and can be identified. 
As in the case of critical dense polymers ${\cal LM}(1,2)$~\cite{PRT14}, the Kac tables extended to $s\le 0$ are expected to encode the $su(2)$ fusion rules. 

\subsection{Logarithmic limit}

The conformal data and Kac characters of the Kac boundary conditions~\cite{PRZ2006,RasKac,PRV,PTC,logKac} can be  understood in terms of taking a logarithmic limit~\cite{LogLimit} of the conformal data of the rational nonunitary minimal models $ {\cal M}(m,m')$. Symbolically, this limit of the minimal CFTs is given  by
\bea
\lim_{m,m'\to\infty, \  {m'\over m}\to {p'\over p}+} {\cal M}(m,m')={\cal LM}(p,p'),\qquad 1\le p<p',\quad \mbox{$p,p'$ coprime}\label{logLimit}
\eea
The (one-sided) limit is taken through coprime pairs $(m,m')$ with ${m'\over m}>{p'\over p}$ to ensure the correct limiting ground states. The logarithmic limit is taken in the continuum scaling limit, after the thermodynamic limit. Since $p\ge 1$, the limit must ultimately be taken through a sequence of nonunitary models with \mbox{$m'\!-\!m>1$}.
The equality indicates the identification of the spectra of these CFTs. In principle, with $(r,s)$ Robin boundary conditions on both sides of the strip, the Jordan cells appearing in the reducible yet indecomposable representations of the logarithmic minimal models should emerge in this limit but there are subtleties~\cite{LogLimit}.
Here we consider the limit of chiral spectra, corresponding to a single character, for which purpose the logarithmic limit is robust in the sense that there are no Jordan cells and the limit is independent of the choice of the sequence.

Taking the logarithmic limit of the conformal data of the rational minimal models ${\cal M}(m,m')$ yields the conformal  data of the logarithmic minimal models ${\cal LM}(p,p')$ including the central charges, conformal weights (\ref{KacFormula}) and Kac characters
\bea
\chi^{p,p'}_{r,s}(q)= q^{-c/24+\Delta_{r,s}^{p,p'}}
{(1-q^{rs})\over (q)_\infty},\qquad  r,s=1,2,3,\ldots
\eea
Explicitly, using $|q|<1$, the limiting CFT data for $r,s=1,2,3,\ldots$ are given by
\begin{align}
 &\qquad\qquad  c^{m,m'}=1-{6(m'-m)^2\over mm'}\ {\to}\  1-{6(p'-p)^2\over pp'}=c^{p,p'}\\
&\Delta_{r,s}^{m,m'}={(rm'-sm)^2-(m'-m)^2\over 4mm'}\ {\to}\ {(rp'-sp)^2-(p'-p)^2\over 4pp'}=\Delta_{r,s}^{p,p'}\\
\mbox{ch}^{m,m'}_{r,s}\!(q)&={q^{-{c\over 24}+\Delta_{r,s}^{m,m'}}\over (q)_\infty}\!\!\! 
\sum_{k=-\infty}^\infty \!\!\!\big[q^{k(k m m'+rm'-sm)}\!-\!q^{(km+r)(km'+s)}\big] {\to}\,\disp q^{-{c\over 24}+\Delta_{r,s}^{p,p'}}\,
{(1\!-\!q^{rs})\over (q)_\infty}=\chi^{p,p'}_{r,s}\!(q)
\end{align}

The characters associated with half-integer Kac labels can also be obtained using the logarithmic limit. 
Since $m$ and $m'$ are coprime, at most one can be even. We fix these parities and take three different logarithmic limits. The half-integer boundary conditions, correspond to the logarithmic limit of free boundary conditions for the minimal models labelled~\cite{BehrendP} by either \mbox{$(r+\frac{m}{2},s+\frac{m'-1}{2})$}, \mbox{$(r+\frac{m-1}{2},s+\frac{m'}{2})$} or \mbox{$(r+\frac{m-1}{2},s+\frac{m'-1}{2})$} depending on the parities of $m,m'$. For $m,m'$ large with $r,s\in{\Bbb Z}$ finite, these sit at the center of a very large but finite rational Kac table. Explicitly, the three logarithmic limits give infinitely extended half-integer Kac tables with
\begin{subequations}
\begin{align}
\lim_{m,m'\to\infty, \  {m'\over m}\to {p'\over p}+\atop \mbox{\tiny $m$ even, $m'$ odd}}\ch^{m,m'}_{r+\frac{m}{2},s+\frac{m'-1}{2}}\!(q)&
\ =\  {q^{-{c\over 24}+\Delta_{r,s-\frac{1}{2}}^{p,p'}}\over (q)_\infty}=\ch^{p,p'}_{r,s-\frac{1}{2}}\!(q),\qquad r,s\in{\Bbb Z}
\label{halfchar1}\\[-6pt]
\lim_{m,m'\to\infty, \  {m'\over m}\to {p'\over p}+\atop \mbox{\tiny $m$ odd, $m'$ even}}\ch^{m,m'}_{r+\frac{m-1}{2},s+\frac{m'}{2}}\!(q)&
\ =\  {q^{-{c\over 24}+\Delta_{r-\frac{1}{2},s}^{p,p'}}\over (q)_\infty}=\ch^{p,p'}_{r-\frac{1}{2},s}\!(q),\qquad r,s\in{\Bbb Z}\label{halfchar2}\\[-6pt]
\lim_{m,m'\to\infty, \  {m'\over m}\to {p'\over p}+\atop \mbox{\tiny $m$ odd, $m'$ odd}}\ch^{m,m'}_{r+\frac{m-1}{2},s+\frac{m'-1}{2}}\!(q)&
\ =\  {q^{-{c\over 24}+\Delta_{r-\frac{1}{2},s-\frac{1}{2}}^{p,p'}}\over (q)_\infty}=\ch^{p,p'}_{r-\frac{1}{2},s-\frac{1}{2}}\!(q),\qquad r,s\in{\Bbb Z}\label{halfchar3}
\end{align}
\label{halfchar}
\end{subequations}
The notation $\ch_{a,b}$ refers to Virasoro characters if the indices $a,b$ are both integers and to Robin Verma characters if either $a$ or $b$ is half-integer.

From Bezout's lemma, since $p,p'$ are coprime, there exists a Bezout pair $(r_0,s_0) \in {\Bbb N}^2$ such that $r_0 p'-s_0 p = 1$. 
Suppose $(p,p')=\mbox{(even,odd)}$. Then for $(r,s) \in {\Bbb Z}^2$ we see that
\begin{subequations}
\bea
&rp'-(s-\half)p = rp'-sp+\half p = rp'-sp+\half p(r_0 p'-s_0 p) = (r+\half p r_0)p' -(s+\half ps_0)p&\qquad\\[6pt]
&\Delta^{p,p'}_{r,s-\frac{1}{2}}=\Delta^{p,p'}_{r+\frac{1}{2}pr_0,s+\frac{1}{2}ps_0},\quad (r+\half pr_0,s+\half ps_0)\in {\Bbb Z}^2&\qquad
\eea
\end{subequations}
So, in this parity case, the Kac conformal weights $\Delta^{p,p'}_{r,s}$ and Robin conformal weights $\Delta^{p,p'}_{r,s-\frac{1}{2}}$ coincide up to a shift in $(r,s)$, even though the characters are different. 
Similar results hold in the other parity cases. To summarize, we find that
\begin{subequations}
\begin{align}
&(p,p')=\mbox{(even,odd)}\!:&&\\ 
&\qquad\{\Delta^{p,p'}_{r,s-\frac{1}{2}}\!: r,s\in{\Bbb Z}\}\equiv \{\Delta^{p,p'}_{r,s}\!: r,s\in{\Bbb Z}\}, 
&\{\Delta^{p,p'}_{r-\frac{1}{2},s-\frac{1}{2}}\!: r,s\in{\Bbb Z}\}\equiv \{\Delta^{p,p'}_{r-\frac{1}{2},s}\!: r,s\in{\Bbb Z}\}&
\nonumber\\
&(p,p')=\mbox{(odd,even)}\!:&&\\
&\qquad\{\Delta^{p,p'}_{r-\frac{1}{2},s}\!: r,s\in{\Bbb Z}\}\equiv \{\Delta^{p,p'}_{r,s}\!: r,s\in{\Bbb Z}\}, 
&\{\Delta^{p,p'}_{r-\frac{1}{2},s-\frac{1}{2}}\!: r,s\in{\Bbb Z}\}\equiv \{\Delta^{p,p'}_{r,s-\frac{1}{2}}\!: r,s\in{\Bbb Z}\}&
\nonumber\\
&(p,p')=\mbox{(odd,odd)}\!:&&\\
&\qquad\{\Delta^{p,p'}_{r-\frac{1}{2},s-\frac{1}{2}}\!: r,s\in{\Bbb Z}\}\equiv \{\Delta^{p,p'}_{r,s}\!: r,s\in{\Bbb Z}\}, 
&\{\Delta^{p,p'}_{r-\frac{1}{2},s}\!: r,s\in{\Bbb Z}\}\equiv \{\Delta^{p,p'}_{r,s-\frac{1}{2}}\!: r,s\in{\Bbb Z}\}\quad&
\nonumber
\end{align}
\end{subequations}

%
%

\subsection{Finite-size corrections}

Consider the ${\cal LM}(p,p')$ lattice models on a strip with $N$ columns and $M$ double rows with Neumann boundary conditions applied on the left edge and $(r,s)$ Robin boundary conditions on the right edge. 
The lattice partition function is 
\be 
Z_{(1,1)|(r,s)}^{(N,M)}=\mbox{Tr}\,\vec D(u,\xi)^{M}=\sum_j D_j(u,\xi)^{M} =\sum_j e^{-M E_j(u,\xi)},\qquad j=0,1,2,3,\ldots
\ee
where $D_j(u,\xi)$ are the eigenvalues of $\drtm(u,\xi)$ and $E_j(u,\xi)$ are their associated energies. In the thermodynamic limit, only the ground state eigenvalue $D_0(u,\xi)$ of the double row transfer matrix in each $(r,s)$ sector contributes to the lattice partition function. 

The conformal data of interest is accessible~\cite{BCN,Aff} through the finite-size corrections to the eigenvalues of the transfer matrix or associated Hamiltonian. 
For the double row transfer matrix eigenvalues, the leading finite-size corrections for large $N$ take the form
\be 
E_j(u,\xi)=-\log D_j(u,\xi)=2Nf_{\text{bulk}}(u)+f_{\text{bdy}}(u,w,\xi)+\frac{2\pi\sin\vartheta}{N}\Big(\!-\frac{c}{24}+\Delta_{r,s-\frac{1}{2}}^{p,p'}+k\Big)+...,\quad k=0,1,2,... \label{fsdtm}
\ee
where $k$ labels the level in the conformal tower. 
The anisotropy angle $\vartheta$~\cite{KimP} and modular nome $q$ are
\bea
\vartheta=\frac{\pi u}{\lambda},\qquad \lambda=\frac{(p'-p)\pi}{p'},\qquad 
q=\exp\!\Big(\!\!-2\pi\,\frac{M}{N}\sin\vartheta\Big)
\eea
Referring to (\ref{KacFormula}), the central charge of the CFT is $c=c^{p,p'}$ while the 
spectrum of conformal weights is given by the possible values of $\Delta_{r,s-\frac{1}{2}}^{p,p'}$ in the Kac table with excitations or descendants labelled by the non-negative integers $k$.
Similarly, for the associated quantum Hamiltonian $\mathcal{H}^{w,d}$, the finite-size corrections of the eigenenergies take the form
\be 
E_j=E_j(\xi)=N\calE_{\text{bulk}}+\calE_{\text{bdy}}(w,\xi)+\frac{\pi v_s}{N}\Big(\!-\frac{c}{24}+\Delta_{r,s-\frac{1}{2}}^{p,p'}+k\Big)+...,\quad k=0,1,2,... 
\label{eq:finite_size_corr_ham}
\ee
where $v_s=\frac{\pi \sin\lambda}{\lambda}$ is the velocity of sound. The Hamiltonian energies $\calE_{\text{bulk}}$ and $\calE_{\text{bdy}}(w,\xi)$ are determined, up to the shift of the ground state energy, by evaluating the derivative at $u=0$ of $f_{\text{bulk}}(u)$ and $f_{\text{bdy}}(u,w,\xi)$ respectively.

\begin{figure}[p]
   \centering
   \includegraphics[width=6.in]{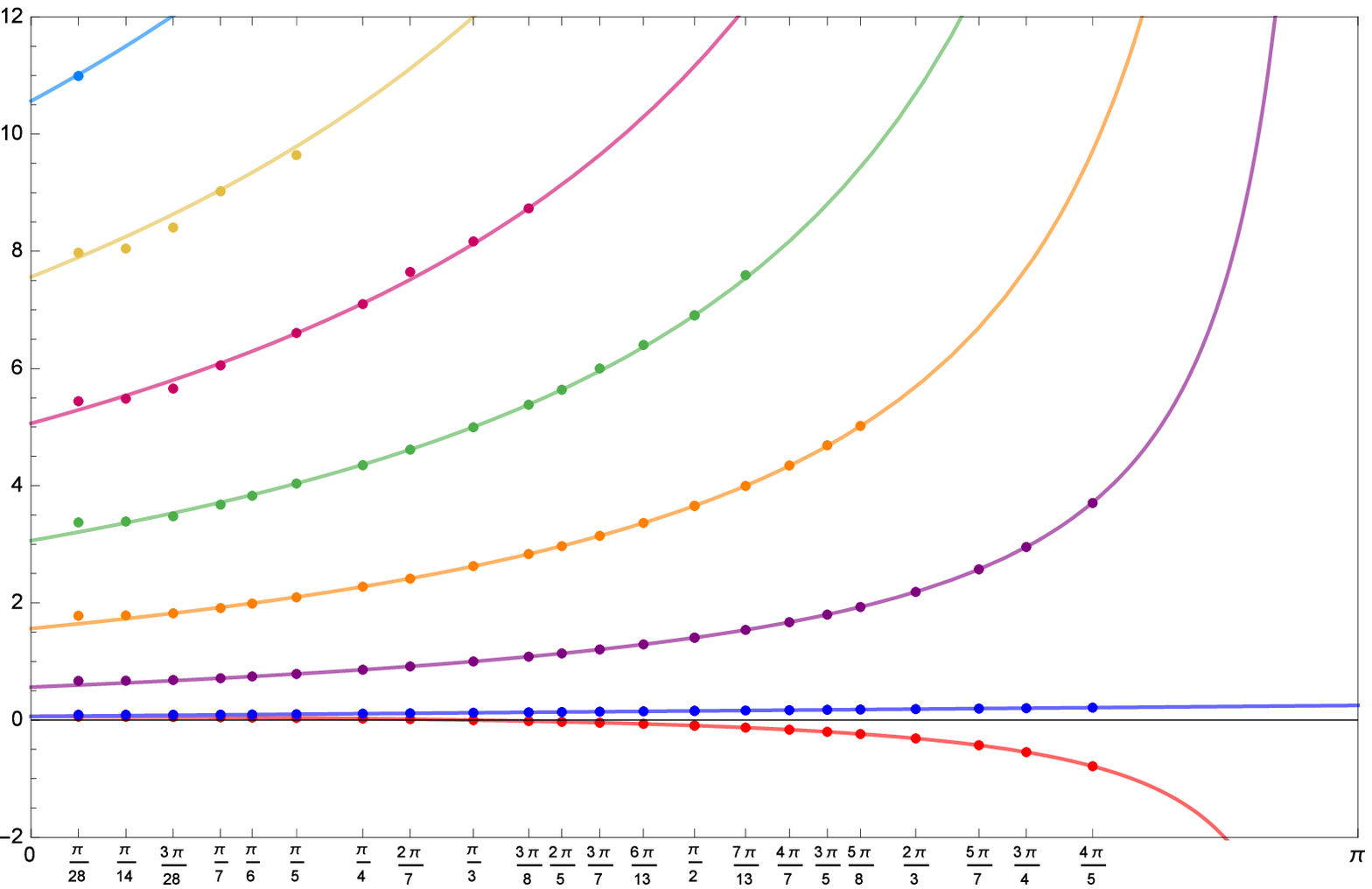} 
   \caption{Plot of the numerical conformal weights $\Delta^{p,p'}_{r,\frac{1}{2}}$ as a function of $\lambda$  for $r=0,1,2,\ldots,6,7$. 
    \label{fig:ConfWtsPos}}
\end{figure}
\begin{figure}[p] 
   \centering
   \includegraphics[width=6.in]{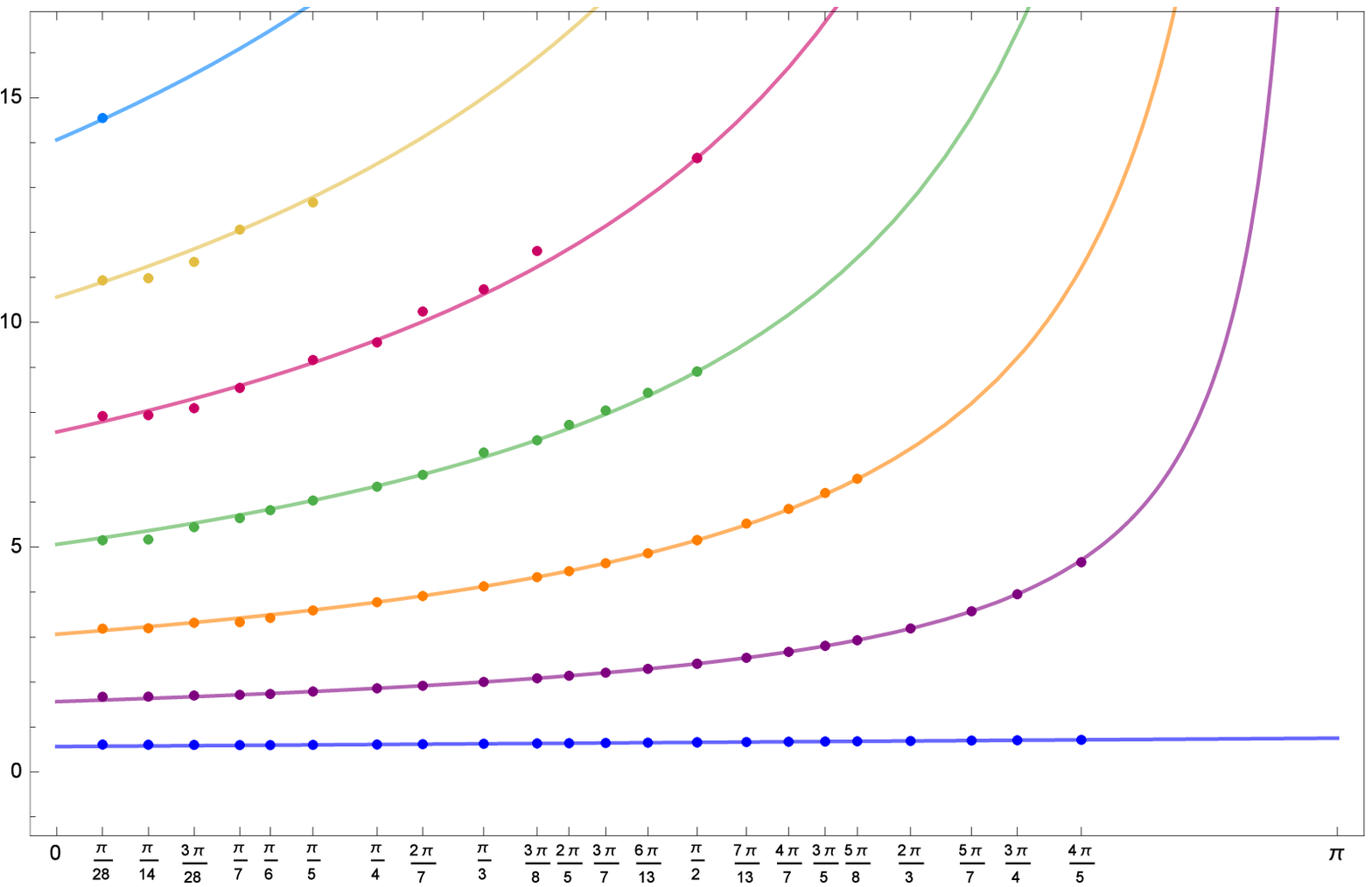} 
   \caption{Plot of the numerical conformal weights $\Delta^{p,p'}_{r,\frac{1}{2}}$ as a function of $\lambda$  for $r=-1,-2,\ldots,-6,-7$.
   \label{fig:ConfWtsNeg}}
\end{figure}

\begin{figure}[p]
   \centering
   \includegraphics[width=6.in]{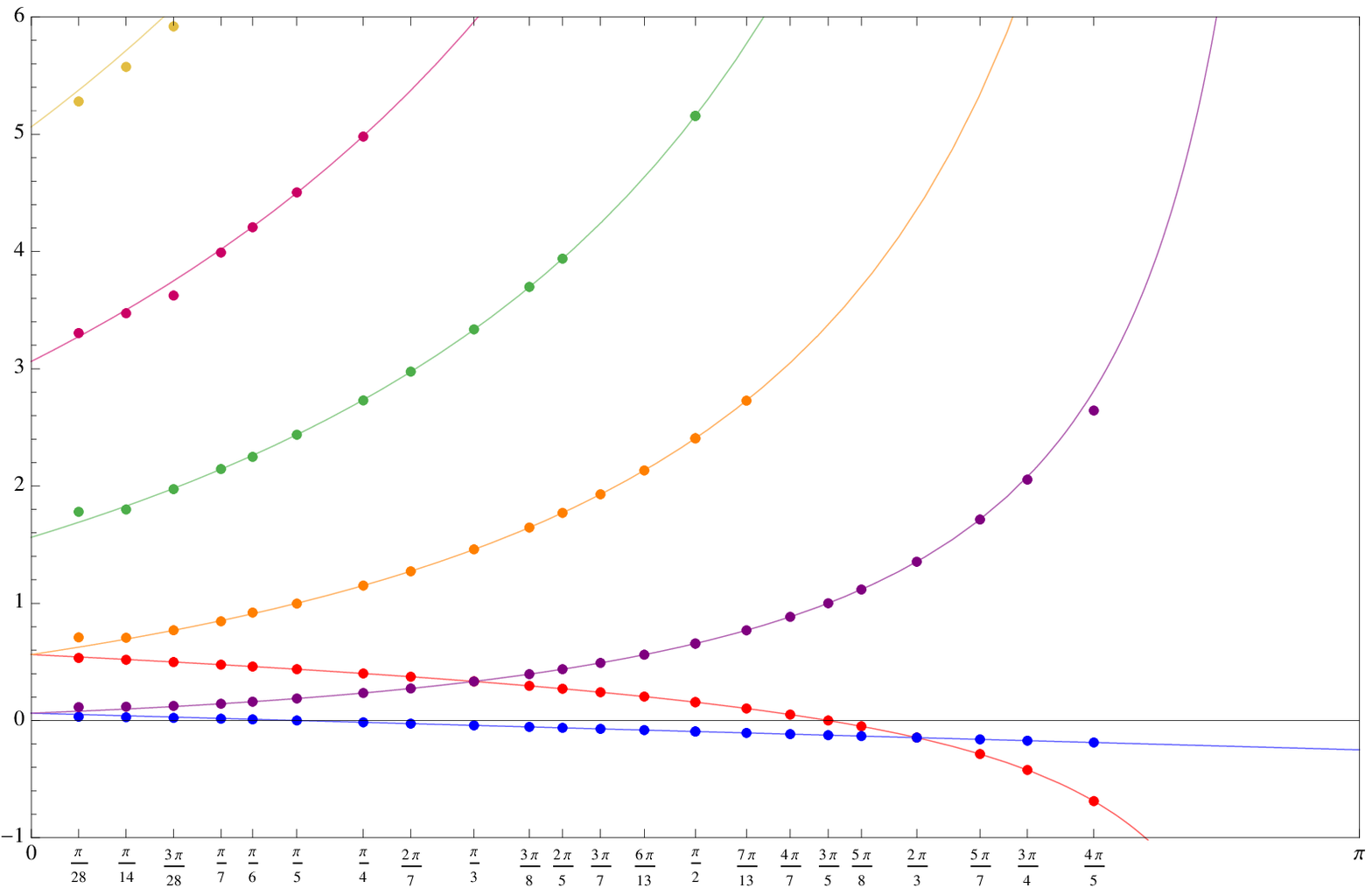} 
   \caption{Plot of the numerical conformal weights $\Delta^{p,p'}_{r,\frac{3}{2}}$ as a function of $\lambda$  for $r=0,1,2,\ldots,5,6$. 
    \label{fig:s2ConfWtsPos}}
\end{figure}
\begin{figure}[p] 
   \centering
   \includegraphics[width=6.in]{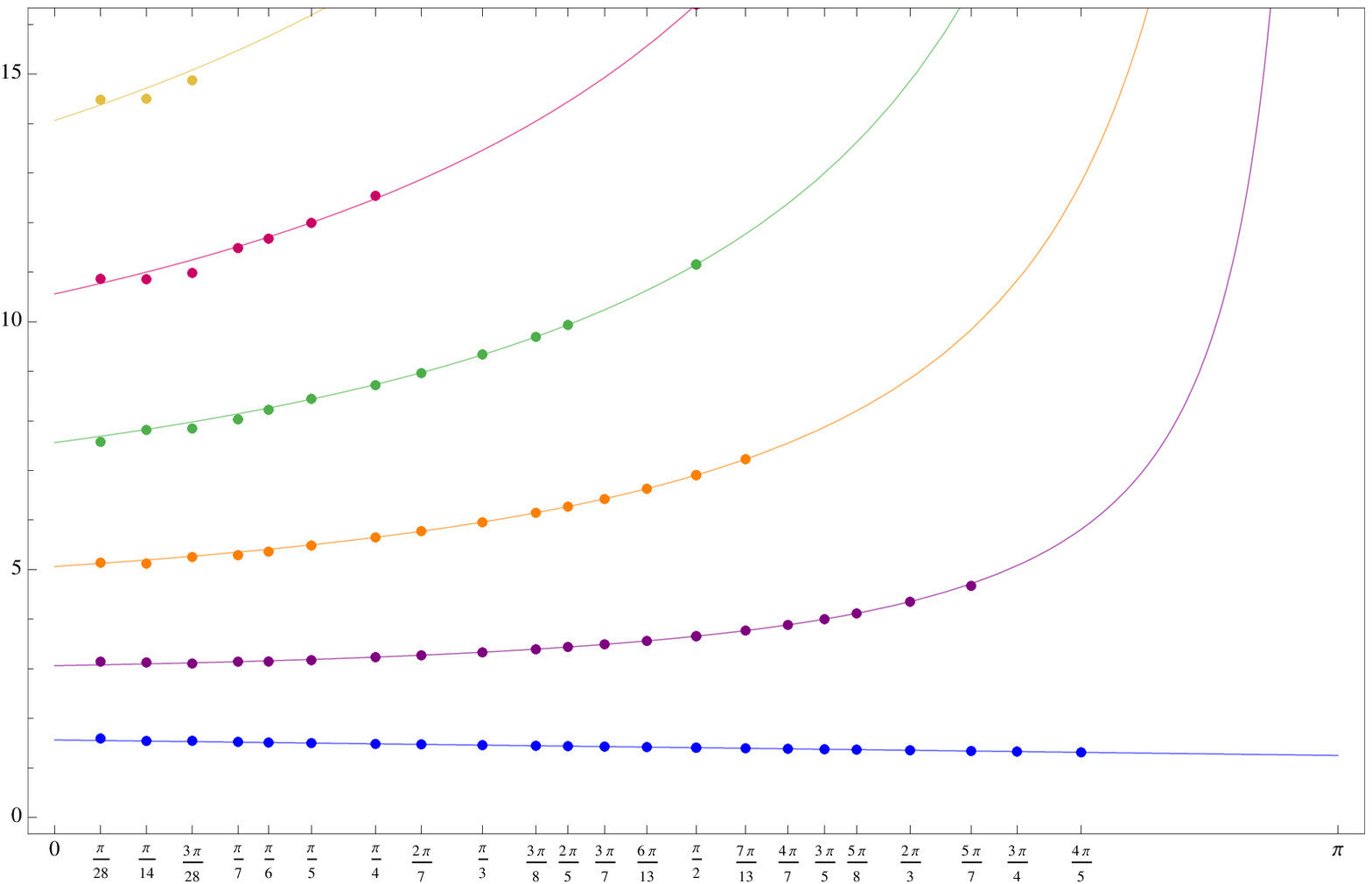} 
   \caption{Plot of the numerical conformal weights $\Delta^{p,p'}_{r,\frac{3}{2}}$ as a function of $\lambda$  for $r=-1,-2,\ldots,-5,-6$.
   \label{fig:s2ConfWtsNeg}}
\end{figure}

\begin{figure}[htb]
   \centering
   \includegraphics[width=6.0in]{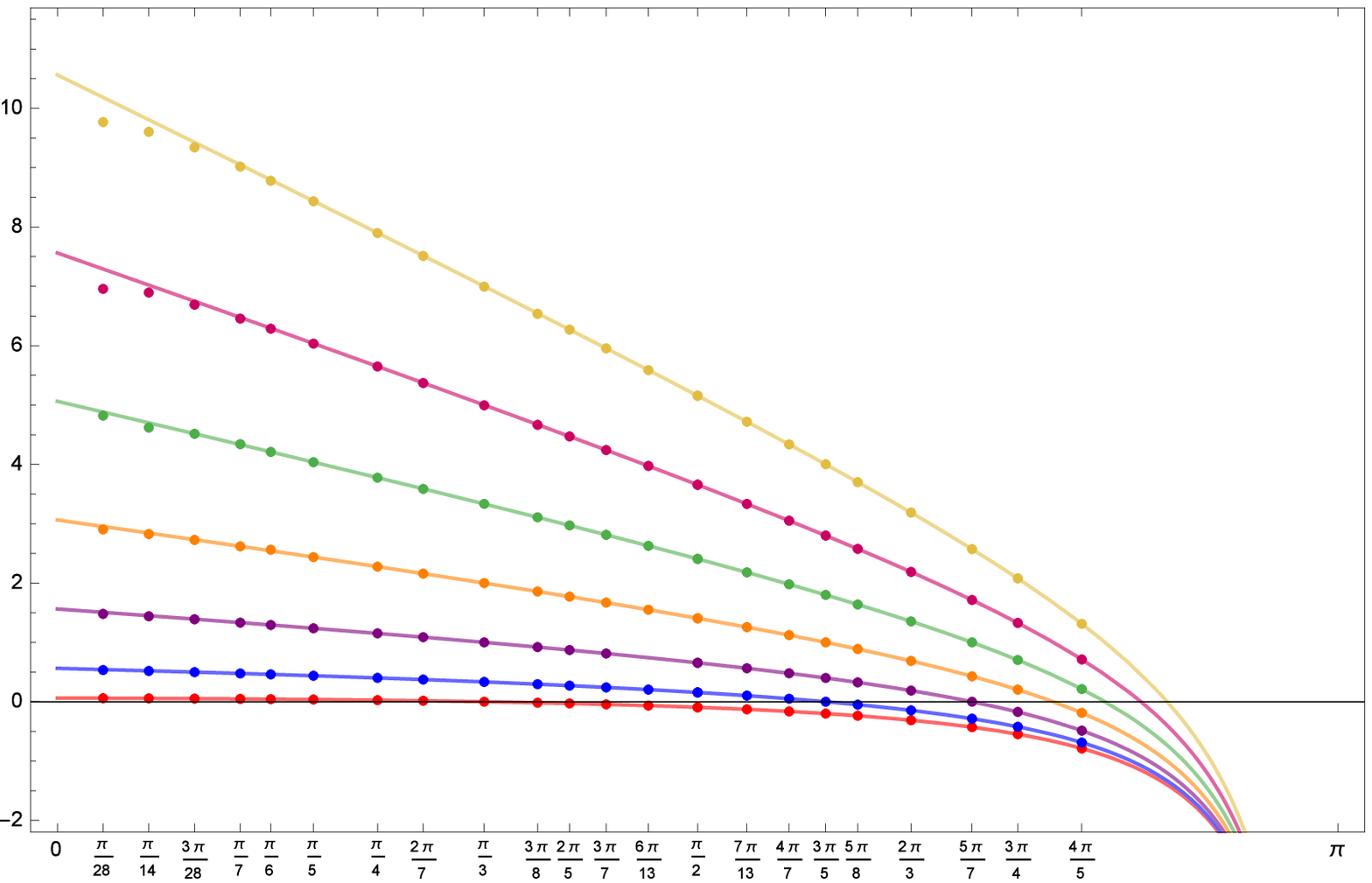} 
   \caption{Plot of the numerical conformal weights $\Delta^{p,p'}_{0,s-\frac{1}{2}}$ as a function of $\lambda$  for $s=1,2,\ldots,6,7$. 
    \label{fig:cdr0}}
\end{figure}

\begin{figure}[p]
   \centering
   \includegraphics[width=6.in]{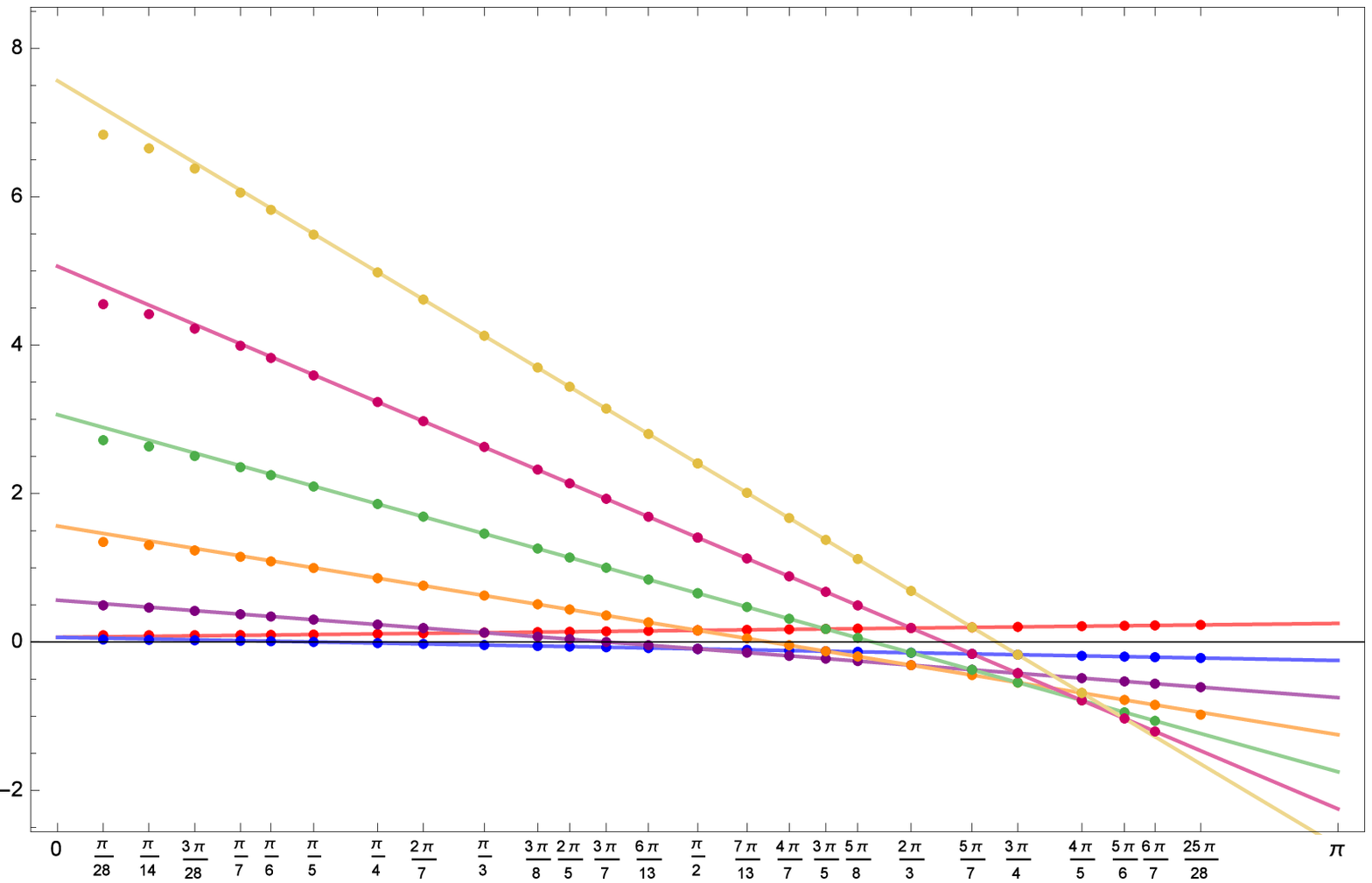} 
   \caption{Plot of the numerical conformal weights $\Delta^{p,p'}_{1,s-\frac{1}{2}}$ as a function of $\lambda$  for $s=1,2,\ldots,6,7$. 
    \label{fig:cdr1}}
\end{figure}
\begin{figure}[p] 
   \centering
   \includegraphics[width=6.in]{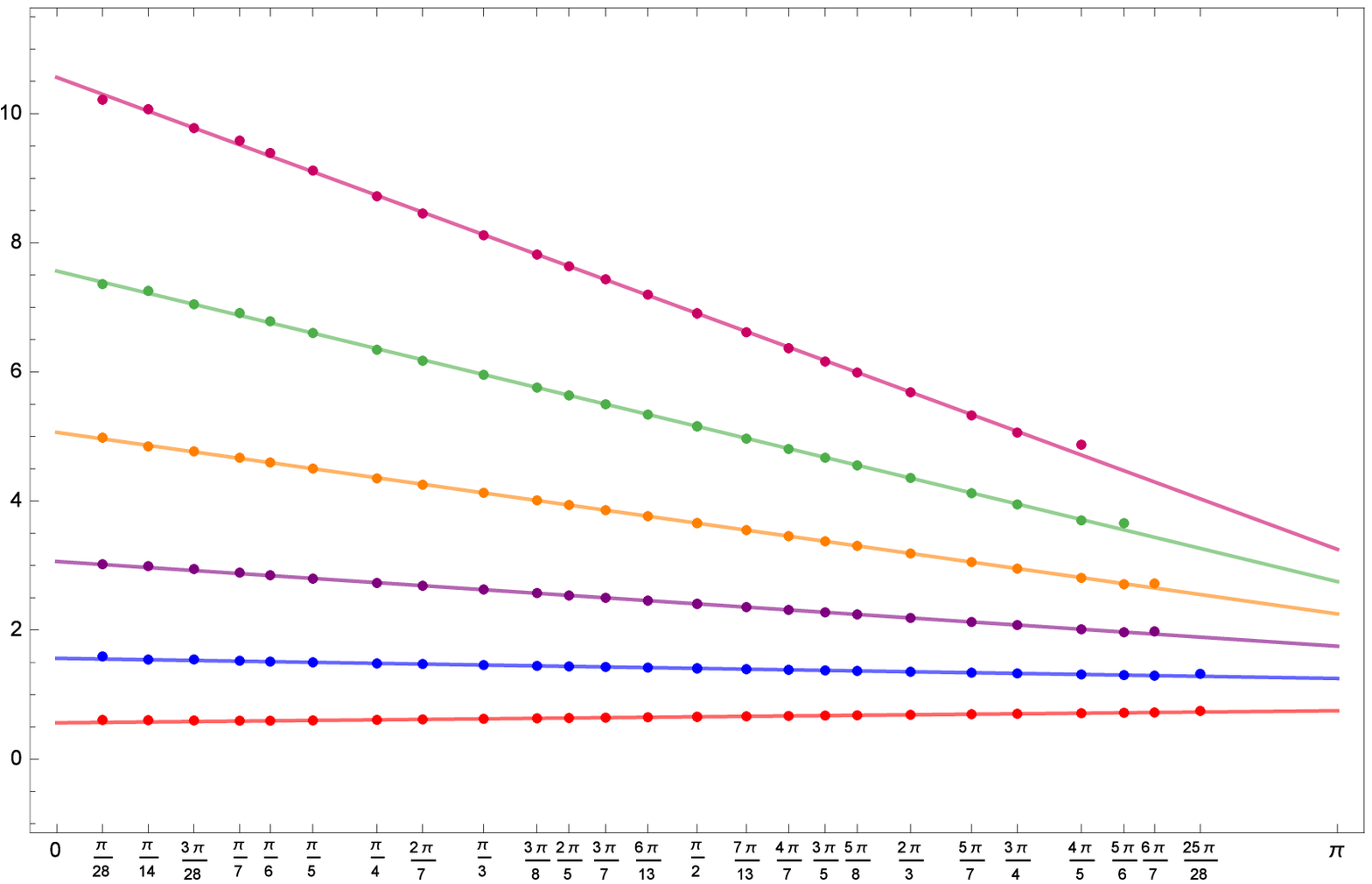} 
   \caption{Plot of the numerical conformal weights $\Delta^{p,p'}_{-1,s-\frac{1}{2}}$ as a function of $\lambda$  for $s=1,2,\ldots,6,7$.
   \label{fig:cdrneg1}}
\end{figure}

\begin{figure}[p]
   \centering
   \includegraphics[width=6.in]{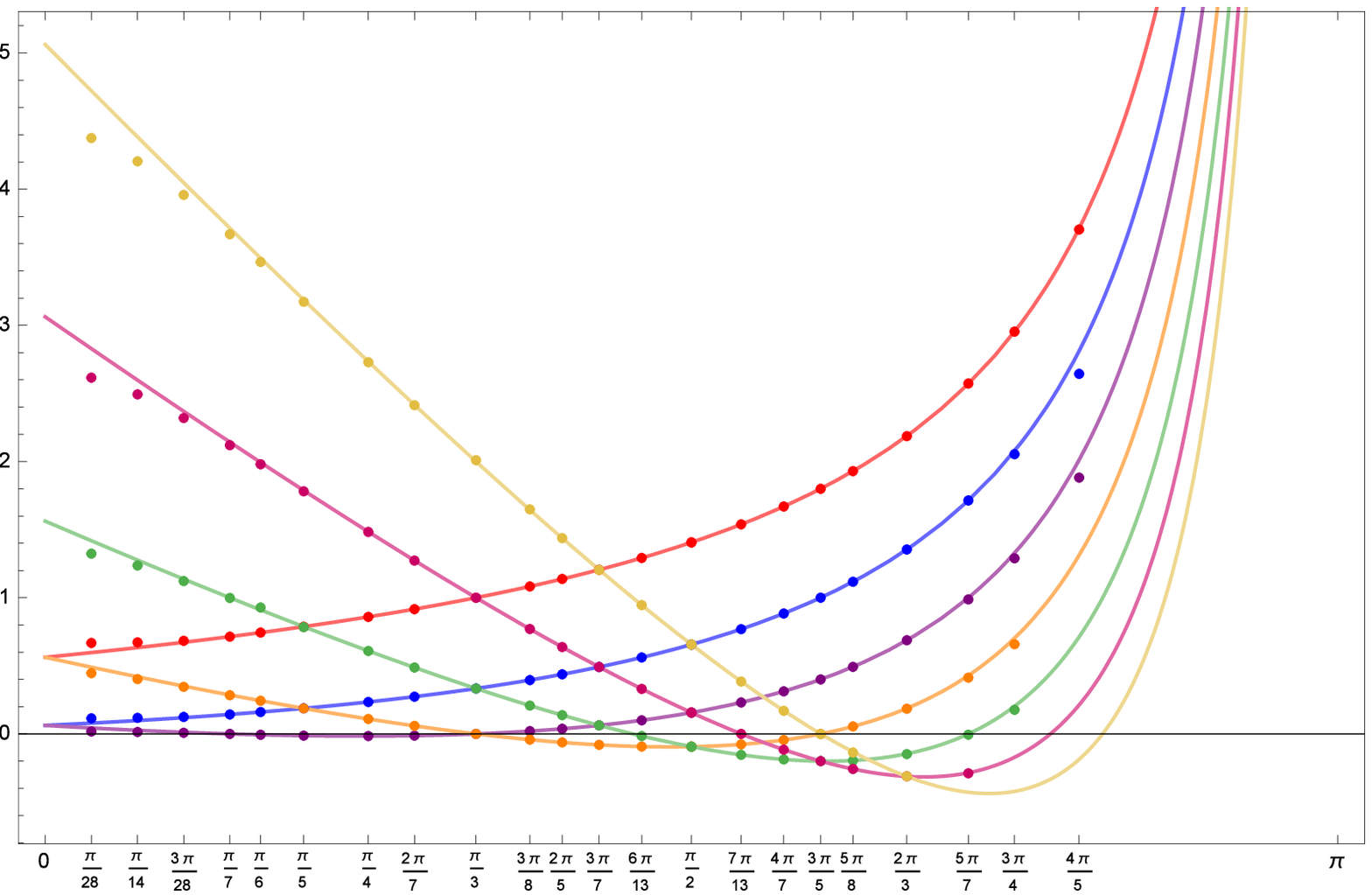} 
   \caption{Plot of the numerical conformal weights $\Delta^{p,p'}_{2,s-\frac{1}{2}}$ as a function of $\lambda$  for $s=1,2,\ldots,6,7$. 
    \label{fig:cdr2}}
\end{figure}
\begin{figure}[p] 
   \centering
   \includegraphics[width=6.in]{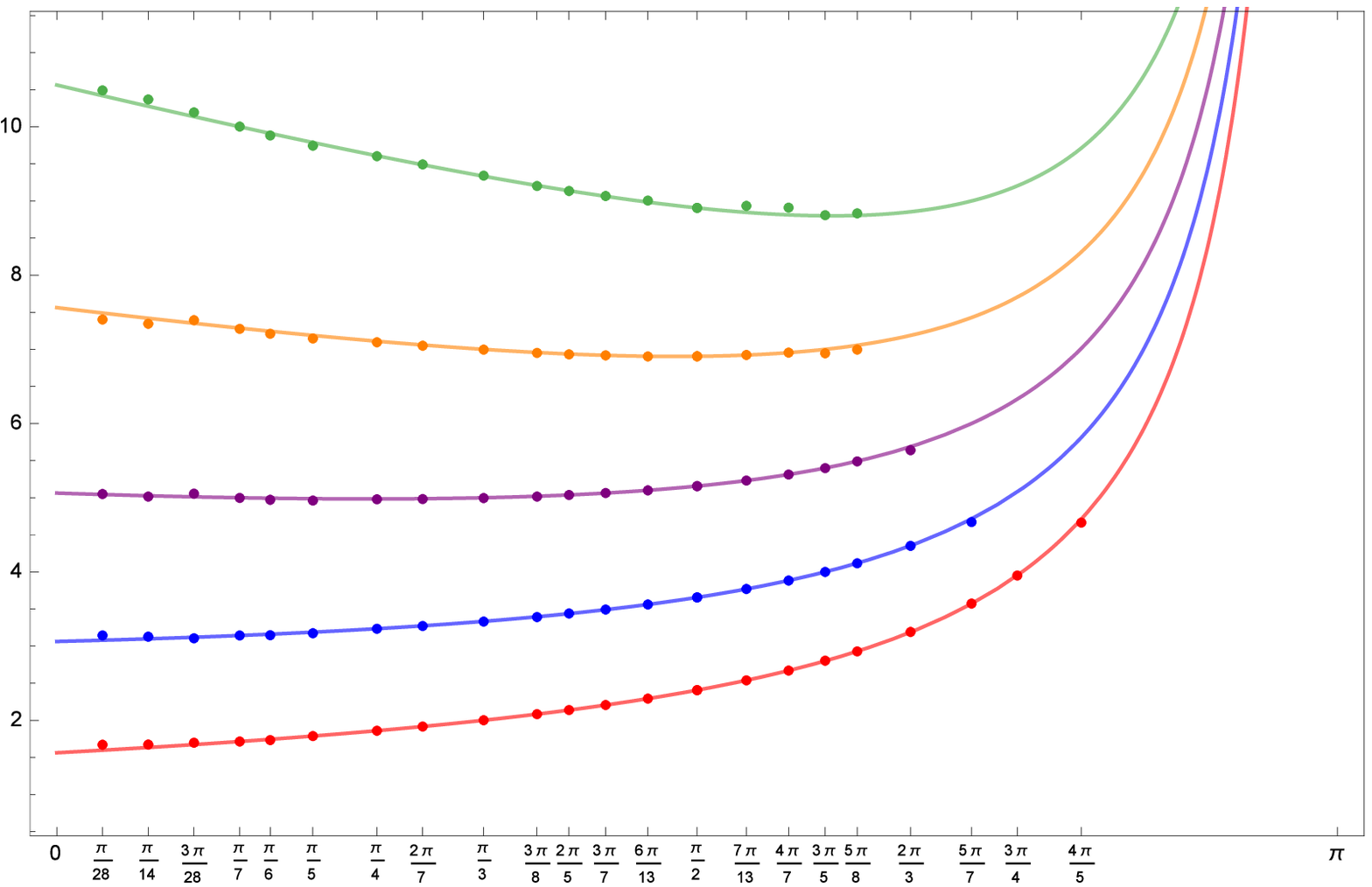} 
   \caption{Plot of the numerical conformal weights $\Delta^{p,p'}_{-2,s-\frac{1}{2}}$ as a function of $\lambda$  for $s=1,2,\ldots,6,7$.
   \label{fig:cdrneg2}}
\end{figure}

\subsection{Numerical conformal weights}

In this section, we present details of the numerical calculations for the conformal weights of the ${\cal LM}(p,p')$ models with $(r,s)$ Robin boundary conditions. 
Since it is numerically more efficient, we calculate the conformal spectra using the Hamiltonians rather than the double row transfer matrices. 
In Mathematica~\cite{Wolfram}, it is convenient to represent a link state as an ordered list of the unique site connected to the $N+w+d$  nodes numbered sequentially from left to right. The defects are folded, without introducing crossings, onto an $s$-type seam on the left. A node from which a boundary link emanates is considered connected to itself. For example, from (\ref{LinkStates}),
\bea
{\cal V}_1^{(3,1)}=\mbox{span}\Big\{\{2|1,3,5|4\}\Big\},\qquad {\cal V}_1^{(3,2)}=\mbox{span}\Big\{\{2|1,6,5|4,3\}, \{6|3,2,5|4,1\}, \{6|5,4,3|2,1\}\Big\}
\eea
The action of the TL generators on the vector space ${\cal V}_d^{(N,w)}$ of link states is implemented succinctly in Mathematica using transformation rules. The matrix representatives are economically encoded as sparse matrices using dispatch tables. The additional generators in the Hamiltonian (\ref{Hbeta}) are similarly encoded directly as the matrix representatives of the operators mapping from ${\cal V}_d^{(N,w)}$ to itself. 

For a given sector, labelled by $w,d$ or the quantum numbers $(r,s)$, the eigenvalues of the quantum Hamiltonian 
${\cal H}^{w,d}$ (\ref{Hbeta}) are obtained numerically using Mathematica for increasing system sizes 
$N$ out to $N+w+d\le 26$. For $r=0$, the parity of $N$ is not restricted. For $r\ne 0$, the parity of $N$ is fixed by
\bea
\mbox{sgn}(r)=(-1)^{N+w+d}
\eea
Although the real transfer matrices are not symmetric, the transfer matrices in all sectors appear to be diagonalizable with real eigenvalues. Presumably, applying non-trivial boundary conditions on both the left and the right edge of the strip would lead to real but non-diagonalizable transfer matrices and reducible yet indecomposable representations with non-trivial Jordan cells.
The first 10 to 20 dominant eigenvalues are obtained numerically using the Arnoldi method~\cite{Arnoldi}. Estimates of the conformal eigenenergies $E_j$, and hence the conformal weights $\Delta^{p,p'}_{r,s-\frac{1}{2}}$, are extrapolated to infinite system size from the finite-size sequences (\ref{eq:finite_size_corr_ham}) using a combination of Vanden Broeck-Schwartz~\cite{vBS} sequence acceleration and polynomial fits in $1/N$. There are no free parameters to fit and, since the bulk and boundary energies are known analytically, these extrapolation methods lead to accurate numerical results.
The most accurate estimates occur for $r=0$ and $s$ small for values of $\lambda$ away from the endpoints at $\lambda=0,\pi$. In such cases, the typical absolute error is of the order of $10^{-7}$ with relative errors in the range $10^{-8}$ to $10^{-6}$. Absolute and relative errors gradually increase to be of the order of $10^{-1}$ for the plotted points with larger values of $r,s$. For larger seam widths, the errors increase and the accuracy decreases because, for given  maximum total width, the maximum bulk width is decreased giving fewer data points for numerical extrapolation. Likewise, the accuracy decreases as $\lambda$ approaches $0$ or $\pi$.

Our numerical data is presented in a series of plots. Plots of the numerical conformal weights $\Delta^{p,p'}_{r,1/2}$ as a function of $\lambda$ for $0\le r\le 7$ and $-1\ge r\ge -7$ are shown in Figures~\ref{fig:ConfWtsPos} and \ref{fig:ConfWtsNeg}. 
Plots of the numerical conformal weights $\Delta^{p,p'}_{r,3/2}$ as a function of $\lambda$ for $0\le r\le 6$ and $-1\ge r\ge -6$ are shown in Figures~\ref{fig:s2ConfWtsPos} and \ref{fig:s2ConfWtsNeg}. 
Plots of the numerical conformal weights $\Delta^{p,p'}_{0,s-1/2}$ for $1\le s\le 7$ are shown in Figure~\ref{fig:cdr0}. 
Plots of the numerical conformal weights $\Delta^{p,p'}_{\pm1,s-1/2}$ and $\Delta^{p,p'}_{\pm2,s-1/2}$ as a function of $\lambda$ are shown in Figures~\ref{fig:cdr1}, \ref{fig:cdrneg1}, \ref{fig:cdr2} and \ref{fig:cdrneg2} respectively. 
The sequences of curves start with red, blue or blue, violet. The points in these plots at $\lambda=\frac{\pi}{2}$, related to critical dense polymers, are the values obtained analytically in \cite{PRT14}.

\subsection{Numerical conformal partition functions}

\allowdisplaybreaks
A consequence of the conjecture that the $(r,s)$ Robin representations correspond to reducible or irreducible modules, with characters (\ref{VirChars}), is that the level degeneracies are determined by the partition numbers $P_n$~\cite{Sloane} independent of the values of $p,p', r$ and $s$
\bea
\frac{1}{(q)_\infty}=\frac{1}{\prod_{n=1}^\infty (1-q^n)}=\sum_{n=0}^\infty P_n q^n=1+q+2q^2+3q^3+5q^4+7q^5+11q^6+\cdots
\eea
This expectation is well supported by the first 7-20 degeneracy levels obtained numerically. Typical results for ${\cal LM}(3,5)$ are
\begin{align}
\widehat{\ch}_{0,\frac{1}{2}}^{3,5}(q)&= 1 + q +2 q^2 + 3q^3 + 5 q^4 + 7 q^5 + \cdots\\
\widehat{\ch}_{0,\frac{3}{2}}^{3,5}(q)&= 1 + q +2 q^2 + 3q^3 + 5 q^4 + \cdots\\
\widehat{\ch}_{2,\frac{1}{2}}^{3,5}(q)&= 1 + q +2 q^2 + 3q^3 + 5 q^4 + \cdots\\
\widehat{\ch}_{-2,\frac{1}{2}}^{3,5}(q)&= 1 + q +2 q^2 + 3q^3 + \cdots\\
\widehat{\ch}_{2,\frac{3}{2}}^{3,5}(q)&= 1 + q +2 q^2 + 3q^3 + \cdots\\
\widehat{\ch}_{-2,\frac{3}{2}}^{3,5}(q)&= 1 + q +2 q^2 + 3q^3 + \cdots
\end{align}
We have analyzed many other cases and the results for other logarithmic minimal models are similar.
\allowdisplaybreaks[0]

\section{Conclusion}

In this paper, we have implemented the $(r,s)$ Robin boundary conditions of Pearce, Rasmussen and Tipunin~\cite{PRT14} for the general logarithmic minimal models ${\cal LM}(p,p')$~\cite{PRZ2006} on the strip. The $(r,s)$ boundary conditions are built from boundary $r$- and $s$-type seams of width $w=\bfloor{\frac{|r|p'}{p}}$ and $d=s-1$ columns respectively. This system is Yang-Baxter integrable in the presence of the boundary. We consider the commuting double row transfer matrices and the associated one-dimensional quantum Hamiltonians which are described algebraically by the action of the one-boundary Temperley-Lieb algebra on suitable spaces of link states. The eigenvalues of the Hamiltonians are calculated numerically with Mathematica out to bulk system sizes $N$ with $N+w+d\le 26$. Estimates of the conformal weights in the various $(r,s)$ Robin sectors are obtained numerically using finite-size corrections to the Hamiltonian eigenenergies. For $r\ne 0$, the extrapolated conformal weights depend on the parity of $N$. The results are neatly encoded by allowing $r$ to be negative and fixing the parity of $N$ by $\mbox{sgn}(r)=(-1)^{N+w+d}$. The bulk free energies are known and the boundary free energies are calculated analytically by solving the boundary inversion relations. Knowing the exact values of these non-universal quantities allows us to obtain accurate numerical estimates of the conformal weights which take the values 
$\Delta^{p,p'}_{r,s-\frac12}$, $r\in{\Bbb Z}$, $s\in{\Bbb N}$ where $\Delta^{p,p'}_{r,s}$ is given by the usual Kac formula. The $(r,s)$ Robin boundary conditions are thus conjugate to scaling operators with half-integer values for the Kac label $s-\half$. Extensive numerical investigation of the level degeneracies support the conjecture that the characters of the associated reducible or irreducible modules are given by  Virasoro Verma characters.

In this paper, we have implemented Robin boundary conditions on just one edge of the strip with the Kac vacuum boundary condition on the other edge leading to transfer matrices that are diagonalizable. 
It would be of interest to implement Robin boundary conditions on both edges of the strip in accord with Cardy fusion~\cite{Cardy1989} on the lattice. This situation is described by the two-boundary Temperley-Lieb algebra~\cite{2bdyTL,DJS} and, in this case, the transfer matrices are expected to exhibit Jordan blocks and lead to reducible yet indecomposable representations of the Virasoro algebra.  It would be interesting to know the complete closed set of fusion rules for these representations.

Lastly, there are strong indications~\cite{Delfino} that there exist conformal weights for the logarithmic minimal models with half-integer values $r-\half$ for the first Kac label. It is an open question as to whether there are boundary conditions associated to these representations. But the application of the logarithmic limit (\ref{halfchar}) suggests that the associated boundary conditions cannot be far away.

\goodbreak

\section*{Acknowledgments} This work was initiated at the Asia Pacific
Center for Theoretical Physics (APCTP). JEB
acknowledges the Korea Ministry of Education, Science and Technology
(MEST) for the support of the Young Scientist Training Program. He
further thanks INFN for his post-doctoral fellowship within the
grant GAST, which has also partially supported this project, together
with the UniTo-SanPaolo research  grant number TO-Call3-2012-0088 {\it
``Modern Applications of String Theory'' (MAST)}, the ESF Network {\it
``Holographic methods for strongly coupled systems'' (HoloGrav)}
(09-RNP-092 (PESC)) and the MPNS--COST Action MP1210.
PAP thanks the APCTP for kind hospitality and the ICTP for support through a Visiting Scholar Award at APCTP. ET is supported by an Australian Postgraduate Award. We thank Jorgen Rasmussen and David Ridout for comments on the manuscript.

\goodbreak
\appendix

\section{Derivation of the Inversion Relation}\label{AppA}

\subsection{Inversion relation for the Robin vacuum boundary}
In this appendix, we derive the inversion relation (\ref{InvRel}) for the case $w=0$. Diagrammatically, the product of the two transfer matrices on the left of the inversion identity (\ref{InvId}) is
\bea
\psset{unit=.8cm}
\setlength{\unitlength}{.8cm}
\frac{\b^2\,\G(0)^2}{\tilde\kappa_0(u,\xi)\tilde\kappa_0(u+\lambda,\xi)}\, \Db(u)\Db(u+\l)=\
\begin{pspicture}[shift=-2.55](0.4,-.5)(6,4)
\facegrid{(1,0)}{(5,4)}
\pspolygon[fillstyle=solid,fillcolor=lightlightblue](5,1)(6,2)(6,0)(5,1)
\pspolygon[fillstyle=solid,fillcolor=lightlightblue](5,3)(6,4)(6,2)(5,3)
\psarc[linewidth=0.025,linecolor=red](1,0){0.16}{0}{90}
\psarc[linewidth=0.025,linecolor=red](1,1){0.16}{0}{90}
\psarc[linewidth=0.025,linecolor=red](1,2){0.16}{0}{90}
\psarc[linewidth=0.025,linecolor=red](1,3){0.16}{0}{90}
\psarc[linewidth=0.025,linecolor=red](2,0){0.16}{0}{90}
\psarc[linewidth=0.025,linecolor=red](2,1){0.16}{0}{90}
\psarc[linewidth=0.025,linecolor=red](2,2){0.16}{0}{90}
\psarc[linewidth=0.025,linecolor=red](2,3){0.16}{0}{90}
\psarc[linewidth=0.025,linecolor=red](4,0){0.16}{0}{90}
\psarc[linewidth=0.025,linecolor=red](4,1){0.16}{0}{90}
\psarc[linewidth=0.025,linecolor=red](4,2){0.16}{0}{90}
\psarc[linewidth=0.025,linecolor=red](4,3){0.16}{0}{90}
\rput(1.5,0.5){$_{u}$}
\rput(2.5,0.5){$_{u}$}
\rput(4.5,0.5){$_{u}$}
\rput(1.5,1.5){$_{\l-u}$}
\rput(2.5,1.5){$_{\l-u}$}
\rput(4.5,1.5){$_{\l-u}$}
\rput(1.5,2.5){$_{u+\l}$}
\rput(2.5,2.5){$_{u+\l}$}
\rput(4.5,2.5){$_{u+\l}$}
\rput(1.5,3.5){$_{-u}$}
\rput(2.5,3.5){$_{-u}$}
\rput(4.5,3.5){$_{-u}$}
\rput(3.5,0.5){$\ldots$}
\rput(3.5,1.5){$\ldots$}
\rput(3.5,2.5){$\ldots$}
\rput(3.5,3.5){$\ldots$}
\rput(5.65,1){$_{u}$}
\rput(5.6,3){$_{u+\l}$}
\psarc[linewidth=1.5pt,linecolor=blue](1,1){.5}{90}{270}
\psarc[linewidth=1.5pt,linecolor=blue](1,3){.5}{90}{270}
\psline[linecolor=blue,linewidth=1.5pt]{-}(5,0.5)(5.5,0.5)
\psline[linecolor=blue,linewidth=1.5pt]{-}(5,1.5)(5.5,1.5)
\psline[linecolor=blue,linewidth=1.5pt]{-}(5,2.5)(5.5,2.5)
\psline[linecolor=blue,linewidth=1.5pt]{-}(5,3.5)(5.5,3.5)
\rput(3,-0.3){\scriptsize $\underbrace{\hspace{4\unitlength}}_N$}
\end{pspicture}
\vspace{1.6cm}
\label{DDuu}
\eea
The first step is to insert, somewhere in the diagram, the identity
\bea
\psset{unit=0.7cm}
\begin{pspicture}[shift=-1.5](0.4,.3)(2,3.6)
\psline[linecolor=blue,linewidth=1.5pt]{-}(0,0.5)(1.5,0.5)
\psline[linecolor=blue,linewidth=1.5pt]{-}(0,1.5)(1.5,1.5)
\psline[linecolor=blue,linewidth=1.5pt]{-}(0,2.5)(1.5,2.5)
\psline[linecolor=blue,linewidth=1.5pt]{-}(0,3.5)(1.5,3.5)
\end{pspicture}
=\;\frac{1}{s_1(2u)s_1(-2u)}\qquad
\begin{pspicture}[shift=-1.5](0.4,.3)(2,3.6)
\pspolygon[fillstyle=solid,fillcolor=lightlightblue](0,2)(1,1)(2,2)(1,3)(0,2)
\pspolygon[fillstyle=solid,fillcolor=lightlightblue](2,2)(3,1)(4,2)(3,3)(2,2)
\psline[linecolor=blue,linewidth=1.5pt]{-}(0,0.5)(4,0.5)
\psline[linecolor=blue,linewidth=1.5pt]{-}(0,1.5)(0.5,1.5)
\psline[linecolor=blue,linewidth=1.5pt]{-}(0,2.5)(0.5,2.5)
\psline[linecolor=blue,linewidth=1.5pt]{-}(1.5,1.5)(2.5,1.5)
\psline[linecolor=blue,linewidth=1.5pt]{-}(1.5,2.5)(2.5,2.5)
\psline[linecolor=blue,linewidth=1.5pt]{-}(3.5,1.5)(4,1.5)
\psline[linecolor=blue,linewidth=1.5pt]{-}(3.5,2.5)(4,2.5)
\psline[linecolor=blue,linewidth=1.5pt]{-}(0,3.5)(4,3.5)
\psarc[linewidth=0.025,linecolor=red](0,2){0.16}{-45}{45}
\psarc[linewidth=0.025,linecolor=red](2,2){0.16}{-45}{45}
\rput(1,2){$2u$}
\rput(3,2){$-2u$}
\end{pspicture}
\vspace{0.3cm}
\label{pushinv}
\eea
Using the Yang-Baxter equation, we push the inserted faces to either end
\be
\psset{unit=.8cm}
\setlength{\unitlength}{.8cm}
\frac{\b^2\,\G(0)^2}{\tilde\kappa_0(u,\xi)\tilde\kappa_0(-u,\xi)}\,\Db(u)\Db(u+\l)=\frac{1}{s_1(2u)s_1(-2u)}\quad\
\begin{pspicture}[shift=-1.9](0.4,-.4)(8,4)
\facegrid{(2,0)}{(6,4)}
\pspolygon[fillstyle=solid,fillcolor=lightlightblue](7,1)(8,2)(8,0)(7,1)
\pspolygon[fillstyle=solid,fillcolor=lightlightblue](7,3)(8,4)(8,2)(7,3)
\pspolygon[fillstyle=solid,fillcolor=lightlightblue](0,2)(1,1)(2,2)(1,3)(0,2)
\pspolygon[fillstyle=solid,fillcolor=lightlightblue](6,2)(7,1)(8,2)(7,3)(6,2)
\psarc[linewidth=0.025,linecolor=red](2,0){0.16}{0}{90}
\psarc[linewidth=0.025,linecolor=red](2,1){0.16}{0}{90}
\psarc[linewidth=0.025,linecolor=red](2,2){0.16}{0}{90}
\psarc[linewidth=0.025,linecolor=red](2,3){0.16}{0}{90}
\psarc[linewidth=0.025,linecolor=red](3,0){0.16}{0}{90}
\psarc[linewidth=0.025,linecolor=red](3,1){0.16}{0}{90}
\psarc[linewidth=0.025,linecolor=red](3,2){0.16}{0}{90}
\psarc[linewidth=0.025,linecolor=red](3,3){0.16}{0}{90}
\psarc[linewidth=0.025,linecolor=red](5,0){0.16}{0}{90}
\psarc[linewidth=0.025,linecolor=red](5,1){0.16}{0}{90}
\psarc[linewidth=0.025,linecolor=red](5,2){0.16}{0}{90}
\psarc[linewidth=0.025,linecolor=red](5,3){0.16}{0}{90}
\rput(2.5,0.5){$_{u}$}
\rput(3.5,0.5){$_{u}$}
\rput(5.5,0.5){$_{u}$}
\rput(2.5,1.5){$_{u+\l}$}
\rput(3.5,1.5){$_{u+\l}$}
\rput(5.5,1.5){$_{u+\l}$}
\rput(2.5,2.5){$_{\l-u}$}
\rput(3.5,2.5){$_{\l-u}$}
\rput(5.5,2.5){$_{\l-u}$}
\rput(2.5,3.5){$_{-u}$}
\rput(3.5,3.5){$_{-u}$}
\rput(5.5,3.5){$_{-u}$}
\rput(4.5,0.5){$\ldots$}
\rput(4.5,1.5){$\ldots$}
\rput(4.5,2.5){$\ldots$}
\rput(4.5,3.5){$\ldots$}
\rput(7.65,1){$_{u}$}
\rput(7.6,3){$_{u+\l}$}
\rput(1,2){$2u$}
\rput(7,2){$-2u$}
\rput(6.5,1){$I$}
\psarc[linewidth=0.025,linecolor=red](0,2){0.16}{-45}{45}
\psarc[linewidth=0.025,linecolor=red](6,2){0.16}{-45}{45}
\psarc[linewidth=1.5pt,linecolor=blue](0.5,1){.5}{90}{270}
\psarc[linewidth=1.5pt,linecolor=blue](0.5,3){.5}{90}{270}
\psline[linecolor=blue,linewidth=1.5pt]{-}(0.5,0.5)(2,0.5)
\psline[linecolor=blue,linewidth=1.5pt]{-}(0.5,3.5)(2,3.5)
\psline[linecolor=blue,linewidth=1.5pt]{-}(1.5,1.5)(2,1.5)
\psline[linecolor=blue,linewidth=1.5pt]{-}(1.5,2.5)(2,2.5)
\psline[linecolor=blue,linewidth=1.5pt]{-}(6,0.5)(7.5,0.5)
\psline[linecolor=blue,linewidth=1.5pt]{-}(6,1.5)(6.5,1.5)
\psline[linecolor=blue,linewidth=1.5pt]{-}(6,2.5)(6.5,2.5)
\psline[linecolor=blue,linewidth=1.5pt]{-}(6,3.5)(7.5,3.5)
\rput(4,-0.3){\scriptsize$\underbrace{\hspace{4\unitlength}}_N$}
\end{pspicture}
\label{A3}
\ee

To obtain the coefficient $\phi(\lambda-u)\phi(\lambda+u)$ we insert, at the position indicated by $I$ in (\ref{A3}), the identity decomposed into orthogonal projectors
\bea
\psset{unit=.475cm}
I\;=\;
\begin{pspicture}[shift=-.92](0,0)(2,2)
\pspolygon[linewidth=0.8pt,linecolor=black,fillstyle=solid,fillcolor=lightlightblue](1,0)(2,1)(1,2)(0,1)
\psarc[linewidth=1.5pt,linecolor=blue](1,2){.72}{-135}{-45}
\psarc[linewidth=1.5pt,linecolor=blue](1,0){.72}{45}{135}
\end{pspicture}\;=\;
\frac{1}{\beta}\ 
\begin{pspicture}[shift=-.92](0,0)(2,2)
\pspolygon[linewidth=0.8pt,linecolor=black,fillstyle=solid,fillcolor=lightlightblue](1,0)(2,1)(1,2)(0,1)
\rput(1,1){\small $\lambda$}
\psarc[linewidth=.5pt,linecolor=red](0,1){.17}{-45}{45}
\end{pspicture}
\ +\ \frac{1}{\beta}\ 
\begin{pspicture}[shift=-.92](0,0)(2,2)
\pspolygon[linewidth=0.8pt,linecolor=black,fillstyle=solid,fillcolor=lightlightblue](1,0)(2,1)(1,2)(0,1)
\rput(1,1){\small $-\lambda$}
\psarc[linewidth=.5pt,linecolor=red](0,1){.17}{-45}{45}
\end{pspicture}
\label{projdecomp}
\eea
The second term leads to the exponentially small second term on the right side of (\ref{InvId}) which is neglected. The first term gives
\bea
\psset{unit=.75cm}
\setlength{\unitlength}{.75cm}
\frac{\b^2\,\G(0)^2}{\tilde\kappa_0(u,\xi)\tilde\kappa_0(-u,\xi)}\,\phi(\lambda\!-\!u)\phi(\lambda\!+\!u)=\frac{1}{\b s_1(2u)s_1(-2u)}\ 
\begin{pspicture}[shift=-2.39](0,-0.5)(9.2,4)
\facegrid{(2,0)}{(6,4)}
\pspolygon[fillstyle=solid,fillcolor=lightlightblue](8,1)(9,2)(9,0)(8,1)
\pspolygon[fillstyle=solid,fillcolor=lightlightblue](8,3)(9,4)(9,2)(8,3)
\pspolygon[fillstyle=solid,fillcolor=lightlightblue](0,2)(1,1)(2,2)(1,3)(0,2)
\pspolygon[fillstyle=solid,fillcolor=lightlightblue](6,1)(7,0)(8,1)(7,2)(6,1)
\pspolygon[fillstyle=solid,fillcolor=lightlightblue](7,2)(8,1)(9,2)(8,3)(7,2)
\psarc[linewidth=0.025,linecolor=red](2,0){0.16}{0}{90}
\psarc[linewidth=0.025,linecolor=red](2,1){0.16}{0}{90}
\psarc[linewidth=0.025,linecolor=red](2,2){0.16}{0}{90}
\psarc[linewidth=0.025,linecolor=red](2,3){0.16}{0}{90}
\psarc[linewidth=0.025,linecolor=red](3,0){0.16}{0}{90}
\psarc[linewidth=0.025,linecolor=red](3,1){0.16}{0}{90}
\psarc[linewidth=0.025,linecolor=red](3,2){0.16}{0}{90}
\psarc[linewidth=0.025,linecolor=red](3,3){0.16}{0}{90}
\psarc[linewidth=0.025,linecolor=red](5,0){0.16}{0}{90}
\psarc[linewidth=0.025,linecolor=red](5,1){0.16}{0}{90}
\psarc[linewidth=0.025,linecolor=red](5,2){0.16}{0}{90}
\psarc[linewidth=0.025,linecolor=red](5,3){0.16}{0}{90}
\rput(2.5,0.5){$_{u}$}
\rput(3.5,0.5){$_{u}$}
\rput(5.5,0.5){$_{u}$}
\rput(2.5,1.5){$_{u+\l}$}
\rput(3.5,1.5){$_{u+\l}$}
\rput(5.5,1.5){$_{u+\l}$}
\rput(2.5,2.5){$_{\l-u}$}
\rput(3.5,2.5){$_{\l-u}$}
\rput(5.5,2.5){$_{\l-u}$}
\rput(2.5,3.5){$_{-u}$}
\rput(3.5,3.5){$_{-u}$}
\rput(5.5,3.5){$_{-u}$}
\rput(4.5,0.5){$\ldots$}
\rput(4.5,1.5){$\ldots$}
\rput(4.5,2.5){$\ldots$}
\rput(4.5,3.5){$\ldots$}
\rput(8.65,1){$_{u}$}
\rput(8.6,3){$_{u+\l}$}
\rput(1,2){$2u$}
\rput(8,2){$-2u$}
\rput(7,1){$\l$}
\psarc[linewidth=0.025,linecolor=red](0,2){0.16}{-45}{45}
\psarc[linewidth=0.025,linecolor=red](7,2){0.16}{-45}{45}
\psarc[linewidth=0.025,linecolor=red](6,1){0.16}{-45}{45}
\psarc[linewidth=1.5pt,linecolor=blue](0.5,1){.5}{90}{270}
\psarc[linewidth=1.5pt,linecolor=blue](0.5,3){.5}{90}{270}
\psline[linecolor=blue,linewidth=1.5pt]{-}(0.5,0.5)(2,0.5)
\psline[linecolor=blue,linewidth=1.5pt]{-}(0.5,3.5)(2,3.5)
\psline[linecolor=blue,linewidth=1.5pt]{-}(1.5,1.5)(2,1.5)
\psline[linecolor=blue,linewidth=1.5pt]{-}(1.5,2.5)(2,2.5)
\psline[linecolor=blue,linewidth=1.5pt]{-}(6,0.5)(6.5,0.5)
\psline[linecolor=blue,linewidth=1.5pt]{-}(7.5,0.5)(8.5,0.5)
\psline[linecolor=blue,linewidth=1.5pt]{-}(6,1.5)(6.5,1.5)
\psline[linecolor=blue,linewidth=1.5pt]{-}(6,2.5)(7.5,2.5)
\psline[linecolor=blue,linewidth=1.5pt]{-}(6,3.5)(8.5,3.5)
\rput(4,-0.3){\scriptsize $\underbrace{\hspace{4\unitlength}}_N$}
\end{pspicture}
\eea
Using the identity 
\bea
\psset{unit=.7cm}
\ \ 
\begin{pspicture}[shift=-0.89](0,0)(3,2)
\facegrid{(0,0)}{(1,2)}
\pspolygon[fillstyle=solid,fillcolor=lightlightblue](1,1)(2,2)(3,1)(2,0)(1,1)
\psarc[linewidth=0.025,linecolor=red]{-}(0,0){0.16}{0}{90}
\psarc[linewidth=0.025,linecolor=red]{-}(0,1){0.16}{0}{90}
\psarc[linewidth=0.025,linecolor=red]{-}(1,1){0.16}{-45}{45}
\rput(.5,.5){$_u$}
\rput(.5,1.5){$_{u+\l}$}
\rput(2,1){$_{\l}$}
\psline[linecolor=blue,linewidth=1.5pt]{-}(1,0.5)(1.5,0.5)
\psline[linecolor=blue,linewidth=1.5pt]{-}(1,1.5)(1.5,1.5)
\end{pspicture}\ \ = \ \ s_1(u)s_1(-u)\ \ 
\begin{pspicture}[shift=-0.89](0,0)(3,2)
\facegrid{(0,0)}{(1,2)}
\pspolygon[fillstyle=solid,fillcolor=lightlightblue](1,1)(2,2)(3,1)(2,0)(1,1)
\rput[bl](0,0){\loopa}
\rput[bl](0,1){\loopb}
\psarc[linewidth=1.5pt,linecolor=blue](1,1){.5}{270}{90}
\psarc[linewidth=1.5pt,linecolor=blue](3,1){.5}{90}{270}
\end{pspicture}\ \ = \ \ s_1(u)s_1(-u)\ \ 
\begin{pspicture}[shift=-0.89](0,0)(3,2)
\facegrid{(2,0)}{(3,2)}
\pspolygon[fillstyle=solid,fillcolor=lightlightblue](0,1)(1,2)(2,1)(1,0)(0,1)
\rput[bl](2,0){\loopb}
\rput[bl](2,1){\loopa}
\psarc[linewidth=1.5pt,linecolor=blue](0,1){.5}{270}{90}
\psarc[linewidth=1.5pt,linecolor=blue](2,1){.5}{90}{270}
\end{pspicture}
\label{A6}
\eea
the projector can be pushed through to the left leaving only the identity behind. On the left boundary, the contribution is
\be
\psset{unit=.77cm}
\begin{pspicture}[shift=-1.9](0.4,0)(3.5,4.2)
\pspolygon[fillstyle=solid,fillcolor=lightlightblue](0,0)(1,1)(0,2)(0,0)
\pspolygon[fillstyle=solid,fillcolor=lightlightblue](0,2)(1,3)(0,4)(0,2)
\pspolygon[fillstyle=solid,fillcolor=lightlightblue](0,2)(1,1)(2,2)(1,3)(0,2)
\pspolygon[fillstyle=solid,fillcolor=lightlightblue](1,1)(2,0)(3,1)(2,2)(1,1)
\rput(1,2){$2u$}
\rput(2,1){$\l$}
\psarc[linewidth=0.025,linecolor=red](0,2){0.16}{-45}{45}
\psarc[linewidth=0.025,linecolor=red](1,1){0.16}{-45}{45}
\psarc[linewidth=1.5pt,linecolor=blue](1,1){.5}{135}{270}
\psarc[linewidth=1.5pt,linecolor=blue](1,3){.5}{90}{225}
\psline[linecolor=blue,linewidth=1.5pt]{-}(1,0.5)(1.5,0.5)
\psline[linecolor=blue,linewidth=1.5pt]{-}(2.5,0.5)(3,0.5)
\psline[linecolor=blue,linewidth=1.5pt]{-}(2.5,1.5)(3,1.5)
\psline[linecolor=blue,linewidth=1.5pt]{-}(1.5,2.5)(3,2.5)
\psline[linecolor=blue,linewidth=1.5pt]{-}(1,3.5)(3,3.5)
\end{pspicture}
= s_2(-2u) \quad
\begin{pspicture}[shift=-1.9](0,0)(1,4.2)
\pspolygon[fillstyle=solid,fillcolor=lightlightblue](0,0)(1,1)(0,2)(0,0)
\pspolygon[fillstyle=solid,fillcolor=lightlightblue](0,2)(1,3)(0,4)(0,2)
\psarc[linewidth=1.5pt,linecolor=blue](1,1){.5}{135}{225}
\psarc[linewidth=1.5pt,linecolor=blue](1,3){.5}{135}{225}
\end{pspicture}
\label{A7}
\ee
where the boundary triangles have weight 1. We thus have
\be
\frac{\phi(\lambda\!-\!u)\phi(\lambda\!+\!u)}{\tilde\kappa_0(u,\xi)\tilde\kappa_0(-u,\xi)}\,=\frac{[s_1(-u)s_1(u)]^N s_2(-2u)}{\b^3\G(0)^2 s_1(2u)s_1(-2u)}\quad\ \ \ 
\psset{unit=1cm}
\begin{pspicture}[shift=-1.9](0.4,0)(7,4)
\facegrid{(1,2)}{(5,4)}
\pspolygon[fillstyle=solid,fillcolor=lightlightblue](0,2)(1,3)(0,4)(0,2)
\pspolygon[fillstyle=solid,fillcolor=lightlightblue](6,1)(7,2)(7,0)(6,1)
\pspolygon[fillstyle=solid,fillcolor=lightlightblue](6,3)(7,4)(7,2)(6,3)
\pspolygon[fillstyle=solid,fillcolor=lightlightblue](5,2)(6,1)(7,2)(6,3)(5,2)
\psarc[linewidth=0.025,linecolor=red](1,2){0.16}{0}{90}
\psarc[linewidth=0.025,linecolor=red](1,3){0.16}{0}{90}
\psarc[linewidth=0.025,linecolor=red](2,2){0.16}{0}{90}
\psarc[linewidth=0.025,linecolor=red](2,3){0.16}{0}{90}
\psarc[linewidth=0.025,linecolor=red](4,2){0.16}{0}{90}
\psarc[linewidth=0.025,linecolor=red](4,3){0.16}{0}{90}
\rput(1.5,2.5){$_{\l-u}$}
\rput(2.5,2.5){$_{\l-u}$}
\rput(4.5,2.5){$_{\l-u}$}
\rput(1.5,3.5){$_{-u}$}
\rput(2.5,3.5){$_{-u}$}
\rput(4.5,3.5){$_{-u}$}
\rput(3.5,2.5){$\ldots$}
\rput(3.5,3.5){$\ldots$}
\rput(6.65,1){$_{u}$}
\rput(6.6,3){$_{u+\l}$}
\rput(6,2){$-2u$}
\psarc[linewidth=0.025,linecolor=red](5,2){0.16}{-45}{45}
\psarc[linewidth=1.5pt,linecolor=blue](1,3){.5}{90}{270}
\psarc[linewidth=1.5pt,linecolor=blue](5.5,1){.5}{90}{270}
\psline[linecolor=blue,linewidth=1.5pt]{-}(1.5,0)(1.5,2)
\psline[linecolor=blue,linewidth=1.5pt]{-}(2.5,0)(2.5,2)
\rput(3.5,1){$\cdots$}
\psline[linecolor=blue,linewidth=1.5pt]{-}(4.5,0)(4.5,2)
\psline[linecolor=blue,linewidth=1.5pt]{-}(5.5,0.5)(6.5,0.5)
\psline[linecolor=blue,linewidth=1.5pt]{-}(5,2.5)(5.5,2.5)
\psline[linecolor=blue,linewidth=1.5pt]{-}(5,3.5)(6.5,3.5)
\end{pspicture}
\ee

On the left, the Neumann boundary acts as a projector due to the drop-down (or push-through) property (equivalent to its definition (3.23) of \cite{PRT14})
\be
\psset{unit=.7cm}
\ \ 
\begin{pspicture}[shift=-.9](0,2)(2.2,4)
\facegrid{(1,2)}{(2,4)}
\pspolygon[fillstyle=solid,fillcolor=lightlightblue](0,2)(1,3)(0,4)(0,2)
\psarc[linewidth=0.025,linecolor=red](1,2){0.16}{0}{90}
\psarc[linewidth=0.025,linecolor=red](1,3){0.16}{0}{90}
\rput(1.5,3.5){$_{u}$}
\rput(1.5,2.5){$_{u+\l}$}
\psarc[linewidth=1.5pt,linecolor=blue](1,3){.5}{90}{270}
\end{pspicture}\ =\ \  s_1(u)s_1(-u)\ \ \ 
\begin{pspicture}[shift=-.9](0,2)(2.2,4)
\facegrid{(1,2)}{(2,4)}
\pspolygon[fillstyle=solid,fillcolor=lightlightblue](0,2)(1,3)(0,4)(0,2)
\psarc[linewidth=0.025,linecolor=red](1,2){0.16}{0}{90}
\psarc[linewidth=0.025,linecolor=red](1,3){0.16}{0}{90}
\psarc[linewidth=1.5pt,linecolor=blue](1,3){.5}{90}{270}
\psarc[linewidth=1.5pt,linecolor=blue](1,4){.5}{270}{360}
\psarc[linewidth=1.5pt,linecolor=blue](1,2){.5}{0}{90}
\psarc[linewidth=1.5pt,linecolor=blue](2,3){.5}{90}{270}
\end{pspicture}
\ee
It can be pushed through to the right to give
\be
\frac{\phi(\lambda-u)\phi(\lambda+u)}{\tilde\kappa_0(u,\xi)\tilde\kappa_0(-u,\xi)}\;=\;\frac{[s_1(u)s_1(-u)]^{2N}s_2(-2u)}{\b^3\G(0)^2 s_1(2u)s_1(-2u)}
\psset{unit=.8cm}
\begin{pspicture}[shift=-1.9](1,0)(7,4)
\pspolygon[fillstyle=solid,fillcolor=lightlightblue](6,1)(7,2)(7,0)(6,1)
\pspolygon[fillstyle=solid,fillcolor=lightlightblue](6,3)(7,4)(7,2)(6,3)
\pspolygon[fillstyle=solid,fillcolor=lightlightblue](5,2)(6,1)(7,2)(6,3)(5,2)
\rput(6.65,1){$_{u}$}
\rput(6.6,3){$_{u+\l}$}
\rput(6,2){$-2u$}
\psarc[linewidth=0.025,linecolor=red](5,2){0.16}{-45}{45}
\psarc[linewidth=1.5pt,linecolor=blue](5.5,3){.5}{90}{270}
\psarc[linewidth=1.5pt,linecolor=blue](5.5,1){.5}{90}{270}
\psline[linecolor=blue,linewidth=1.5pt]{-}(1.5,0)(1.5,4)
\psline[linecolor=blue,linewidth=1.5pt]{-}(2.5,0)(2.5,4)
\rput(3.5,2){$\cdots$}
\psline[linecolor=blue,linewidth=1.5pt]{-}(4.5,0)(4.5,4)
\psline[linecolor=blue,linewidth=1.5pt]{-}(5.5,0.5)(6.5,0.5)
\psline[linecolor=blue,linewidth=1.5pt]{-}(5.5,3.5)(6.5,3.5)
\end{pspicture}
\label{A10}
\ee
The remaining diagrammatic boundary term is evaluated by computing independently the eight configurations and summing over the results to give $\beta s_2(2u)\Gamma(u)\Gamma(-u)$.
The relation thus simplifies to give the right side of (\ref{InvRel}) with $w=0$.

\subsection{Inversion relation for a Robin boundary with an $r$-type seam}
The derivation of the inversion relation for an $r$-type Kac boundary seam is presented in Appendix~C of \cite{PTC}. The extension to the $r$-type Robin boundary seam presents no further difficulties. Setting
\bea
{\cal N}=\mathcal{N}^{(w)}(u,\xi)\,\mathcal{N}^{(w)}(u\!+\!\l,\xi)
\eea
the first step is to insert two faces using  the local inversion relation (\ref{pushinv}) and to push the two faces to the ends. No extra factors arise from the presence of the $r$-type seam
\begin{equation}
\psset{unit=.9cm}
\setlength{\unitlength}{.9cm}
\hspace{-14pt}{\cal N}\Db(u)\Db(u+\l)=\frac{1}{s_1(2u)s_1(-2u)}\ 
\begin{pspicture}[shift=-1.85](0,-.2)(9,4.2)
\multiput(0,0)(0,2){2}{\pspolygon[linewidth=0.8pt,linecolor=black,fillstyle=solid,fillcolor=lightlightblue](0,0)(1,1)(0,2)}
\multiput(0,0)(8,0){2}{\pspolygon[linewidth=0.8pt,linecolor=black,fillstyle=solid,fillcolor=lightlightblue](0,2)(1,1)(2,2)(1,3)}
\multiput(9,0)(0,2){2}{\pspolygon[linewidth=0.8pt,linecolor=black,fillstyle=solid,fillcolor=lightlightblue](0,1)(1,0)(1,2)}
\facegrid{(2,0)}{(8,4)}
\rput(2.5,.5){\small $u$}
\rput(2.5,1.5){\small $\lambda\!+\!u$}
\rput(4.5,.5){\small $u$}
\rput(4.5,1.5){\small $\lambda\!+\!u$}
\rput(3.5,.5){\small ...}
\rput(3.5,1.5){\small ...}
\rput(5.5,.5){\scriptsize$u\!\!-\!\!\xi_{w}$}
\rput(5.5,1.5){\scriptsize$u\!\!-\!\!\xi_{w\!-\!1}$}
\rput(6.5,.5){\small ...}
\rput(6.5,1.5){\small ...}
\rput(7.5,.5){\scriptsize $u\!\!-\!\!\xi_{1}$}
\rput(7.5,1.5){\scriptsize $u\!\!-\!\!\xi_{0}$}
\rput(2.5,2.5){\small $\lambda\!-\!u$}
\rput(2.5,3.5){\small $-u$}
\rput(4.5,2.5){\small $\lambda\!-\!u$}
\rput(4.5,3.5){\small $-u$}
\rput(3.5,2.5){\small ...}
\rput(3.5,3.5){\small ...}
\rput(5.5,2.5){\scriptsize$-\!u\!\!-\!\!\xi_{w\!-\!1}$}
\rput(5.5,3.5){\scriptsize$-\!u\!\!-\!\!\xi_{w}$}
\rput(6.5,2.5){\small ...}
\rput(6.5,3.5){\small ...}
\rput(7.5,2.5){\scriptsize $-\!u\!\!-\!\!\xi_{0}$}
\rput(7.5,3.5){\scriptsize $-\!u\!\!-\!\!\xi_{1}$}
\rput(1,2){\small $2u$}
\rput(9,2){\small $-2u$}
\rput(9.65,1){$_{u}$}
\rput(9.6,3){$_{u+\l}$}
\multirput(2,0)(3,0){2}{\multirput(0,0)(2,0){2}{\multirput(0,0)(0,1){4}{\psarc[linewidth=.5pt,linecolor=red](0,0){.1}{0}{90}}}}
\multiput(0.65,1)(0,2){2}{\psarc[linewidth=1.5pt,linecolor=blue](0,0){.525}{106}{254}}
\multiput(0,0)(7.5,0){2}{\multiput(0,0)(0,3){2}{\psline[linewidth=1.5pt,linecolor=blue](0.5,0.5)(2,0.5)}}
\multiput(1,1)(6.5,0){2}{\multiput(0,0)(0,1){2}{\psline[linewidth=1.5pt,linecolor=blue](0.5,0.5)(1,0.5)}}
\multiput(0,2)(8,0){2}{\psarc[linewidth=.5pt,linecolor=red](0,0){.17}{-45}{45}}
\psline[linewidth=1pt,linecolor=red,linestyle=dashed](5,-0.4)(5,4.4)
\end{pspicture}
\end{equation}
Next, the second projector on the right side of (\ref{projdecomp})  is inserted in the bottom two rows on the boundary between the seam and the bulk. The projector is pushed through the bottom rows of the bulk using (\ref{A6}), reflected at the boundary as in (\ref{A7}) and finally pushed back through the top rows of the bulk using (\ref{A6}) again
\begin{equation}
\psset{unit=.8cm}
\setlength{\unitlength}{.8cm}
{\cal N}\phi(\l-u)\phi(\l+u)=\dfrac{[s_1(u)s_1(-u)]^{2N}s_2(-2u)}{\b s_1(2u)s_1(-2u)}\times\hspace{-.23in}
\begin{pspicture}[shift=-1.85](2.5,0)(12,4)
\multiput(11,0)(10,0){1}{\pspolygon[linewidth=0.8pt,linecolor=black,fillstyle=solid,fillcolor=lightlightblue](-1,2)(0,1)(1,2)(0,3)}
\multiput(11,0)(0,2){2}{\pspolygon[linewidth=0.8pt,linecolor=black,fillstyle=solid,fillcolor=lightlightblue](0,1)(1,0)(1,2)}
\multiput(10,2)(10,0){1}{\psarc[linewidth=.5pt,linecolor=red](0,0){.17}{-45}{45}}
\rput(11,2){\small $-2u$}
\rput(11.65,1){$_{u}$}
\rput(11.6,3){$_{u+\l}$}
\facegrid{(7,0)}{(10,4)}
\multirput(7,0)(2,0){2}{\multirput(0,0)(0,1){4}{\psarc[linewidth=.5pt,linecolor=red](0,0){.1}{0}{90}}}
\rput(7.5,2.5){\scriptsize$-\!u\!\!-\!\!\xi_{w\!-\!1}$}
\rput(7.5,3.5){\scriptsize$-\!u\!\!-\!\!\xi_{w}$}
\rput(8.5,2.5){\small ...}
\rput(8.5,3.5){\small ...}
\rput(9.5,2.5){\scriptsize $-\!u\!\!-\!\!\xi_{0}$}
\rput(9.5,3.5){\scriptsize $-\!u\!\!-\!\!\xi_{1}$}
\rput(7.5,.5){\scriptsize$u\!\!-\!\!\xi_{w}$}
\rput(7.5,1.5){\scriptsize$u\!\!-\!\!\xi_{w\!-\!1}$}
\rput(8.5,.5){\small ...}
\rput(8.5,1.5){\small ...}
\rput(9.5,.5){\scriptsize $u\!\!-\!\!\xi_{1}$}
\rput(9.5,1.5){\scriptsize $u\!\!-\!\!\xi_{0}$}
\psline[linewidth=1.5pt,linecolor=blue](10,0.5)(11.5,0.5)
\psline[linewidth=1.5pt,linecolor=blue](10,1.5)(10.5,1.5)
\psline[linewidth=1.5pt,linecolor=blue](10,2.5)(10.5,2.5)
\psline[linewidth=1.5pt,linecolor=blue](10,3.5)(11.5,3.5)
\psarc[linewidth=1.5pt,linecolor=blue](7,1){0.5}{90}{270}
\psline[linewidth=1.5pt,linecolor=blue](5.5,0)(5.5,4)
\psline[linewidth=1.5pt,linecolor=blue](4.5,0)(4.5,4)
\psline[linewidth=1.5pt,linecolor=blue](3.5,0)(3.5,4)
\psarc[linewidth=1.5pt,linecolor=blue](7,3){0.6}{90}{-90}
\end{pspicture}
\end{equation}
The results from Appendix~C of \cite{PTC} can be borrowed to push the projectors on the boundary through the $r$-type seam. The remaining boundary term corresponds to (\ref{A10}) so the final result is 
\begin{align}
\begin{split}
{\cal N}\phi(\lambda+u)\phi(\lambda-u)&=[s_1(-u)s_1(u)]^{2N}\,\frac{s_2(-2u)s_2(2u)}{s_1(2u)s_1(-2u)}\,\G(u)\G(-u)\\
&\quad\times 
\prod_{j=1}^{w}s_1(u+\xi_j)s_1(-u-\xi_j)s_1(u-\xi_j)s_1(-u+\xi_j)
\end{split}
\label{rTypeInv}
\end{align}
The extra product is the contribution from the $r$-type seam which coincides with the same quantity computed in \cite{PTC}.
Using the identity 
\bea
\frac{\prod_{j=1}^{w}s_1(u\!+\!\xi_j)s_1(-u\!-\!\xi_j)s_1(u\!-\!\xi_j)s_1(-u\!+\!\xi_j)}
{\mathcal{N}^{(w)}(u,\xi)\mathcal{N}^{(w)}(u\!+\!\l,\xi)}
=\dfrac{\sin(\xi\!+\!u)\sin(\xi\!-\!u)\sin(\xi_{w+1}\!+\!u)\sin(\xi_{w+1}\!-\!u)}{\b^2\,\Gamma(0)^2 \sin^2\xi\sin^2\xi_{w+1}}\quad\label{bigId}
\eea
 with $\mathcal{N}^{(w)}(u,\xi)$ defined as in (\ref{normalization})
this reduces to precisely (\ref{InvRel}).

\section{Solution of the Inversion Relation}
\label{AppB}

\subsection{Solution of the inversion relation for the Robin vacuum}
Taking the logarithm of the inversion relations (\ref{eqs_kappa}) gives functional equations that can be solved by Fourier/Lapace transforms.
In this appendix, we solve (\ref{eqs_kappa3}) for the Robin vacuum boundary free energy $\k_R(u,\xi)$. The right side of the inversion relation decomposes as
\begin{equation}
\frac{\G(u)\G(-u)}{\G(0)^2}=\dfrac{\sin(\xi+\l+u)\sin(\xi+\l-u)}{\sin^2(\xi+\l)}\, \dfrac{\sin(\xi+\a+u)\sin(\xi+\a-u)}{\sin^2(\xi+\a)}
\end{equation}
As a result, the Robin vacuum boundary free energy is a sum of two contributions
\begin{equation}\label{f_R}
-f_R(u,\xi)=\log\k_R(u,\xi)=g_{\xi+\l}(u)+g_{\xi+\a}(u)
\end{equation}
where the function $g_x(u)$ is the solution of the fundamental inversion relation
\begin{equation}\label{equ_shift}
g_x(u)+g_x(u+\l)=\log \dfrac{\sin(x+u)\sin(x-u)}{\sin^2x},\qquad g_x(u)=g_x(\l-u)
\end{equation}
In this formulation, the problem is invariant under the translation of the variable $x$ by a period $\pi$. It is also invariant under reversal of the sign of $x$. So, for simplicity, we can assume $x>0$. 

The general solution of the fundamental inversion relation takes the form
\begin{equation}
g_x(u)=\int_{-\infty}^\infty\dfrac{\sinh ut\sinh(\l-u)t\cosh(2x+k_x\pi)t}{t\sinh\pi t\cosh\l t}\,dt
\end{equation}
where the integer $k_x\in{\Bbb Z}$ is chosen to ensure the convergence of the integral in the physical strip \mbox{$0<\Re u<\l$}. The integral converges if $|2x+k_x\pi|<\pi$, leading us to choose $k_x$ such that $\th_x+\pi=2x+(k_x+1)\pi=2x$ mod $2\pi$. 
As a consequence, the function $g_x(u)$ is expressed in terms of the angle $\th_x$ with $\th_x+\pi=2x$ mod $2\pi$ as introduced in (\ref{thx}) 
\begin{equation}\label{sol_g_k}
g_x(u)=\int_{-\infty}^\infty\dfrac{\sinh ut\sinh(\l-u)t\cosh \th_x t}{t\sinh\pi t\cosh\l t}\,dt,\qquad x\in{\Bbb R}
\end{equation}
This solution is $\pi$-periodic and even in the variable $x$. The result (\ref{sol_fR}) for the boundary free energy $f_R(u,\xi)$ is given by (\ref{f_R}) with  $g_x(u)$ given by ({\ref{sol_g_k}).

\subsection{Solution of the inversion relation with an $r$-type seam}

Referring to (\ref{kappaDecomp}) and suppressing the dependence on $\alpha$, we have
\bea
\kappa_R(u,w,\xi)=\kappa_R(u,\xi)\,\kappa_w(u,\xi),\qquad w>0
\eea
From (\ref{InvRel}) and using
\bea
\Gamma(u)=\Gamma(u|\xi,\alpha)=\disp\Gamma(u|\xi_w,\alpha_{-w})\,\frac{s(\xi_1-u)}{s(\xi_{w+1}-u)}
\eea 
the combined inversion relation is
\begin{align}
\kappa_R(u,w,\xi)\kappa_R(u+\lambda,w,\xi)&=\frac{\Gamma(u)\Gamma(-u)}{\Gamma(0)^2}\,\frac{s(\xi_1)^2s(\xi_{w+1}+u)s(\xi_{w+1}-u)}{s(\xi_{w+1})^2s(\xi_1+u)s(\xi_1-u)}
\nonumber\\
&=\frac{\Gamma(u|\xi_w,\alpha_{-w})\Gamma(-u|\xi_w,\alpha_{-w})}{\Gamma(0|\xi_w,\alpha_{-w})^2}
\end{align}

It follows that the inversion relation for $w=0$ is
\bea
\kappa_R(u,0,\xi)\kappa_R(u+\lambda,0,\xi)=\frac{\Gamma(u|\xi,\alpha)\Gamma(-u|\xi,\alpha)}{\Gamma(0|\xi,\alpha)^2}
\eea
We conclude that $\kappa(u,w,\xi)$ satisfies the same inversion relation as $\kappa(u,0,\xi)$ but with the replacements $(\xi,\alpha)\mapsto (\xi_w,\alpha_{-w})$. 
This is in accord with the algebraic equivalence, found in Appendix~\ref{AppC2}, between the $w=0$ and $w\ge 1$ Robin $K$ matrices as representations of the one-boundary TL algebra.

\section{Boundary Algebra}
\label{AppC}

\def\B{{\mathrm B}}
\subsection{Robin boundaries as representations of one-boundary TL algebra}
\label{AppC1}
In this section we show that, for $\b_1=\b_2=1$ and $w\ge 0$, the TL generators $e_j$, $j=1,2,\ldots, N\!-\!1$ supplemented with the identity $I$ and the Robin boundary generator
\bea
F_N(w) =\sum_{k=0}^{w} (-1)^k \frac{c_{2w-2k-1}}{c_1}\,e_N^{(k)},\ \ 
c_k=2\cos\tfrac{k\lambda}{2},\ \  e_N^{(k)}=e_Ne_{N+1}\cdots e_{N+k},\ \  e_{N+w}=f_{N+w}\label{FNw}
\eea
form a representation of the one-boundary TL algebra (\ref{1bdyTL}) acting from the vector space ${\cal V}^{(N,w)}_0$ to itself. 
It is then straightforward to allow for defects. 
The first few Robin boundary generators are
\bea
F_N(0)=f_{N},\qquad F_N(1)=e_{N}-e_{N}f_{N+1},\qquad 
F_N(2)=\frac{c_3}{c_1}\,e_N-e_Ne_{N+1}+e_Ne_{N+1}f_{N+2}
\eea
Specifically, we show that
\bea
e_{N-1}F_N(w) e_{N-1}=\B_1(w) e_{N-1},\qquad F_N(w)^2=\B_2(w) F_N(w)
\eea
where 
\bea
\B_1(w)=\frac{c_{2w-1}}{c_1},\qquad \B_2(w)=\frac{c_{2w+1}}{c_1}
\eea

The first relation follows easily since, in expanding $F_N(w)$ in $e_{N-1} F_N(w) e_{N-1}$, only the first term
\bea
\frac{c_{2w-1}}{c_1}\, e_{N-1}e_Ne_{N-1}=\frac{c_{2w-1}}{c_1}\, e_{N-1}
\eea
survives with $e_N=f_N$ for $w=0$. All of the other terms vanish since the $e_{N-1}$ on the right can be pushed to the left until it sits immediately to the right of $e_N$. Using $e_{N-1}e_Ne_{N-1}=e_{N-1}$ leaves a word that starts with $e_{N-1}e_{N+1}$ or $e_{N-1}f_{N+1}$ and is killed because a half-arc occurs on the lower edge of the boundary seam between $j=N+1$ and $j=N+2$.

Assuming $w\ge 1$ and squaring the expanded $F_N(w)$ gives
\begin{align}
&F_N(w)^2=\big[\frac{c_{2w-1}}{c_1}\,e_N-\frac{c_{2w-3}}{c_1}\,e_Ne_{N+1}\big]^2
+\frac{c_{2w-1}}{c_1}\,e_N\sum_{k=2}^{w} (-1)^k \frac{c_{2w-2k-1}}{c_1}\,e_N^{(k)}\nonumber\\
&\qquad\qquad\qquad\qquad -\frac{c_{2w-3}}{c_1}\,e_Ne_{N+1}\sum_{k=2}^{w} (-1)^{k} \frac{c_{2w-2k-1}}{c_1}\,e_N^{(k)}\\
&=\frac{c_2c_{2w-1}-c_{2w-3}}{c_1^2}\Big[{c_{2w-1}}e_N-{c_{2w-3}}e_Ne_{N+1}
 +\sum_{k=2}^{w} (-1)^{k} c_{2w-2k-1}e_N^{(k)}\Big]
 \,=\,\frac{c_{2w+1}}{c_1}\,F_N(w)\nonumber
\end{align}
where we have used the TL algebra with $e_N^2=c_2e_N$. In the first step, we omitted all the terms that are killed because a half-arc occurs on the lower edge of the boundary seam. In the second step, we only use the TL algebra.

\subsection{Algebraic equivalence of Robin boundary conditions}
\label{AppC2}

In Appendix~\ref{AppC1}, the parameter $\a$ of the Robin boundary was fixed to enforce the equality $\b_1=\b_2=1$ of the boundary loop fugacities. However, useful relations can be derived by exploiting this extra degree of freedom. In this manner, the Robin boundary with an $r$-type seam of width $w$ can be algebraically related to a vacuum ($w=0$) Robin boundary with shifted parameters $\xi_w=\xi+w\l$, $\a_{-w}=\a-w\l$. This algebraic relation holds at the level of transfer matrices considered as elements of the (linear or planar) one-boundary TL  algebra. 
Note that, since the two transfer matrices act on different spaces of link states, this relation cannot be used to deduce the conformal dimensions of the $(r,s)$ Robin boundaries. On the other hand, the boundary energy is a common factor of the eigenvalues in every representation and is determined by algebraic relations. As such, it is constrained to satisfy the algebraic relation between Robin boundaries.

To show the algebraic equivalence of an $r$-type seam of width $w\ge 1$ and parameters $(\xi,\a)$ to a vacuum Robin boundary with parameters $(\xi_w,\alpha_{-w})=(\xi+w\l,\a-w\l)$ we consider the boundary $K$ matrix as given in (5.12) of \cite{PRT14} and show that
\bea
\psset{unit=1.2cm}
\qquad\begin{pspicture}[shift=-0.89](1,0)(6.2,2)
\facegrid{(1,0)}{(5,2)}
\pspolygon[fillstyle=solid,fillcolor=lightlightblue](5,1)(6,2)(6,0)(5,1)
\psarc[linewidth=0.02,linecolor=red](1,0){0.12}{0}{90}
\psarc[linewidth=0.02,linecolor=red](1,1){0.12}{0}{90}
\psarc[linewidth=0.02,linecolor=red](3,0){0.12}{0}{90}
\psarc[linewidth=0.02,linecolor=red](3,1){0.12}{0}{90}
\psarc[linewidth=0.02,linecolor=red](4,0){0.12}{0}{90}
\psarc[linewidth=0.02,linecolor=red](4,1){0.12}{0}{90}
\rput(1.55,0.5){$_{u-\xi_{w}}$}
\rput(3.55,0.5){$_{u-\xi_2}$}
\rput(4.55,0.5){$_{u-\xi_1}$}
\rput(1.5,1.5){$_{-\!u\!-\!\xi_{w\!-\!1}}$}
\rput(3.5,1.5){$_{-u-\xi_1}$}
\rput(4.5,1.5){$_{-u-\xi_0}$}
\rput(2.5,0.5){$\ldots$}
\rput(2.5,1.5){$\ldots$}
\rput(5.55,1){$_{u,\xi,\a}$}
\psline[linecolor=blue,linewidth=1.5pt](0.5,0.5)(1,0.5)
\psline[linecolor=blue,linewidth=1.5pt](0.5,1.5)(1,1.5)
\psline[linecolor=blue,linewidth=1.5pt](5,0.5)(5.5,0.5)
\psline[linecolor=blue,linewidth=1.5pt](5,1.5)(5.5,1.5)
\end{pspicture}
\;=\sum_{k=0}^{w+1} \alpha_k^{(w)} e_N^{(k-1)}
\cong\;\eta^{(w)}(u,\xi)\,\dfrac{s(\xi_1-u)}{s(\xi_{w+1}-u)}\ \ 
\begin{pspicture}[shift=-0.89](0,0)(1,2)
\pspolygon[fillstyle=solid,fillcolor=lightlightblue](0,1)(1.1,2)(1.1,0)(0,1)
\rput(0.6,1){$_{u,\xi_w,\a_{-w}}$}
\psline[linecolor=blue,linewidth=1.5pt](0,0.5)(0.55,0.5)
\psline[linecolor=blue,linewidth=1.5pt](0,1.5)(0.55,1.5)
\end{pspicture}\quad
\label{uxi}
\eea
where $e_N^{(-1)}=I$, $e_N^{(k)}$ is as in (\ref{FNw}) and the $\a$-independent factor $\eta^{(w)}(u,\xi)$ is given by (\ref{expr_eta}). 
The symbol $\cong$ indicates that the expressions are algebraically equivalent but act on different states. 
Setting $\Gamma(u)=\Gamma(u|\xi,\alpha)=R s(\xi_1-u)s_0(\alpha+\xi+u)$, $\beta_1=R s(\alpha)$, $\beta_2=Rs_{-1}(\alpha)$ and simplifying (5.12) of \cite{PRT14}, the coefficients $\alpha_k^{(w)}$ are given by
\bea
\frac{\alpha_k^{(w)}}{\eta^{(w)}(u,\xi)}=\begin{cases}
\Gamma(u|\xi,\alpha)=\disp\Gamma(u|\xi_w,\alpha_{-w})\,\frac{s(\xi_1-u)}{s(\xi_{w+1}-u)},\quad&k=0\\[8pt]
(-1)^{k-1} R s_{k-1-w}(\alpha)\,\disp\frac{s(2u)s(\xi_1-u)}{s(\xi_{w+1}-u)},\quad&k=1,2,\ldots,w\\[8pt]
(-1)^w\,\disp\frac{s(2u)s(\xi_1-u)}{s(\xi_{w+1}-u)},\quad&k=w+1
\end{cases}
\eea
It follows that 
\bea
\frac{1}{\eta^{(w)}(u,\xi)}\sum_{k=0}^{w+1} \alpha_k^{(w)} e_N^{(k-1)}=\frac{s(\xi_1-u)}{s(\xi_{w+1}-u)}\,\big[\Gamma(u|\xi_w,\alpha_{-w}) I+s(2u)F_N(w)\big]
\eea
where
\bea
F_N(w)=R \sum_{k=0}^{w-1} (-1)^k {s_{k-w}(\alpha)} e_N^{(k)}+(-1)^w e_N^{(w)}
\eea

The TL generators $e_j$, $j=1,2,\ldots, N\!-\!1$ supplemented with the identity $I$ and the Robin boundary generator $F_N(w)$ form a representation of the one-boundary TL algebra (\ref{1bdyTL}) acting from the vector space ${\cal V}^{(N,w)}_0$ to itself. The first few Robin boundary generators are
\bea
\begin{array}{c}
F_N(0)=f_{N},\quad F_N(1)=\beta_2 e_{N}\!-\!e_{N}f_{N+1},\quad 
F_N(2)=(\beta\beta_2\!-\!\beta_1)\,e_N\!-\!\beta_2 e_Ne_{N+1}\!+\!e_Ne_{N+1}f_{N+2}\\[6pt]
F_N(3)=(\beta^2\beta_2\!-\!\beta_2\!-\!\beta\beta_1)e_N\!-\!(\beta\beta_2\!-\!\beta_1)\,e_Ne_{N+1}\!+\!\beta_2 e_Ne_{N+1}e_{N+2}\!-\!e_Ne_{N+1}e_{N+2}f_{N+3}
\end{array}
\eea
In particular, following the arguments of Appendix~\ref{AppC1}, it can be shown that
\bea
e_{N-1}F_N(w) e_{N-1}=\B_1(w) e_{N-1},\qquad F_N(w)^2=\B_2(w) F_N(w)
\eea
where 
\bea
\B_1(w)=Rs(\alpha_{-w}),\qquad \B_2(w)=Rs_{-1}(\alpha_{-w})
\eea
are related to the expressions for $\beta_1,\beta_2$ by replacing $\alpha$ with $\alpha_{-w}$. Indeed, setting $\beta_1=\beta_2=1$, or equivalently $R=2\sin\tfrac{\lambda}{2}$ and $\alpha=\tfrac{\lambda+\pi}{2}$, the results of this section reduce to those of Appendix~\ref{AppC1}.


\end{document}